\documentclass[draftclsnofoot, onecolumn]{IEEEtran}
\IEEEoverridecommandlockouts
\usepackage{cite}
\usepackage{amsmath,amssymb,amsfonts}
\usepackage{mathtools}
\usepackage{algorithmic}
\usepackage{graphicx}
\usepackage{textcomp}
\usepackage{xcolor}
\usepackage{cite}
\usepackage{bm}
\usepackage{amsthm}
\usepackage{booktabs}
\usepackage{multirow}
\usepackage{makecell}
\usepackage{subcaption}
\usepackage{colortbl}
\usepackage[colorlinks,linkcolor=red!50!yellow, citecolor=magenta!60!black, urlcolor=pink]{hyperref}
\usepackage[capitalize,noabbrev]{cleveref}
\usepackage[normalem]{ulem}
\newtheorem{theorem}{Theorem}
\newtheorem{lemma}{Lemma}
\newtheorem{proposition}{Proposition}
\newtheorem{corollary}{Corollary}
\theoremstyle{definition}

\def\BibTeX{{\rm B\kern-.05em{\sc i\kern-.025em b}\kern-.08em
    T\kern-.1667em\lower.7ex\hbox{E}\kern-.125emX}}
\begin{document}

\title{
A Gaussian Perspective for Distributional Discrepancy in Generative Diffusion Models
}
\author{Qiang Sun, H. Vincent Poor, Wenyi Zhang\thanks{Q. Sun and W. Zhang are with Department of Electronic Engineering and Information Science, University of Science and Technology of China, Hefei, China (qiangsun@mail.ustc.edu.cn, wenyizha@ustc.edu.cn), and H. V. Poor is with Department of Electrical Engineering, Princeton University, Princeton, NJ, USA (poor@princeton.edu).}}

\maketitle
\begin{abstract}
This paper introduces an analytical approach to quantifying and optimizing the distributional discrepancy in generative diffusion models. For a multivariate Gaussian source, we explicitly derive the closed-form evolution trajectory and the resulting Kullback-Leibler (KL) divergence between the distributions of the source data and the reversely sampled data. Asymptotic analysis via the Euler-Maclaurin expansion characterizes the convergence behavior of this KL divergence, extracting its dominant term as an explicit functional of the noise schedule. Minimizing this dominant term via the calculus of variations yields a noise schedule described by a tangent law, inherently determined by the source covariance spectrum. We further prove that the Gaussian source exhibits an extremal property for the KL divergence among general source distributions with a given covariance. We also utilize the analytical KL divergence as a principled metric to identify efficient time discretization strategies for pretrained diffusion models, and demonstrate via experiments over diverse datasets that the identified strategies consistently outperform established baselines, particularly under constrained function evaluation budgets.
\end{abstract}


\section{Introduction}

Diffusion models (DMs)~\cite{sohl2015deep, song2019generative, ho2020denoising, song2020score, kingma2021variational} have emerged as a prevailing paradigm in generative modeling, achieving remarkable success across a wide range of applications, including high-fidelity image~\cite{rombach2022high, hoogeboom2025simpler}, audio~\cite{kong2020diffwave, liu2023audioldm}, and video synthesis~\cite{ho2022video, blattmann2023stable}, as well as planning~\cite{janner2022planning, chi2025diffusion}, solving inverse problems~\cite{kawar2022denoising, chung2022diffusion}, and scientific modeling~\cite{watson2023novo, abramson2024accurate}. 
Furthermore, theoretical foundations underlying the standard data-to-noise diffusion trajectory have inspired related paradigms, such as flow matching~\cite{liu2022letus, lipman2023flow} and Schr\"{o}dinger bridge\cite{de2021schdiffusion, zhou2023denoising} methods, which generalize the framework by extending the standard trajectory to more flexible transport dynamics between two general probability distributions.

The fundamental mechanism of a standard DM intrinsically consists of two processes: the forward process and the reverse sampling process~\cite{ho2020denoising}, and a critical component underpinning both processes is the noise schedule, which dictates the precise pacing at which information is progressively destroyed and subsequently recovered. The forward process gradually corrupts data samples from the source distribution into tractable standard Gaussian noise, explicitly governed by a predefined noise schedule. 
Substantial efforts have been devoted to identifying effective noise schedule design choices for training a superior denoising estimator, ranging from handcrafted engineering heuristics~\cite{nichol2021improved, kingma2021variational, karras2022elucidating, kingma2023understanding, jabri2023sca, chen2023imp, hoogeboom2023simple, hang2025improved} to more principled theoretical analyses, such as those based on Fisher information~\cite{santos2023using, zhang2025cosine}. Although effective training configurations have been explored, there remains a lack of explicit characterization of how noise schedules impact the discrepancy between the true source distribution and the one produced by reverse sampling, leaving this fundamental relationship largely unexplored.

The reverse sampling process aims to recover the source distribution from a noise prior by numerically solving a reverse-time stochastic differential equation (SDE)~\cite{song2020score} or its equivalent probability flow ordinary differential equation (ODE)~\cite{song2020score, song2020denoising} using a learned denoising estimator. Unlike earlier generative paradigms~\cite{kingma2013auto, goodfellow2014generative}, DMs achieve exceptional generation quality at the expense of a multi-step iterative sampling process. This process requires repeatedly evaluating a large estimator network to solve corresponding differential equations, with the cost typically quantified by the number of function evaluations (NFEs). Consequently, in practical deployment of pretrained DMs, the time discretization strategy, specifically, the allocation of time steps during the reverse sampling process, plays a pivotal role. Under a stringent budget of NFEs, optimizing this discrete grid can substantially improve generation quality~\cite{lu2022dpm, karras2022elucidating}. While a growing body of literature has sought to optimize the time discretization strategy~\cite{wang2023learning,li2023autodiffusion,sabour2024align,xue2024accelerating, williams2024score, chen2024trajectory, xu2025variance, zhu2025hierarchical, benita2025spectral,lee2024beta,tong2025ltd,min2025b}, the majority of extant methods rely on auxiliary training phases or searching procedures to find efficient discretization
strategies, limiting their plug-and-play applicability.

In this paper, to understand the interplay between noise scheduling and sampling efficiency from first principles, we develop our approach based on the Gaussian setting in which the source distribution is multivariate Gaussian with a given covariance matrix. This formulation leads to a linear optimal posterior estimator, and the reverse sampling process can be analyzed step by step, thereby permitting an explicit closed-form characterization of the generative distribution's evolution trajectory. Leveraging this analytical tractability, we investigate the Kullback-Leibler (KL) divergence between the source distribution and the one produced by reverse sampling, as a distributional discrepancy metric.

Rather than being a reductive simplification, the Gaussian setting serves as a powerful analytical testbed that has yielded substantial insights across several aspects of DMs, ranging from error analysis~\cite{pierret2024diffusion,hurault2026} and stopping-time analysis~\cite{wu2025optimal} to practical advances such as accelerated sampling~\cite{sabour2024align,wang2024unreasonable,benita2025spectral}. Moreover, some works suggest that Gaussian analyses can reveal fundamental mechanisms underlying DMs, including their generalization behavior and implicit inductive biases~\cite{li2024generalizability}. Taken together, these findings establish the Gaussian setting as a useful paradigm for probing the inner workings of practical DMs, thereby motivating  a deeper investigation of Gaussian source analysis in this work.

Building upon this analytical KL divergence metric of the Gaussian setting, we apply the Euler-Maclaurin expansion~\cite{zwillinger2002crc} to characterize its asymptotic convergence rate as the number of reverse sampling steps increases, ultimately isolating the dominant term of the KL divergence. Meanwhile, the analytical tractability of the Gaussian setting permits extending the asymptotic properties to the Wasserstein distance, thereby implying a direct functional link between the noise schedule and the generative quality. To  minimize the dominant KL divergence term, we formulate the noise schedule design as a variational problem. Since the conventional scalar noise schedule couples the contributions of all covariance eigenmodes, we first consider an anisotropic relaxation, where each eigenmode is allowed to follow its own noise trajectory. This relaxation decouples the variational problem and leads to an exact eigenmode-specific tangent law. Inspired by this solution, we further propose a global cross-eigenmode tangent law, whose optimal coefficient is uniquely determined by the covariance spectrum of the source distribution. To validate the effectiveness of the tangent law, we conduct numerical experiments for Gaussian and Gaussian mixture model (GMM) sources, for which the optimal score estimator admits a closed-form expression~\cite{wang2024unreasonable}, allowing us to isolate the effect of noise scheduling from network approximation error. In our experimental results, the tangent law noise schedule consistently outperforms widely adopted baselines~\cite{ho2020denoising, chen2023imp, nichol2021improved}.

To connect the insights gleaned under the Gaussian setting with DMs under arbitrary source distributions, we establish an extremal property of the Gaussian setting; that is, for sufficiently fine time discretizations, the Gaussian setting achieves the lowest KL divergence between the source distribution and the one produced by reverse sampling, under optimal denoising. This result is built upon reformulating the DM in a continuous-time SDE framework, which is equivalent to the discrete-time ODE framework in terms of the transition probability law. The reverse-time SDE is further modified by a piecewise approximation in order to mirror the time discretization effect. Inspecting the KL divergence between the path measures induced by the exact reverse-time SDE and its piecewise counterpart, under the SDE framework, we characterize its dominant term leveraging standard tools in stochastic calculus such as Girsanov’s theorem and It\^o’s lemma~\cite{oksendal2003stochastic}. The extremal property of the Gaussian setting is thus established by applying information-theoretic arguments including the maximum entropy principle and the multivariate Cram\'er-Rao bound (CRB) to the dominant KL divergence term.

Finally, we consider the practically relevant scenario of deploying pretrained DMs and repurpose the analytical KL divergence as a metric to evaluate arbitrary time discretization strategies. This provides a principled and computationally lightweight criterion for selecting discrete grids, bypassing the overhead of auxiliary retraining or search procedures. Guided by our analytical KL divergence, we obtain improved time discretization strategies and evaluate them on CIFAR-10~\cite{krizhevsky2009learning}, FFHQ-64~\cite{karras2019style}, and AFHQv2~\cite{Choi2020CVPR} datasets under various model configurations~\cite{karras2022elucidating, song2020score}. Experimental results confirm that our selected strategy consistently outperforms established baselines and state-of-the-art search-based methods, particularly when the budget on NFEs is stringent.

The remaining part of this paper is organized as follows. 
In \cref{sec:pre}, we establish the mathematical foundations of DMs and elucidate several essential properties of the Gaussian source. Building upon these, \cref{sec:mainKL} presents a comprehensive analysis of the terminal KL divergence under the Gaussian source setting. Subsequently, we extend these findings to general source distributions and establish the extremal property in \cref{sec:mainLB}. Numerical validation and the practical utility of our framework across diverse source distributions are explored in \cref{sec:exp}. Finally, \cref{sec:conclu} concludes the paper.

\textit{Notation:} Throughout this paper, $\log(\cdot)$ denotes the natural logarithm. For time-dependent functions, we omit the explicit temporal argument when there is no ambiguity, writing $f_t$ for $f(t)$ and $f_{t_j}$ for its value at the discretization point $t_j$. First- and second-order time derivatives are denoted by $\dot f \coloneqq \mathrm{d}f/\mathrm{d}t$ and $\ddot f \coloneqq \mathrm{d}^2 f/\mathrm{d}t^2$, respectively. For a vector field $\boldsymbol{\phi}$, $\mathbf{J}_{\boldsymbol{\phi}}$ denotes its Jacobian matrix. We use $\|\cdot\|^2$ for the squared Euclidean norm of a vector, and $\|\cdot\|_F$ for the Frobenius norm of a matrix. The relation $A \succeq B$ denotes the standard L\"owner partial order, meaning that $A-B$ is positive semidefinite. A multivariate Gaussian distribution with mean $\boldsymbol{m}$ and covariance $\boldsymbol{\Sigma}$ is denoted by $\mathcal{N}(\boldsymbol{m},\boldsymbol{\Sigma})$, while $\mathcal{N}(\mathbf{x};\boldsymbol{m},\boldsymbol{\Sigma})$ denotes the corresponding probability density evaluated at $\mathbf{x}$. 
Path measures induced by stochastic processes are denoted by blackboard symbols such as $\mathbb{P}$. Quantities accented with a hat, such as $\hat{\mathbf{x}}$, $\hat{p}$, and $\hat{\mathbb{P}}$, refer to objects associated with the reverse sampling process. Finally, the subscript $(\cdot)_G$ is reserved for quantities evaluated under the Gaussian setting.

\section{Preliminaries} \label{sec:pre}
This section briefly reviews the standard DM paradigm, detailing the forward diffusion process, noise scheduling, and the corresponding reverse sampling dynamics. Subsequently, we introduce the multivariate Gaussian setup and its basic properties. This setup serves as the primary analytical framework in this paper, enabling tractable closed-form derivations.

\subsection{Basics of Diffusion Models} \label{sub:pre_dms}
The fundamental mechanism underlying a standard DM consists of two phases: a forward diffusion process for training the denoising estimator, and a reverse sampling process for iteratively recovering data from noise. The diffusion mechanism is explicitly driven by the forward process, which applies the following Gaussian perturbation~\cite{ho2020denoising, kingma2021variational}
\begin{equation} \label{eq:forward}
    \mathbf{x}_t = \alpha_{t} \mathbf{x}_{0} + \sigma_{t} \boldsymbol{\epsilon},
    \quad \boldsymbol{\epsilon} \sim \mathcal{N}(\mathbf{0}, \mathbf{I}),
\end{equation}
where $\mathbf{x}_0 \in \mathbb{R}^L$ denotes a source data sample drawn from the distribution $q_0$, $t \in [0, 1]$ denotes the normalized diffusion time, and the nonnegative scalar functions $\alpha_t$ and $\sigma_t$ parameterize the noise schedule. 
Consequently, the forward transition kernel is given by
\begin{equation} \label{eq:q_forwardtransit}
    q_{t|0}(\mathbf{x}_t | \mathbf{x}_0) = \mathcal{N}(\mathbf{x}_t; \alpha_t \mathbf{x}_0, \sigma_t^2 \mathbf{I}).
\end{equation}
The noise schedule is explicitly designed to make the signal-to-noise ratio (SNR) $\mathrm{SNR}_t = \alpha_t^2 / \sigma_t^2$ monotonically decrease with respect to $t$. Ideally, the continuous noise schedule satisfies the boundary conditions of $\lim_{t\to0} \mathrm{SNR}_t = \infty$ (pure data) and $\lim_{t\to1} \mathrm{SNR}_t = 0$ (pure noise). In practical implementations, the two most prevalent types of noise schedule are variance-preserving (VP) and variance-exploding (VE)~\cite{song2020score}. A VP schedule enforces $\alpha_t^2 + \sigma_t^2 = 1$, whereas a VE schedule fixes $\alpha_t \equiv 1$ and progressively increases $\sigma_t$ to a predefined maximum value $\sigma_{\max} \gg 1$. These practical designs ensure that the forward diffusion process approximately diffuses the source distribution into a tractable Gaussian prior $p_{\mathrm{prior}}$ (i.e., $q_1 \approx p_{\mathrm{prior}}$).

To invert the forward diffusion process for generative tasks, DMs employ a time-conditioned neural network, $\boldsymbol{\epsilon}_{\theta}(\mathbf{x}_t, t)$, to predict the injected noise. The network is trained by minimizing the mean squared error (MSE) objective
\begin{equation} \label{eq:dmloss}
    \mathcal{L}_{\theta} = \mathbb{E}_{t, \mathbf{x}_0 \sim q_0, \boldsymbol{\epsilon} \sim \mathcal{N}(\mathbf{0}, \mathbf{I})} \left[ \| \boldsymbol{\epsilon} - \boldsymbol{\epsilon}_{\theta}(\alpha_{t} \mathbf{x}_{0} + \sigma_{t} \boldsymbol{\epsilon}, t) \|^2 \right].
\end{equation}
This objective implies that the optimal denoising estimator is the conditional expectation, i.e., $\boldsymbol{\epsilon}^{*}_{\theta}(\mathbf{x}_t,t)=\mathbb{E}[\boldsymbol{\epsilon}|\mathbf{x}_t]$. Inverting the linear forward relation in~\eqref{eq:forward} demonstrates that learning to predict the noise $\boldsymbol{\epsilon}$ is equivalent to estimating the posterior mean of the source data $\mathbf{x}_0$:
\begin{equation} \label{eq:optepi} 
    \boldsymbol{\epsilon}^{*}_{\theta}(\mathbf{x}_t, t) = \mathbb{E}\left[ \frac{\mathbf{x}_t - \alpha_t \mathbf{x}_0}{\sigma_t} \middle| \mathbf{x}_t \right] = \frac{\mathbf{x}_t - \alpha_t \mathbb{E}[\mathbf{x}_0 | \mathbf{x}_t]}{\sigma_t}. 
\end{equation} 

Once the noise predictor is trained, the reverse sampling process proceeds by iteratively denoising an initial state drawn from the prior distribution $p_{\mathrm{prior}}$.
Deterministic reverse sampling is often adopted to reduce stochastic variance and improve computational efficiency~\cite{song2020denoising,lu2022dpm}.
Specifically, given a monotonically decreasing time discretization sequence $\{t_i\}_{i=0}^N$ from $t_N=1$ to $t_0=0$, 
the deterministic reverse sampling process evolves according to the following rule for $j = N, N-1, \dots, 1$~\cite{song2020denoising,lu2022dpm}:
\begin{equation} \label{eq:ddim}
    \hat{\mathbf{x}}_{t_{j-1}} = \frac{\alpha_{t_{j-1}}}{\alpha_{t_{j}}} \hat{\mathbf{x}}_{t_j} + \left( \sigma_{t_{j-1}} - \frac{\alpha_{t_{j-1}}}{\alpha_{t_{j}}} \sigma_{t_j} \right) \boldsymbol{\epsilon}_{\theta}(\hat{\mathbf{x}}_{t_j}, t_j).
\end{equation}
Under the forward transition kernel in \eqref{eq:q_forwardtransit}, the reverse sampling process is initialized at the terminal time $t_N$ with $\hat{\mathbf{x}}_{t_N} \sim p_{\mathrm{prior}} = \mathcal{N}(0, c^2 \mathbf{I})$ 
where $c = 1$ for the VP setting and $c = \sigma_{\max}$ for the VE setting.

Inevitably, the $N$-step discretization in \eqref{eq:ddim} introduces an accumulated error from the prior, causing the generated distribution at $t_0$ to deviate from the true source distribution. Motivated by this issue, we analyze the dynamics of the reverse sampling process in \cref{sec:mainKL} and use the KL divergence to characterize the discretization-induced discrepancy between the source and generated distributions, thereby providing a principled basis for noise schedule design.

We further extend our analysis to an anisotropic noise schedule~\cite{sahoo2024diffusion}, which enables the flexible assignment of distinct noise levels across different components. Specifically, we replace the scalar schedule coefficients with simultaneously diagonalizable matrices $\boldsymbol{\alpha}_t$ and $\boldsymbol{\sigma}_t$ in a time-independent orthonormal basis.
Under this formulation, the forward diffusion process and the corresponding deterministic reverse sampling process naturally generalize as follows:
\begin{align} 
    \mathbf{x}_t &= \boldsymbol{\alpha}_{t} \mathbf{x}_{0} + \boldsymbol{\sigma}_{t} \boldsymbol{\epsilon}, \quad \boldsymbol{\epsilon} \sim \mathcal{N}(\mathbf{0}, \mathbf{I}), \label{eq:ani_forward} \\
    \hat{\mathbf{x}}_{t_{j-1}} &= \boldsymbol{\alpha}_{t_{j-1}} \boldsymbol{\alpha}_{t_j}^{-1} \hat{\mathbf{x}}_{t_j} + \left( \boldsymbol{\sigma}_{t_{j-1}} - \boldsymbol{\alpha}_{t_{j-1}} \boldsymbol{\alpha}_{t_j}^{-1} \boldsymbol{\sigma}_{t_j} \right) \boldsymbol{\epsilon}_{\theta}(\hat{\mathbf{x}}_{t_j}, t_j).
    \label{eq:ani_ddim}
\end{align}
This anisotropic formulation provides a natural extension of the scalar noise schedule design problem. It enables an eigenmode-wise analysis of the resulting generative discrepancy, as developed in \cref{sec:mainKL}.

\subsection{Stochastic Differential Equation Formulation} \label{sub:pre_sde}
In this subsection, we introduce the equivalent continuous-time SDE formulation for DMs, encompassing both the exact form and its piecewise approximation. These forms will be utilized to derive the theoretical lower bound for general source distributions in \cref{sec:mainLB}. 

First, the discrete-time forward evolution that yields the marginal distributions in~\eqref{eq:q_forwardtransit} can be equivalently described by the following linear continuous-time SDE~\cite{song2020score, kingma2021variational}:
\begin{equation} \label{eq:sde}
    \mathrm{d} \mathbf{x}_t = f_t \mathbf{x}_t \mathrm{d}t + g_t \mathrm{d} \mathbf{w}_t, \quad \mathbf{x}_0 \sim q_0(\mathbf{x}_0),
\end{equation}
where $\mathbf{w}_t$ is the standard Wiener process. The drift coefficient $f_t$ and diffusion coefficient $g_t$ are explicitly determined by the corresponding noise schedule components $\alpha_t$ and $\sigma_t$ via~\cite{kingma2021variational}
\begin{equation} \label{eq:sde_coeffs}
    f_t = \frac{\dot{\alpha}_t}{\alpha_t}, \quad g_t = \sqrt{2 \sigma_t \left( \dot{\sigma}_t - \frac{\dot{\alpha}_t}{\alpha_t}\sigma_t \right)}.
\end{equation}
According to the classical theory of stochastic processes~\cite{anderson1982reverse}, the corresponding reverse continuous-time SDE flowing from $t=1$ to $0$ is given by
 \begin{equation} \label{eq:sde_reverse}
    \mathrm{d}\mathbf{x}_t = \left[ f_t \mathbf{x}_t - g_t^2 \nabla_{\mathbf{x}_t} \log q_t(\mathbf{x}_t) \right]\mathrm{d}t + g_t \mathrm{d} \bar{\mathbf{w}}_t,
\end{equation} 
where $\nabla_{\mathbf{x}_t} \log q_t(\mathbf{x}_t)$ is the score function and $\bar{\mathbf{w}}_t$ is a standard Wiener process in reverse time.
When initialized at the terminal distribution $\mathbf{x}_1 \sim q_1(\mathbf{x}_1)$, this continuous-time process perfectly reverses the forward diffusion trajectory, enabling exact sampling from the source distribution $q_0(\mathbf{x}_0)$. Notably, the exact reverse SDE in~\eqref{eq:sde_reverse} shares the same marginal distributions as the probability flow ODE, which directly yields the deterministic numerical scheme presented in~\eqref{eq:ddim}.

To render the reverse SDE tractable, the unknown score function $\nabla_{\mathbf{x}_t} \log q_t(\mathbf{x}_t)$ in~\eqref{eq:sde_reverse} must be empirically estimated. By Tweedie's formula~\cite{efron2011tweedie}, the optimal denoising estimator \eqref{eq:optepi} introduced in \cref{sub:pre_dms} is equivalent to the scaled true score. This equivalence allows us to substitute the intractable score with our denoising estimator approximated by the time-conditioned neural network minimizing $\mathcal{L}_\theta$ \eqref{eq:dmloss}, yielding $\nabla_{\mathbf{x}_t} \log q_t(\mathbf{x}_t) \approx - \boldsymbol{\epsilon}_{\theta}(\mathbf{x}_t, t) / \sigma_t$. Substituting this approximation into~\eqref{eq:sde_reverse}, we obtain a tractable reverse SDE.

However, even with an estimated score function, integrating the reverse SDE requires continuous-time evaluation of a large neural denoising network, which is computationally infeasible. In implementation, a piecewise constant approximation of the score function on a discrete time grid $\{t_i\}_{i=0}^N$ is adopted, leading to the following piecewise approximate SDE~\cite{chen2023sampling, liu2022letus}
\begin{equation} \label{eq:approx_sde}
    \mathrm{d}\hat{\mathbf{x}}_t = \left[ f_t \hat{\mathbf{x}}_t - h_t \boldsymbol{\epsilon}_{\theta}(\hat{\mathbf{x}}_{t_j},t_j) \right]\mathrm{d}t + g_t \mathrm{d} \bar{\mathbf{w}}_t,
\end{equation}
where we introduce $h_t = g_t^2 / (-\sigma_t)$ according to Tweedie's formula.
In other words, over the integration interval $t \in (t_{j-1}, t_{j}]$, the continuous score component is approximated by freezing the neural estimator at the right endpoint $t_j$. This significantly reduces the required NFEs.

These well-defined continuous-time and piecewise approximate reverse SDE forms provide the necessary foundation for deriving the lower bound of discretization error in \cref{sec:mainLB}.

\subsection{Gaussian Source and Its Basic Properties} \label{sec:pre_gauss}
For an arbitrary source distribution $q_0(\mathbf{x}_0)$,\footnote{Without loss of generality, we assume that all sources have zero mean vector.} we introduce a reference multivariate Gaussian source distribution $q_{G,0}(\mathbf{x}_0) = \mathcal{N}(\mathbf{0}, \boldsymbol{\Sigma}_{\mathbf{x}})$ with covariance matching that of $q_0$. To avoid unnecessary technicalities, we assume that the covariance matrix $\boldsymbol{\Sigma}_{\mathbf{x}} \in \mathbb{R}^{L \times L}$ is positive definite. It admits the eigen-decomposition $\boldsymbol{\Sigma}_{\mathbf{x}} = \mathbf{U} \mathbf{\Lambda} \mathbf{U}^\top$, where $\mathbf{U}$ is an orthogonal matrix and $\mathbf{\Lambda} = \text{diag}(\mu_1, \dots, \mu_L)$ contains strictly positive eigenvalues $\mu_i > 0$ for $i = 1, \ldots, L$.

Owing to its analytical tractability, the Gaussian source $q_{G, 0}$ yields the following exact closed-form properties, which form the basis of our derivations. Under the linear forward diffusion process in~\eqref{eq:forward}, the marginal distribution at any intermediate time $t$ remains Gaussian, given by 
\begin{equation} \label{eq:forward_prob}
    q_{G, t}(\mathbf{x}_t) = \mathcal{N}(\mathbf{0}, \boldsymbol{\Sigma}_t), \quad \boldsymbol{\Sigma}_t \coloneqq \alpha_t^2 \boldsymbol{\Sigma}_{\mathbf{x}} + \sigma_t^2 \mathbf{I}.
\end{equation} 
Moreover, for the Gaussian source, the optimal posterior estimator, i.e., the minimum mean squared error (MMSE) estimator, admits a linear closed-form solution~\cite{poor2013introduction}
\begin{equation} \label{eq:mmse}
    \mathbb{E}[\mathbf{x}_0|\mathbf{x}_t] = \alpha_t \boldsymbol{\Sigma}_{\mathbf{x}} \boldsymbol{\Sigma}_t^{-1} \mathbf{x}_t = \alpha_t \boldsymbol{\Sigma}_{\mathbf{x}} (\alpha_t^2 \boldsymbol{\Sigma}_{\mathbf{x}} + \sigma_t^2 \mathbf{I})^{-1} \mathbf{x}_t.
\end{equation}
Furthermore, the corresponding Hessian matrix and Fisher information matrix (FIM) also admit exact closed forms
\begin{eqnarray}
    \nabla_{\mathbf{x}_t}^2 \log q_{G, t}(\mathbf{x}_t) &=& - \boldsymbol{\Sigma}_t^{-1}, \label{eq:gauss_hessian}\\
    \mathcal{I}(q_{G, t}) &=& \mathbb{E} \left[ \left( \nabla_{\mathbf{x}_t} \log q_{G, t}(\mathbf{x}_t) \right) \left( \nabla_{\mathbf{x}_t} \log q_{G, t}(\mathbf{x}_t) \right)^\top \right] = \boldsymbol{\Sigma}_t^{-1}. \label{eq:gauss_fim}
\end{eqnarray}

Notably, if an arbitrary source distribution shares the same initial covariance matrix $\boldsymbol{\Sigma}_{\mathbf{x}}$, the linear nature of the forward process explicitly preserves this covariance matching property. That is, the covariance matrix of the marginal $q_t(\mathbf{x}_t)$ is still $\mathrm{Cov}[\mathbf{x}_t] = \boldsymbol{\Sigma}_t$. Consequently, this covariance matching property guarantees two crucial information-theoretic consequences: the maximum entropy principle and the CRB~\cite{cover1999elements}, which will serve as the cornerstone for our lower bound derivation for arbitrary source distributions in \cref{sec:mainLB}. Specifically, the Gaussian distribution maximizes the differential entropy subject to a given covariance constraint, while the CRB dictates that the FIM is bounded below by the inverse of the covariance matrix. These extremal properties are formally expressed as
\begin{eqnarray}
    h(q_t) &\leq& h(q_{G, t}), \label{eq:max_entropy}\\
    \mathcal{I}(q_t) &\succeq& \mathcal{I}(q_{G, t}) = \boldsymbol{\Sigma}_t^{-1}, \label{eq:crb}
\end{eqnarray}
where $h(q_t)\coloneqq - \int q_t(\mathbf{x}_t) \log q_t(\mathbf{x}_t) \mathrm{d}\mathbf{x}_t$ defines the differential entropy, and equality in both bounds holds if and only if $q_t$ is Gaussian.

Finally, to quantify the distributional discrepancy, we employ the KL divergence and the 2-Wasserstein distance ($W_2$). For Gaussian distributions $\mathcal{N}_1(\mathbf{0}, \boldsymbol{\Sigma}_1)$ and $\mathcal{N}_2(\mathbf{0}, \boldsymbol{\Sigma}_2)$, these metrics admit the following closed-form expressions
\begin{align}
    D_{\text{KL}}(\mathcal{N}_1 \parallel \mathcal{N}_2) &= \frac{1}{2} \left[ \mathrm{tr}(\boldsymbol{\Sigma}_2^{-1} \boldsymbol{\Sigma}_1) - L + \log \frac{\det \boldsymbol{\Sigma}_2}{\det \boldsymbol{\Sigma}_1} \right], \label{eq:kl_div} \\
    W_2^2(\mathcal{N}_1, \mathcal{N}_2) &= \mathrm{tr}\left( \boldsymbol{\Sigma}_1 + \boldsymbol{\Sigma}_2 - 2 (\boldsymbol{\Sigma}_1^{1/2} \boldsymbol{\Sigma}_2 \boldsymbol{\Sigma}_1^{1/2})^{1/2} \right). \label{eq:w2_dist}
\end{align}
Notably, the $W_2$ distance is related to the Fr\'echet Inception Distance (FID)~\cite{heusel2017fid}, which is a standard metric widely used for evaluating image generation quality.

\section{Terminal Distribution Analysis under Gaussian Source} \label{sec:mainKL}

Leveraging the analytical tractability of Gaussian sources established in \cref{sec:pre_gauss}, this section provides a detailed analysis of the distributional discrepancy induced by the deterministic reverse sampling process in~\eqref{eq:ddim}, driven by the optimal denoising estimator in~\eqref{eq:optepi} and \eqref{eq:mmse}. We first quantify the discrepancy via the KL divergence between the true source distribution and the one produced by reverse sampling. Subsequently, we analyze its dominant term and explore the noise schedule to minimize the KL divergence.

\subsection{KL Divergence Analysis and Dominant Discrepancy Term} \label{subsec:kl_analysis}

In this subsection, we first characterize the KL divergence in closed-form expression at the terminal time $t_0$, obtained by analyzing the evolution trajectory of the reverse sampling steps, and then employ the Euler-Maclaurin expansion to isolate the dominant discrepancy term governing the convergence rate, thereby formulating an optimization problem for the noise schedule.

\begin{proposition}[Terminal KL Divergence] \label{prop:terminal_kl}
    Given a time discretization sequence $\{t_i\}_{i=0}^{N}$, the KL divergence between the source distribution and the reversely sampled distribution produced by~\eqref{eq:ddim} at the terminal step $t_0$ admits the following closed-form expression
    \begin{equation} \label{eq:kl_closed_t0}
        D_{\mathrm{KL}}\!\left(q_{G, t_0} \,\|\, \hat{p}_{G, t_0}\right) = \frac{1}{2} \sum_{\ell=1}^{L} \left( \frac{n_\ell}{m_\ell} - \log \frac{n_\ell}{m_\ell} - 1 \right),
    \end{equation}
    where $q_{G, t_0}(\mathbf{x}_{t_0}) = \mathcal{N}(\mathbf{0}, \boldsymbol{\Sigma}_{t_0})$ with $\boldsymbol{\Sigma}_{t_0} = \mathbf{U}\mathbf{N}\mathbf{U}^{\top}$, and $\hat{p}_{G, t_0}(\hat{\mathbf{x}}_{t_0}) = \mathcal{N}(\mathbf{0}, \mathbf{U}\mathbf{M}\mathbf{U}^{\top})$, respectively. Here, $\mathbf{N}$ and $\mathbf{M}$ are diagonal matrices whose diagonal elements, $\{n_\ell\}_{\ell=1}^L$ and $\{m_\ell\}_{\ell=1}^L$, are explicitly evaluated as
    \begin{equation} \label{eq:eigenvalues_nm}
        n_\ell = \alpha_{t_0}^2\mu_\ell+\sigma_{t_0}^2, \quad \text{and} \quad m_\ell = \prod_{j=1}^{N} \left( \frac{\alpha_{t_{j-1}}\alpha_{t_j}\mu_\ell + \sigma_{t_{j-1}}\sigma_{t_j}}{\alpha_{t_j}^2\mu_\ell + \sigma_{t_j}^2} \right)^{\!2}.
    \end{equation}
\end{proposition}
\textit{Proof sketch:} The optimal denoising estimator admits an exact linear form under the Gaussian source. Consequently, the deterministic reverse sampling update in~\eqref{eq:ddim} reduces to a linear recurrence relation. This linearity guarantees that the generated marginal distribution remains Gaussian at every step. By recursively tracking the covariance evolution through this linear system, we analytically determine the terminal covariance $\mathbf{M}$, which directly yields the closed-form KL divergence in~\eqref{eq:kl_closed_t0}. A detailed derivation is deferred to Appendix~\ref{appendix:termi_kl_proof}.

From \cref{prop:terminal_kl}, we observe that the accumulated error from the reverse sampling process is encapsulated within $m_\ell$, and the KL divergence is determined by each eigenvalue ratio $n_\ell / m_\ell$. In the following, to establish the exact asymptotic behavior of the KL divergence and identify the dominant term, we consider a uniform time discretization sequence $t_i \coloneqq i/N$ on the interval $[0, 1]$, and apply the Euler-Maclaurin expansion to evaluate the asymptotic behavior. This leads to the following corollary:
\begin{proposition}[Asymptotic KL Expansion] \label{prop:asymptotic_kl}
    Under the uniform time discretization setting, as $N \to \infty$, the terminal KL divergence expands as
    \begin{equation} \label{eq:kl_expansion}
        D_{\mathrm{KL}}\!\left(q_{G,0} \,\|\, \hat{p}_{G,0}\right) = \frac{1}{N^2} \sum_{\ell=1}^{L} E_\ell^2 + \mathcal{O}\!\left(\frac{1}{N^{3}}\right),
    \end{equation}
    where the $\ell$-th dominant error coefficient $E_\ell$ as a function of the eigenvalue $\mu_\ell$ is given by the integral:
    \begin{equation} \label{eq:E1_ell}
        E_\ell = -\frac{\mu_{\ell}}{2} \int_{0}^{1} \left(\frac{\alpha_t\dot{\sigma}_t-\sigma_t\dot{\alpha}_t}{\alpha_t^{2}\mu_{\ell}+\sigma_t^{2}}\right)^2 \,\mathrm{d}t.
    \end{equation}
\end{proposition}
\textit{Proof sketch:} Taking the logarithm converts the series product form of $m_\ell$ in~\eqref{eq:eigenvalues_nm} into a summation over the sampling grid. We subsequently apply the Euler-Maclaurin formula to expand this sum, thereby establishing the asymptotic relationship between $n_\ell$ and $m_\ell$. Substituting this relation back into the exact KL divergence in~\eqref{eq:kl_closed_t0} and performing a Taylor expansion isolates the leading $\mathcal{O}(1/N^2)$ term, yielding the asymptotic expression in~\eqref{eq:kl_expansion}. See details in Appendix~\ref{appendix:asy_kl_proof}.

We may also conduct analogous asymptotic analysis for the $W_2$ distance in~\eqref{eq:w2_dist}. This metric is of particular practical interest due to its intrinsic connection with the FID, which serves as the primary metric in our subsequent real-world image experiments in \cref{sub:realimage}. To this end, the following proposition establishes that minimizing the KL divergence provides a theoretically sound surrogate for minimizing the $W_2$ distance, as the former upper bounds the latter.

\begin{proposition}[Asymptotic 2-Wasserstein Distance] \label{prop:w2_asymptotic}
    Under the uniform time discretization setting, as $N \to \infty$, the $W_2$ distance between the source distribution $q_{G,0}$ and the generated distribution $\hat{p}_{G,0}$ admits the following asymptotic expansion:
    \begin{equation} \label{eq:w2_asymptotic}
        W_2^2\!\left(q_{G,0}, \hat{p}_{G,0}\right) = \frac{1}{N^2} \sum_{\ell=1}^{L} \mu_{\ell} E_\ell^2 + \mathcal{O}\!\left(\frac{1}{N^{3}}\right).
    \end{equation}
    Furthermore, this expansion reveals that the KL divergence acts as an upper bound for the $W_2$ distance:
    \begin{equation} \label{eq:w2_kl_bound}
        W_2^2\!\left(q_{G,0}, \hat{p}_{G,0}\right) \le \mu_{\max} D_{\mathrm{KL}}\!\left(q_{G,0} \,\|\, \hat{p}_{G,0}\right) + \mathcal{O}\!\left(\frac{1}{N^{3}}\right),
    \end{equation}
    where $\mu_{\max} \coloneqq \max_{1 \le \ell \le L} \mu_\ell$.
\end{proposition}
\textit{Proof sketch:} Based on the closed-form $W_2$ distance in~\eqref{eq:w2_dist} and the relationship between $n_\ell$ and $m_\ell$ established via the Euler-Maclaurin expansion in the proof of \cref{prop:asymptotic_kl}, the upper bound follows. See details in Appendix~\ref{appendix:asy_w2_proof}.

Building upon the asymptotic behavior of the KL divergence and $W_2$ distance, we observe that as the number of discretization steps $N \to \infty$, the distribution discrepancy governed by $E_\ell^2 / N^2$ vanishes. Consequently, we establish the following corollary regarding the asymptotic consistency of the generative process.

\begin{corollary}[Asymptotic Consistency] \label{cor:consistency}
    Evaluating the asymptotic expansion from \cref{prop:asymptotic_kl} under continuous-time limit $N \to \infty$, we have
    \begin{equation}
        \lim_{N \to \infty} m_\ell = n_\ell.
    \end{equation}
    Applying this limit to the KL divergence yields
    \begin{equation}
        \lim_{N \to \infty} D_{\mathrm{KL}}\!\left(q_{G,0} \,\|\, \hat{p}_{G,0}\right) = 0;
    \end{equation}
    that is, $\hat{p}_{G,0}$ converges to $q_{G,0}$ in distribution. This implies that the reverse sampling process~\eqref{eq:ddim} is asymptotically consistent, perfectly reconstructing the exact source distribution, which is in accordance with classical diffusion theory~\cite{anderson1982reverse}. 
\end{corollary}

Consequently, it becomes evident that the generation quality for any finite number of sampling steps $N$ is dominated exclusively by the magnitude of each coefficient $E_\ell$. To systematically investigate how the noise schedule impacts the reverse sampling dynamics and to minimize this cumulative discretization error, we formulate the following variational optimization problem, hereafter referred to as the variational noise schedule (VNS) problem:
\begin{equation} \label{eq:vns_problem}
    \min_{\alpha_t,\,\sigma_t} \ \mathcal{L}[\alpha_t,\sigma_t] \coloneqq \sum_{\ell=1}^{L} E_\ell^2.
\end{equation}

\subsection{Tangent Law} \label{sub:tan}
The VNS problem in~\eqref{eq:vns_problem} is difficult to solve in closed form, because a single scalar noise schedule must simultaneously balance the contributions from all covariance eigenmodes. To gain analytical tractability, we first relax this isotropic constraint and decouple the objective eigenmode-wise. This relaxation not only isolates the eigenmode-specific optimality but also naturally bridges our analysis with the emerging framework of anisotropic diffusion models~\cite{sahoo2024diffusion} introduced in \cref{sub:pre_dms}.
Specifically, rather than being constrained by a uniform scalar schedule $\boldsymbol{\alpha}_t = \alpha_t \mathbf{I}$, one can assign an independent noise trajectory to each covariance eigenmode. For a Gaussian source with covariance $\boldsymbol{\Sigma}_x = \mathbf{U} \mathbf{\Lambda} \mathbf{U}^{\top}$, we can choose the orthonormal basis of the anisotropic noise schedule to be the eigenbasis of $\boldsymbol{\Sigma}_x$. Concretely, we design the anisotropic noise schedule matrices as
\begin{align}
    \boldsymbol{\alpha}_t &= \mathbf{U} \mathrm{diag}(\alpha_{1,t}, \dots, \alpha_{L,t}) \mathbf{U}^{\top}, \\
    \boldsymbol{\sigma}_t &= \mathbf{U} \mathrm{diag}(\sigma_{1,t}, \dots, \sigma_{L,t}) \mathbf{U}^{\top}.
\end{align}
This choice ensures that the generalized forward process in~\eqref{eq:ani_forward} and the deterministic reverse update in~\eqref{eq:ani_ddim} decouple across the source eigenmodes. Following a derivation analogous to the proofs of \cref{prop:terminal_kl} and \cref{prop:asymptotic_kl}, the original VNS objective~\eqref{eq:vns_problem} naturally generalizes to the following anisotropic formulation:
\begin{equation} \label{eq:vns_problem_ani}
    \min_{\{\alpha_{\ell,t},\,\sigma_{\ell,t}\}_{\ell=1}^L} \sum_{\ell=1}^{L} \left(E_\ell^{\mathrm{(a)}}\right)^2,
\end{equation}
where $E_\ell^{\mathrm{(a)}}$ denotes the dominant error coefficient for the $\ell$-th eigenmode in the anisotropic case, given by
\begin{equation} \label{eq:E1_ell_ani}
    E_\ell^{\mathrm{(a)}} = -\frac{\mu_{\ell}}{2} \int_{0}^{1} \left(\frac{\alpha_{\ell,t}\dot{\sigma}_{\ell,t}-\sigma_{\ell,t}\dot{\alpha}_{\ell,t}}{\alpha_{\ell,t}^{2}\mu_{\ell}+\sigma_{\ell,t}^{2}}\right)^2 \,\mathrm{d}t.
\end{equation}
By solving this anisotropic variational problem, we derive the exact analytical optimal schedule, which we formally establish as the anisotropic tangent law.
\begin{theorem}[Anisotropic Tangent Law] \label{thm:ani_tangent_law}
        Let $\boldsymbol{\eta}(t) \coloneqq \boldsymbol{\alpha}_t^{-1} \boldsymbol{\sigma}_t$ denote the noise-to-signal ratio matrix. The optimal schedule solution for \eqref{eq:vns_problem_ani} follows an anisotropic tangent law, given by
        \begin{equation} \label{eq:matrix_schedule}
            \boldsymbol{\eta}(t) = \mathbf{U} \, \mathrm{diag}\!\left(\sqrt{\mu_{1}} \tan\!\left(\frac{\pi}{2} t \right), \dots, \sqrt{\mu_{L}} \tan\!\left(\frac{\pi}{2} t \right)\right) \mathbf{U}^{\top}.
        \end{equation}
    \end{theorem}
\begin{proof}
    We first simplify the global objective in~\eqref{eq:vns_problem_ani}, which is a sum of strictly non-negative terms $\left(E_\ell^{\mathrm{(a)}}\right)^2$. Since the schedule parameters $\{\alpha_{\ell,t}, \sigma_{\ell,t}\}$ for different eigenmodes are optimized independently, the minimization of the global sum is equivalent to minimizing each term individually. It thus follows that
    \begin{equation}
        \min_{\{\alpha_{\ell,t},\,\sigma_{\ell,t}\}_{\ell=1}^L} \sum_{\ell=1}^{L} \left(E_\ell^{\mathrm{(a)}}\right)^2 
        = \sum_{\ell=1}^{L} \min_{\alpha_{\ell,t},\,\sigma_{\ell,t}} \left(E_\ell^{\mathrm{(a)}}\right)^2.
    \end{equation}

    Next, by introducing the eigenmode-specific noise-to-signal ratio $\eta_\ell(t) \coloneqq \sigma_{\ell,t} / \alpha_{\ell,t}$, we have
    \begin{equation}
        \dot{\eta}_\ell = \frac{\alpha_{\ell,t}\dot{\sigma}_{\ell,t}-\sigma_{\ell,t}\dot{\alpha}_{\ell,t}}{\alpha_{\ell,t}^{2}}.
    \end{equation}
    Therefore, the dominant error coefficient in \eqref{eq:E1_ell_ani} can be rewritten as
    \begin{equation}
       E_\ell^{\mathrm{(a)}}  = -\frac{\mu_\ell}{2} \int_{0}^{1} \left( \frac{\dot{\eta}_\ell}{\mu_\ell+\eta_\ell^{2}} \right)^2 \,\mathrm{d}t.
    \end{equation}
    Substituting this simplified kernel back into the decoupled objective yields
    \begin{equation}
        \sum_{\ell=1}^{L} \min_{\alpha_{\ell,t},\,\sigma_{\ell,t}} \left(E_\ell^{\mathrm{(a)}}\right)^2 = \sum_{\ell=1}^{L} \frac{\mu_\ell^2}{4} \min_{\eta_\ell} \left( \int_{0}^{1} \left( \frac{\dot{\eta}_\ell}{\mu_\ell+\eta_\ell^{2}} \right)^2 \,\mathrm{d}t \right)^2.
    \end{equation}
    
    Let us denote the integral functional as $\mathcal{J}_{\ell}[\eta_\ell] \coloneqq \int_{0}^{1} \left( \frac{\dot{\eta}_\ell}{\mu_\ell+\eta_\ell^{2}} \right)^2 \,\mathrm{d}t$. Because the square function $f(x) = x^2$ is strictly monotonically increasing for positive inputs, and the integral is inherently non-negative, minimizing the square of the integral is strictly equivalent to minimizing the integral itself. Furthermore, omitting the constant positive coefficient $\mu_\ell^2 / 4$ does not alter the optimal schedule trajectory. Consequently, the global optimization problem directly reduces to solving the following independent variational problem for each eigenmode $\mu_\ell$:
    \begin{equation} \label{eq:functional_J_eta}
        \min_{\eta_\ell} \ \mathcal{J}_{\ell}[\eta_\ell] = \min_{\eta_\ell} \int_{0}^{1} \left( \frac{\dot{\eta}_\ell}{\mu_\ell+\eta_\ell^{2}} \right)^2 \,\mathrm{d}t.
    \end{equation}

    To solve \eqref{eq:functional_J_eta}, we introduce a change of variables:
    \begin{equation} \label{eq:Q_ell}
        Q_{\ell}(t) \coloneqq \frac{1}{\sqrt{\mu_{\ell}}}\arctan\!\left(\frac{\eta_\ell(t)}{\sqrt{\mu_{\ell}}}\right),
    \end{equation}
    and we obtain
    \begin{equation}
        \dot{Q}_{\ell}(t) = \frac{1}{\sqrt{\mu_\ell}} \cdot \frac{1}{1 + (\eta_\ell/\sqrt{\mu_\ell})^2} \cdot \frac{\dot{\eta}_\ell}{\sqrt{\mu_\ell}} = \frac{\dot{\eta}_\ell}{\mu_\ell+\eta_\ell^{2}}.
    \end{equation}
    Consequently, the non-linear objective functional transforms into a straightforward quadratic form:
    \begin{equation}
        \mathcal{J}_{\ell}[\eta_\ell] = \int_{0}^{1} \dot{Q}_{\ell}^{2} \,\mathrm{d}t.
    \end{equation}
    The Euler-Lagrange equation~\cite{zwillinger2002crc} for the Lagrangian $L(t, Q_{\ell}, \dot{Q}_{\ell}) = \dot{Q}_{\ell}^2$ is given by
    \begin{equation} \label{eq:euler_equation}
        \frac{\partial L}{\partial Q_{\ell}}-\frac{\mathrm{d}}{\mathrm{d}t}\left(\frac{\partial L}{\partial \dot{Q}_{\ell}}\right) = -2\ddot{Q}_{\ell} = 0,
    \end{equation}
    which directly dictates that the optimal $Q_{\ell}(t)$ must be linear: $Q_{\ell}(t) = c t + d$. 

    Recall the boundary conditions $\lim_{t\to0} \mathrm{SNR}_t = \infty$ and $\lim_{t\to1} \mathrm{SNR}_t = 0$ introduced in \cref{sub:pre_dms}. Within the anisotropic framework, the boundary conditions naturally translate to the limits $\lim_{t\to 0} \eta_\ell(t)=0$ and $\lim_{t\to1} \eta_\ell(t)=\infty$ for each eigenmode, so we have
    \begin{align}
        Q_\ell(0) &= \lim_{\eta_\ell \to 0} \frac{1}{\sqrt{\mu_\ell}}\arctan\!\left(\frac{\eta_\ell}{\sqrt{\mu_\ell}}\right) = 0, \\
            Q_\ell(1) &= \lim_{\eta_\ell \to \infty} \frac{1}{\sqrt{\mu_\ell}}\arctan\!\left(\frac{\eta_\ell}{\sqrt{\mu_\ell}}\right) = \frac{\pi}{2\sqrt{\mu_\ell}}.
    \end{align}
    Substituting these boundary values into the linear trajectory $Q_\ell(t) = ct + d$ yields $Q_{\ell}(t) = \frac{\pi t}{2\sqrt{\mu_\ell}}$. Equating this explicit solution to the original definition of $Q_\ell(t)$ in~\eqref{eq:Q_ell} directly establishes the eigenmode-specific tangent law schedule
\begin{equation} \label{eq:eigmode_tangetlaw}
    \eta_{\ell}(t) = \sqrt{\mu_{\ell}} \tan\!\left(\frac{\pi}{2} t\right).
\end{equation}
    
    Finally, transforming the optimal eigenmode-specific profile back into the anisotropic regime using the shared eigenbasis $\mathbf{U}$ yields $\boldsymbol{\eta}(t) = \mathbf{U} \mathrm{diag}(\eta_1(t), \dots, \eta_L(t)) \mathbf{U}^{\top}$, thereby concluding the proof.
\end{proof}

Returning to the original VNS problem in~\eqref{eq:vns_problem}, this anisotropic solution in \cref{thm:ani_tangent_law} is of theoretical value: by allowing each eigenmode to follow its own optimal trajectory, it establishes an absolute lower bound.

\begin{corollary} \label{cor:vns_lb}
    Let $\mathcal{L}_{\mathrm{lb}}$ denote the absolute theoretical lower bound of the VNS objective~\eqref{eq:vns_problem}, achieved by assigning the eigenmode-specific tangent law $\eta_\ell(t) = \sqrt{\mu_\ell}\tan\!\left(\frac{\pi}{2}t\right)$ to each component independently, which evaluates to
    \begin{equation} \label{eq:vns_lb}
        \mathcal{L}_{\mathrm{lb}} \coloneqq \sum_{\ell=1}^{L} \left(E_{\ell}^{\mathrm{(a)}}\left(\eta_\ell\right)\right)^2 = \frac{\pi^4 L}{64}.
    \end{equation}
\end{corollary}

\begin{proof} 
    According to the \cref{lem:gamma_E_ell_closed_form} in Appendix~\ref{appen:proofofthm2}, substituting the optimal eigenmode-specific profile $\eta_\ell(t) = \sqrt{\mu_\ell}\tan\!\left(\frac{\pi}{2}t\right)$ directly into the dominant error coefficient yields
    \begin{equation} \label{eq:E_mu_ell}
        E_\ell^{\mathrm{(a)}}\left(\eta_\ell\right) = -\frac{\pi^2}{16} \left( \sqrt{\frac{\mu_\ell}{\mu_\ell}} + \sqrt{\frac{\mu_\ell}{\mu_\ell}} \right) = -\frac{\pi^2}{8}.
    \end{equation}
    Finally, squaring this coefficient and summing over all $L$ eigenvalue dimensions directly establishes the lower bound in~\eqref{eq:vns_lb}, thereby concluding the proof.
\end{proof}

From a practical standpoint, contemporary DMs favor a unified scalar noise schedule shared across all eigenvalue components to maintain computational efficiency and architectural simplicity.
To achieve this, we introduce a parameterized unified scalar profile $\eta_\gamma(t)$ governed by a positive constant $\gamma > 0$, which inherits the trigonometric structure of the eigenmode-specific tangent law:
\begin{equation} \label{eq:eta_gamma}
    \eta_\gamma(t) \coloneqq \sqrt{\gamma} \tan\!\left(\frac{\pi}{2} t\right), \quad t \in [0, 1).
\end{equation}
By substituting this parameterized global profile into the VNS objective, the functional reduces to a scalar function $\mathcal{L}(\gamma) \coloneqq \sum_{\ell=1}^{L}  E_\ell^2(\gamma)$. We establish the following theorem to determine the optimal scalar parameter.

\begin{theorem}[Global Cross-Eigenmode Tangent Law] \label{thm:global_opt_gamma}
    For the VNS problem under the parameterized global profile~\eqref{eq:eta_gamma}, the dominant error coefficient $E_\ell(\gamma)$ evaluates to
    \begin{equation} \label{eq:E_gamma}
        E_\ell(\gamma) = -\frac{\pi^2}{16} \left( \sqrt{\frac{\mu_\ell}{\gamma}} + \sqrt{\frac{\gamma}{\mu_\ell}} \right).
    \end{equation}
    Furthermore, the objective function $\mathcal{L}(\gamma)$ admits a unique closed-form minimizer $\gamma^\star$, given by
    \begin{equation} \label{eq:gamma_star}
        \gamma^{\star} = \sqrt{ \frac{\mathrm{tr}(\boldsymbol{\Sigma}_{\mathbf{x}})}{\mathrm{tr}(\boldsymbol{\Sigma}_{\mathbf{x}}^{-1})} }.
    \end{equation}
\end{theorem}

\cref{thm:ani_tangent_law} establishes the optimal eigenmode-specific tangent law, thereby defining a strict theoretical lower bound for the original VNS problem. Conversely, \cref{thm:global_opt_gamma} derives the optimal unified scalar parameter $\gamma^\star$ (proof in Appendix~\ref{appen:proofofthm2}), effectively bridging this theoretical eigenmode-specific profile with practical isotropic diffusion architectures. This naturally raises a critical question regarding the performance penalty incurred by enforcing such a unified scalar constraint. Building upon \cref{cor:vns_lb}, we explicitly quantify this performance gap.

\begin{corollary}[] \label{cor:performgap}
    Let $\mathcal{L}(\gamma^\star)$ denote the VNS objective achieved by the optimal global cross-eigenmode tangent law in~\eqref{eq:gamma_star}. The performance gap $\Delta \mathcal{L} \coloneqq \mathcal{L}(\gamma^\star) - \mathcal{L}_{\mathrm{lb}}$ is given by
    \begin{equation} \label{eq:performgap}
        \Delta \mathcal{L} = \frac{\pi^4}{128} \left[ \sqrt{\mathrm{tr}(\boldsymbol{\Sigma}_{\mathbf{x}})\mathrm{tr}(\boldsymbol{\Sigma}_{\mathbf{x}}^{-1})} - L \right] \geq 0.
    \end{equation}
    Furthermore, equality holds if and only if $\boldsymbol{\Sigma}_{\mathbf{x}} = c\mathbf{I}$ for some constant $c > 0$.
\end{corollary}
\begin{proof}
    For the optimal global cross-eigenmode tangent law, substituting $\gamma^\star = \sqrt{\mathrm{tr}(\boldsymbol{\Sigma}_{\mathbf{x}}) / \mathrm{tr}(\boldsymbol{\Sigma}_{\mathbf{x}}^{-1})}$ into the objective $\mathcal{L}(\gamma)$ yields
    \begin{equation}
        \mathcal{L}(\gamma^\star) = \sum_{\ell=1}^{L} E_{\ell}^2(\gamma^{\star}) = \left(\frac{\pi^2}{16}\right)^2 \left[ 2 \sqrt{\mathrm{tr}(\boldsymbol{\Sigma}_{\mathbf{x}})\mathrm{tr}(\boldsymbol{\Sigma}_{\mathbf{x}}^{-1})} + 2L \right].
    \end{equation}
    Subtracting $\mathcal{L}_{\mathrm{lb}}$ \eqref{eq:vns_lb} from $\mathcal{L}(\gamma^\star)$ directly recovers the gap $\Delta \mathcal{L}$ presented in~\eqref{eq:performgap}.
    
    Then, by the Cauchy-Schwarz inequality, we have
    \begin{equation}
        \mathrm{tr}(\boldsymbol{\Sigma}_{\mathbf{x}})\mathrm{tr}(\boldsymbol{\Sigma}_{\mathbf{x}}^{-1})  \ge \left( \sum_{\ell=1}^L \sqrt{\mu_\ell} \cdot \frac{1}{\sqrt{\mu_\ell}} \right)^2 = L^2,
    \end{equation}
    which guarantees that $\Delta \mathcal{L} \ge 0$. Furthermore, the condition for equality in the Cauchy-Schwarz inequality dictates that all eigenvalues must be identical (i.e., $\mu_\ell = c$ for some positive constant $c$), thereby concluding the proof.
\end{proof}

A critical observation regarding the gap derived in \cref{cor:performgap} is its dependence on the inverse trace $\mathrm{tr}(\boldsymbol{\Sigma}_{\mathbf{x}}^{-1})$. This inverse trace diverges in the ill-posed setting where some eigenvalues approach zero, causing the KL-based performance gap to become unbounded. Fortunately, this singularity can be elegantly resolved by shifting our perspective from the KL divergence to the $W_2$ distance as the generative discrepancy metric.

Recall from \cref{prop:w2_asymptotic} that the asymptotic expansion of the terminal $W_2$ distance is dominated by the weighted sum $\sum_{\ell=1}^L \mu_{\ell} E_\ell^2$. By incorporating the global cross-eigemmode tangent law schedule~\eqref{eq:eta_gamma}, the VNS functional under the $W_2$ metric generalizes to minimizing the weighted objective function $\mathcal{L}_{W_2}(\gamma) \coloneqq \sum_{\ell=1}^{L} \mu_{\ell} E_{\ell}^2(\gamma)$. Following a derivation analogous to \cref{thm:global_opt_gamma} and \cref{cor:performgap}, we establish the following proposition (proof in Appendix~\ref{appen:w2_gamma_proof}).

\begin{proposition}[] \label{prop:w2_optimal_gamma}
    Under the $W_2$ distance, the optimal scalar parameter $\gamma_{W_2}^\star$ that minimizes the weighted objective function $\mathcal{L}_{W_2}(\gamma)$ is given by
    \begin{equation} \label{eq:w2_gamma_star}
        \gamma_{W_2}^\star = \sqrt{ \frac{\mathrm{tr}(\boldsymbol{\Sigma}_{\mathbf{x}}^2)}{L} }.
    \end{equation}
    Furthermore, the corresponding performance gap $\Delta \mathcal{L}_{W_2} \coloneqq \mathcal{L}_{W_2}(\gamma_{W_2}^\star) - \mathcal{L}_{W_2,\mathrm{lb}}$ explicitly evaluates to
    \begin{equation} \label{eq:w2_performgap}
        \Delta \mathcal{L}_{W_2} = \frac{\pi^4}{128} \left[ \sqrt{L \cdot \mathrm{tr}(\boldsymbol{\Sigma}_{\mathbf{x}}^2)} - \mathrm{tr}(\boldsymbol{\Sigma}_{\mathbf{x}}) \right].
    \end{equation}
\end{proposition}

In short, the $W_2$ distance effectively acts as a weighted objective where the intrinsic weight $w_\ell = \mu_\ell$ elegantly resolves the theoretical singularity via the cancellation $\mu_\ell \cdot \mu_\ell^{-1} = 1$. Motivated by this natural weighting mechanism, we generalize our framework to accommodate arbitrary positive weights. This generalization is relevant in practical generative regimes, where different spectral components naturally carry varying degrees of perceptual or semantic importance. For instance, low-frequency modes typically dictate the global structure of an image, while high-frequency modes govern fine-grained textures and local details. By assigning a customized semantic weight $w_\ell > 0$ to each eigenmode, we establish the following corollary for the generalized weighted VNS problem.

\begin{corollary}[] \label{cor:weighted_gamma}
    Let $\gamma^{\star}_{w} \coloneqq \underset{\gamma > 0}{\arg\min} \sum_{\ell=1}^{L} w_{\ell} E_{\ell}^2(\gamma)$. This optimal scalar parameter admits a unique closed-form solution:
    \begin{equation} \label{eq:weighted_gamma_def}
        \gamma^{\star}_{w} = \sqrt{ \frac{\sum_{\ell=1}^{L}w_\ell \mu_\ell}{\sum_{\ell=1}^{L}w_\ell \mu_\ell^{-1}} } = \sqrt{ \frac{\mathrm{tr}(\mathbf{W}\boldsymbol{\Lambda})}{\mathrm{tr}(\mathbf{W}\boldsymbol{\Lambda}^{-1})} },
    \end{equation}
    where $\mathbf{W} = \operatorname{diag}(w_1, \dots, w_L)$ represents the diagonal weight matrix.
    The gap for this weighted formulation explicitly evaluates to the following form:
    \begin{equation} \label{eq:weighted_gap}
        \Delta \mathcal{L}_{w} \coloneqq \sum_{\ell=1}^{L} w_{\ell} \left(E_{\ell}^2(\gamma^{\star}_{w}) - \left(E_{\ell}^{\mathrm{(a)}}(\eta_\ell)\right)^2 \right] = \frac{\pi^4}{128} \left[ \sqrt{\mathrm{tr}(\mathbf{W}\boldsymbol{\Lambda})\mathrm{tr}(\mathbf{W}\boldsymbol{\Lambda}^{-1})} - \mathrm{tr}(\mathbf{W}) \right].
    \end{equation}
\end{corollary}

\section{Extremal Property of the Gaussian Setting} \label{sec:mainLB}
To extend the analytical insights obtained in the Gaussian setting to DM operating on general source distributions, this section establishes a formal theoretical bridge at the level of the stochastic DM dynamics as introduced in \cref{sub:pre_sde}. Each reverse SDE induces a path measure, namely the probability law of the entire trajectory over time. Accordingly, we measure the discrepancy between two reverse processes through the KL divergence between their associated path measures, rather than only through divergences between fixed-time marginals. By combining stochastic analysis of reverse SDEs with information-theoretic arguments, we prove a fundamental extremal property: under optimal denoising, the Gaussian source attains the minimum path-space KL divergence among all source distributions sharing the same covariance.

As introduced in \cref{sub:pre_sde}, let $\mathbb{P}_{\theta}$ denote the path measure induced by the reverse SDE parameterized by the noise estimator, and let $\hat{\mathbb{P}}_{\theta}$ denote the path measure induced by its piecewise approximation~\eqref{eq:approx_sde}. The KL divergence $D_{\mathrm{KL}}(\mathbb{P}_{\theta}\parallel\hat{\mathbb{P}}_{\theta})$ therefore quantifies the discrepancy between these two processes at the trajectory level. Since these two processes share the same diffusion coefficient, Girsanov's theorem yields a decomposition of this path-space KL divergence into two terms: a marginal KL discrepancy at the initialization of the reverse process, i.e., at $t=1$, and a time-integrated quadratic discrepancy between their drift fields.

\begin{proposition} \label{prop:piecewise_sde_kl}
    Consider the path measures $\mathbb{P}_{\theta}$ and $\hat{\mathbb{P}}_{\theta}$ induced by the following reverse SDE and its piecewise approximation:
    \begin{subequations}
    \begin{align} 
        \mathbb{P}_{\theta}: \quad
        & \mathrm{d}\hat{\mathbf{x}}_t = \left( f_t\hat{\mathbf{x}}_t - h_t\boldsymbol{\epsilon}_{\theta}(\hat{\mathbf{x}}_t,t) \right)\mathrm{d}t + g_t\mathrm{d}\bar{\mathbf{w}}_t, \quad \hat{\mathbf{x}}_1\sim q_1, \label{eq:twosdes1} \\
        \hat{\mathbb{P}}_{\theta}: \quad
        & \mathrm{d}\hat{\mathbf{x}}_t = \left( f_t\hat{\mathbf{x}}_t - h_t\boldsymbol{\epsilon}_{\theta}(\hat{\mathbf{x}}_{t_j},t_j) \right)\mathrm{d}t + g_t\mathrm{d}\bar{\mathbf{w}}_t, \quad \hat{\mathbf{x}}_1\sim\mathcal{N}(\mathbf{0},c^2\mathbf{I}). \label{eq:twosdes2}
    \end{align}
    \end{subequations}
    The path-space KL divergence in the asymptotic limit $\delta\coloneqq\max_j\delta_j\to0$ satisfies:
    \begin{equation} \label{eq:lemma_dkl_results2}
        D_{\mathrm{KL}}(\mathbb{P}_{\theta}\parallel\hat{\mathbb{P}}_{\theta}) = D_{\mathrm{KL}}\!\left(q_1\parallel\mathcal{N}(\mathbf{0},c^2\mathbf{I})\right) + \frac{1}{4} \sum_{j=1}^{N} h_{t_j}^2\delta_j^2 \, \mathbb{E}_{p_{\theta,t_j}} \!\left[ \|\mathbf{J}_{\boldsymbol{\epsilon}}(\hat{\mathbf{x}}_{t_j},t_j)\|_F^2 \right] + \mathcal{R}(\delta),
    \end{equation}
    where $\delta_j\coloneqq t_j-t_{j-1}$, the Jacobian is defined as $\mathbf{J}_{\boldsymbol{\epsilon}}\coloneqq\nabla_{\mathbf{x}} \boldsymbol{\epsilon}_{\theta}(\mathbf{x},t)$, and the global remainder is bounded by $\mathcal{R}(\delta)=\mathcal{O}(\delta^{3/2})$.
\end{proposition}

\textit{Proof sketch.}
The two processes share the same diffusion coefficient and differ only in their drift terms. Therefore, by the chain rule for Radon--Nikodym derivatives and Girsanov's theorem, we obtain
\begin{equation*}
    D_{\mathrm{KL}}(\mathbb{P}_{\theta} \parallel \hat{\mathbb{P}}_{\theta})
    =
    D_{\mathrm{KL}}\!\left(q_1 \parallel \mathcal{N}(\mathbf{0}, c^2\mathbf{I})\right)
    + \frac{1}{2} \sum_{j=1}^N \int_{t_{j-1}}^{t_j}
    \frac{h_t^2}{g_t^2}
    \mathbb{E}_{\mathbb{P}_{\theta}} \!\left[
    \left\|
    \boldsymbol{\epsilon}_{\theta}(\hat{\mathbf{x}}_t, t)
    -
    \boldsymbol{\epsilon}_{\theta}(\hat{\mathbf{x}}_{t_j}, t_j)
    \right\|^2
    \right] \mathrm{d}t .
\end{equation*}

Next, applying It\^o's formula to the neural estimator along the reverse trajectory gives the local stochastic dynamics
\begin{equation*}
    \mathrm{d}\boldsymbol{\epsilon}_{\theta}(\hat{\mathbf{x}}_t,t)
    =
    \boldsymbol{\phi}(\hat{\mathbf{x}}_t,t)\mathrm{d}t
    +
    g_t\mathbf{J}_{\boldsymbol{\epsilon}}(\hat{\mathbf{x}}_t,t)
    \mathrm{d}\bar{\mathbf{w}}_t .
\end{equation*}

Integrating this equation over $[t,t_j]$ and applying the It\^o isometry identifies the leading mean-square variation. Under a sufficiently fine time discretization, the dominant contribution comes from the martingale term
$g_t\mathbf{J}_{\boldsymbol{\epsilon}}\,\mathrm{d}\bar{\mathbf{w}}_t$, while the It\^o drift and cross terms contribute only to higher-order residuals. Consequently,
\begin{equation*}
    \int_{t_{j-1}}^{t_j}
    \frac{h_t^2}{g_t^2}
    \mathbb{E}_{\mathbb{P}_{\theta}} \!\left[
    \left\| \boldsymbol{\epsilon}_{\theta}(\hat{\mathbf{x}}_t, t)
    -\boldsymbol{\epsilon}_{\theta}(\hat{\mathbf{x}}_{t_j}, t_j)\right\|^2
    \right] \mathrm{d}t
    = \frac{1}{2} h_{t_j}^2 \delta_j^2
    \mathbb{E}_{p_{\theta,t_j}} \!\left[
    \left\|\mathbf{J}_{\boldsymbol{\epsilon}}(\hat{\mathbf{x}}_{t_j}, t_j)
    \right\|_F^2\right]
    + \mathcal{O}(\delta_j^{5/2}) .
\end{equation*}

Substituting this local result into the Girsanov integral yields the leading coefficient in \cref{eq:lemma_dkl_results2}. The required prerequisite theorems, regularity assumptions, and the full technical derivation are deferred to Appendix~\ref{appen:sdediserror}.

\cref{prop:piecewise_sde_kl} identifies the leading-order contribution to the path-space KL divergence caused by the piecewise approximation for a general source distribution. Furthermore, if the neural denoising estimator perfectly matches the true score function, the learned path measure $\mathbb{P}_\theta$ will exactly coincide with the true path measure $\mathbb{Q}$ that perfectly reverses the forward diffusion trajectory. This allows us to interpret $D_{\mathrm{KL}}(\mathbb{Q}\parallel\hat{\mathbb{P}}_{\theta})$ as the intrinsic trajectory-level error introduced by the piecewise approximation. Building upon this scenario, for the Gaussian source described in \cref{sec:pre_gauss}, we define $\mathbb{Q}_G$ and $\hat{\mathbb{P}}_{G}$ as the path measures induced by the exact reverse SDE and its piecewise approximation, respectively.

The following theorem shows that the Gaussian setting is not merely a special case, but an extremal benchmark for general settings. Among all source distributions with the same covariance, the Gaussian source attains the smallest asymptotic path-space discrepancy induced by the piecewise-constant approximation. Moreover, this path-space lower bound dominates the terminal marginal discrepancy $D_{\mathrm{KL}}(q_{G,0}\parallel \hat p_{G,0})$, precisely the quantity analyzed in the preceding section.

\begin{theorem} \label{thm:lower_bound}
    Assume that the denoising estimator is optimal and that the source distribution has the same covariance as the Gaussian source considered in \cref{sec:pre_gauss}. Then, the Gaussian source yields an asymptotic lower bound on the generative discrepancy induced by the piecewise approximation for a sufficiently fine time discretization:
    \begin{equation}
        D_{\mathrm{KL}}(\mathbb{Q} \parallel \hat{\mathbb{P}}_{\theta})
        \ge
        D_{\mathrm{KL}}(\mathbb{Q}_G \parallel \hat{\mathbb{P}}_{G})
        \ge
        D_{\mathrm{KL}}(q_{G,0} \parallel \hat{p}_{G,0}).
    \end{equation}
\end{theorem}
\begin{proof}
    We first establish the inequality $D_{\mathrm{KL}}(\mathbb{Q} \parallel \hat{\mathbb{P}}_{\theta}) \ge D_{\mathrm{KL}}(\mathbb{Q}_G \parallel \hat{\mathbb{P}}_{G})$.
    Under the assumption of optimal denoising estimator, the spatial Jacobian of the neural denoising estimator exactly matches the scaled Hessian of the log-marginal density: $\nabla_{\mathbf{x}} \boldsymbol{\epsilon}^{\star}_{\theta}(\mathbf{x}_t, t) = -\sigma_t \nabla_{\mathbf{x}}^2 \log q_t(\mathbf{x}_t)$. For the Gaussian source, this Hessian equates to the precision matrix, yielding $\nabla_{\mathbf{x}} \boldsymbol{\epsilon}^{\star}_{G}(\mathbf{x}_t, t) = \sigma_t \boldsymbol{\Sigma}_t^{-1}$. 
    
    Invoking the decomposition of the KL divergence derived in \cref{prop:piecewise_sde_kl}, we have
    \begin{align} \label{eq:delta_decomp}
        D_{\mathrm{KL}}(\mathbb{Q} \parallel \hat{\mathbb{P}}_{\theta}) &- D_{\mathrm{KL}}(\mathbb{Q}_G \parallel \hat{\mathbb{P}}_{G}) \notag \\
        &=  D_{\mathrm{KL}}(q_1 \parallel \mathcal{N}(\mathbf{0}, c^2 \mathbf{I})) - D_{\mathrm{KL}}(q_{1,G} \parallel \mathcal{N}(\mathbf{0}, c^2 \mathbf{I}))  \notag \\
        &\quad\quad + \sum_{j=1}^{N} \frac{1}{4} h_{t_j}^2 \delta_j^2 \left( \mathbb{E}_{q_{t_j}}\!\left[ \| \sigma_{t_j} \nabla_{\mathbf{x}}^2 \log q_{t_j}(\mathbf{x}) \|_F^2 \right] - \| \sigma_{t_j} \boldsymbol{\Sigma}_{t_j}^{-1} \|_F^2 \right) + \mathcal{R}(\delta) \notag \\
        &= \left[h(q_{1,G}) - h(q_1)\right] + \sum_{j=1}^{N} \frac{1}{4} h_{t_j}^2 \sigma_{t_j}^2 \delta_j^2 \left( \mathbb{E}_{q_{t_j}}\!\left[ \| \nabla_{\mathbf{x}}^2 \log q_{t_j}(\mathbf{x}) \|_F^2 \right] - \| \boldsymbol{\Sigma}_{t_j}^{-1} \|_F^2 \right) + \mathcal{R}(\delta).
    \end{align}

    For the first term, we have $h(q_{1,G}) - h(q_1) \ge 0$ following from the maximum entropy principle presented in \eqref{eq:max_entropy} in \cref{sec:pre_gauss}, as both distributions share the same covariance matrix. 
    
    For the second term, we have
    \begin{align}
        \mathbb{E}_{q_{t_j}}\!\left[ \| \nabla_{\mathbf{x}}^2 \log q_{t_j}(\mathbf{x}) \|_F^2 \right] - \| \boldsymbol{\Sigma}_{t_j}^{-1} \|_F^2 
        &\ge \left\| \mathbb{E}_{q_{t_j}}\!\left[ \nabla_{\mathbf{x}}^2 \log q_{t_j}(\mathbf{x}) \right] \right\|_F^2 - \| \boldsymbol{\Sigma}_{t_j}^{-1} \|_F^2 \label{eq:jensen_step} \\
        &= \| \mathcal{I}(q_{t_j}) \|_F^2 - \| \boldsymbol{\Sigma}_{t_j}^{-1} \|_F^2 \label{eq:neg_fisher_step} \\
        &\ge 0, \label{eq:crb_step}
    \end{align}
    where $\mathcal{I}(q_{t_j}) \coloneqq \mathbb{E}_{q_{t_j}}[-\nabla_{\mathbf{x}}^2 \log q_{t_j}(\mathbf{x})]$ defines the FIM. The inequality~\eqref{eq:jensen_step} follows from Jensen's inequality applied to the convex squared Frobenius norm $\| \cdot \|_F^2$. The identity~\eqref{eq:neg_fisher_step} utilizes the fact that the expected Hessian of the log-likelihood equals the negative FIM under mild regularity conditions. Finally, the inequality~\eqref{eq:crb_step} is guaranteed by the multivariate CRB introduced in \eqref{eq:crb} in \cref{sec:pre_gauss}, which establishes that $\mathcal{I}(q_{t_j}) - \boldsymbol{\Sigma}_{t_j}^{-1}$ is positive semi-definite; equality holds if and only if $q_{t_j}$ is also a Gaussian distribution.

    So in summary, we establish the extremal property $D_{\mathrm{KL}}(\mathbb{Q} \parallel \hat{\mathbb{P}}_{\theta}) \ge D_{\mathrm{KL}}(\mathbb{Q}_G \parallel \hat{\mathbb{P}}_{G})$ for sufficiently small $\delta$.
    Furthermore, the inequality $D_{\mathrm{KL}}(\mathbb{Q}_G \parallel \hat{\mathbb{P}}_{G}) \ge D_{\mathrm{KL}}(q_{G,0} \parallel \hat{p}_{G,0})$ is a direct consequence of the chain rule for KL divergence~\cite{cover1999elements}, since the marginal distribution at the terminal time $t=0$ is merely a marginalization of the full generative path measure. This completes the proof.
\end{proof}

Therefore, the explicit Gaussian discrepancy derived in \cref{sec:mainKL} is not an isolated calculation, but a universal baseline for the generative discrepancy among all source distributions with the same covariance. From an information-theoretic perspective, the Gaussian distribution maximizes the differential entropy among all continuous distributions with a prescribed covariance matrix, and therefore represents the least structured distribution at the level of second-order statistics. In the context of DMs, \Cref{thm:lower_bound} shows that the generative discrepancy of this Gaussian source case forms an irreducible floor for the general problem. Hence, the Gaussian analysis in \cref{sec:mainKL} is not only a simplified surrogate, but also a necessary baseline for understanding the fundamental performance limitations of DMs. Practical DMs operating on non-Gaussian data cannot evade this Gaussian lower bound; any additional structural complexity can only appear on top of this baseline.

\section{Experiments} \label{sec:exp}
In this section, we first validate our theoretical findings through experiments on synthetic datasets, and subsequently apply our approach to real-world generation tasks using pretrained diffusion models.

\subsection{Numerical Validation of Tangent Law Schedule} \label{sub:gmm}

To evaluate the effectiveness of the tangent law schedule against representative baselines~\cite{lipman2023flow,nichol2021improved,song2020score,jabri2023sca}, we consider both a multivariate Gaussian setting and a GMM setting for examining robustness under multimodal structures. In both synthetic settings, the ground-truth score function admits exact closed-form expressions determined solely by the underlying distribution parameters, enabling exact evaluation of the reverse sampling dynamics~\eqref{eq:ddim} and the resulting generative discrepancy.

\begin{figure}[htbp]
    \centering
    \begin{subfigure}[t]{0.31\columnwidth} 
        \centering
        \includegraphics[width=1\textwidth]{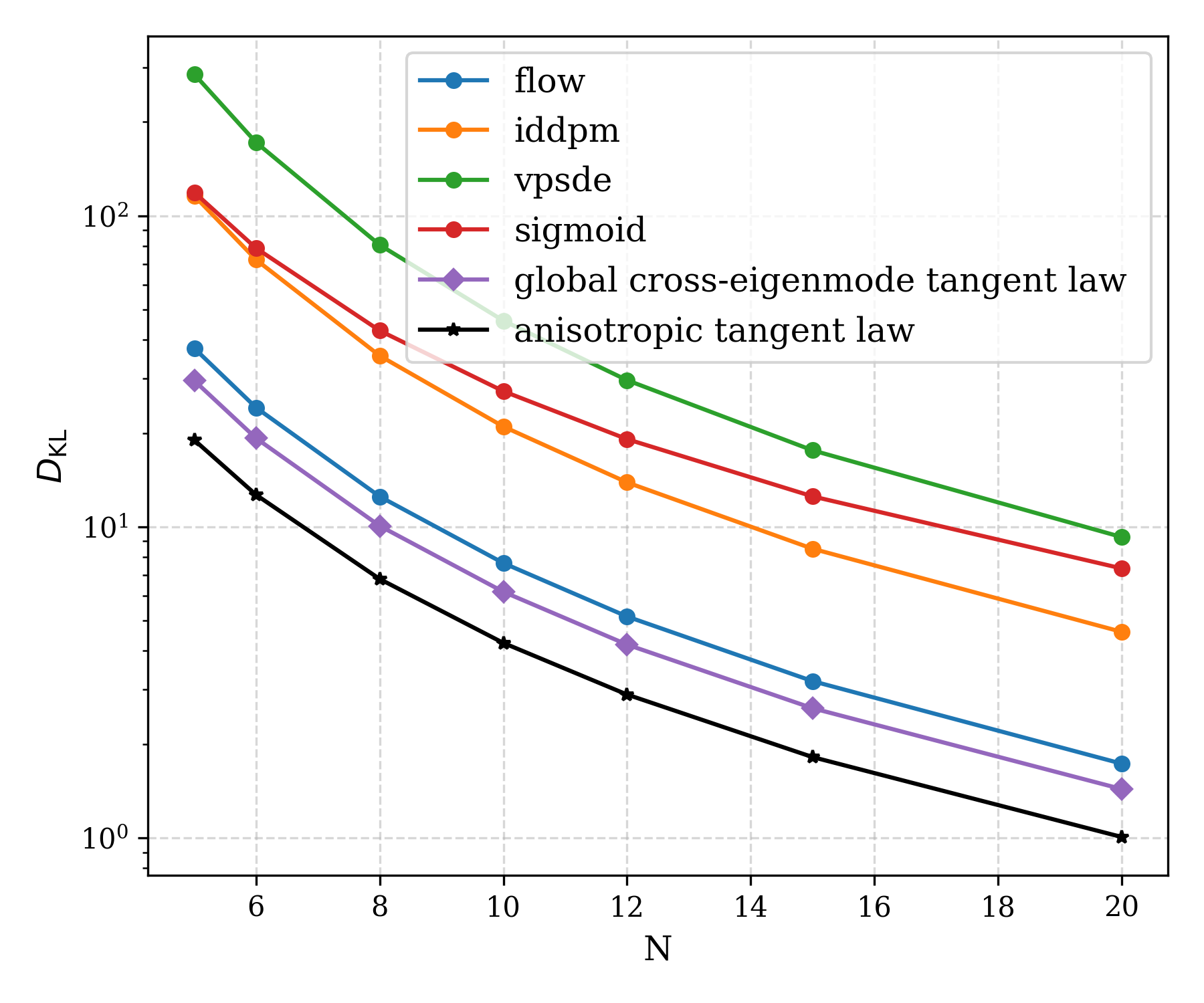} 
        \caption{$D_{\mathrm{KL}}$ vs. reverse sampling steps}
        \label{subfig:gaussian_kl}
    \end{subfigure}
    \hfill 
    \begin{subfigure}[t]{0.31\columnwidth} 
        \centering
        \includegraphics[width=1\textwidth]{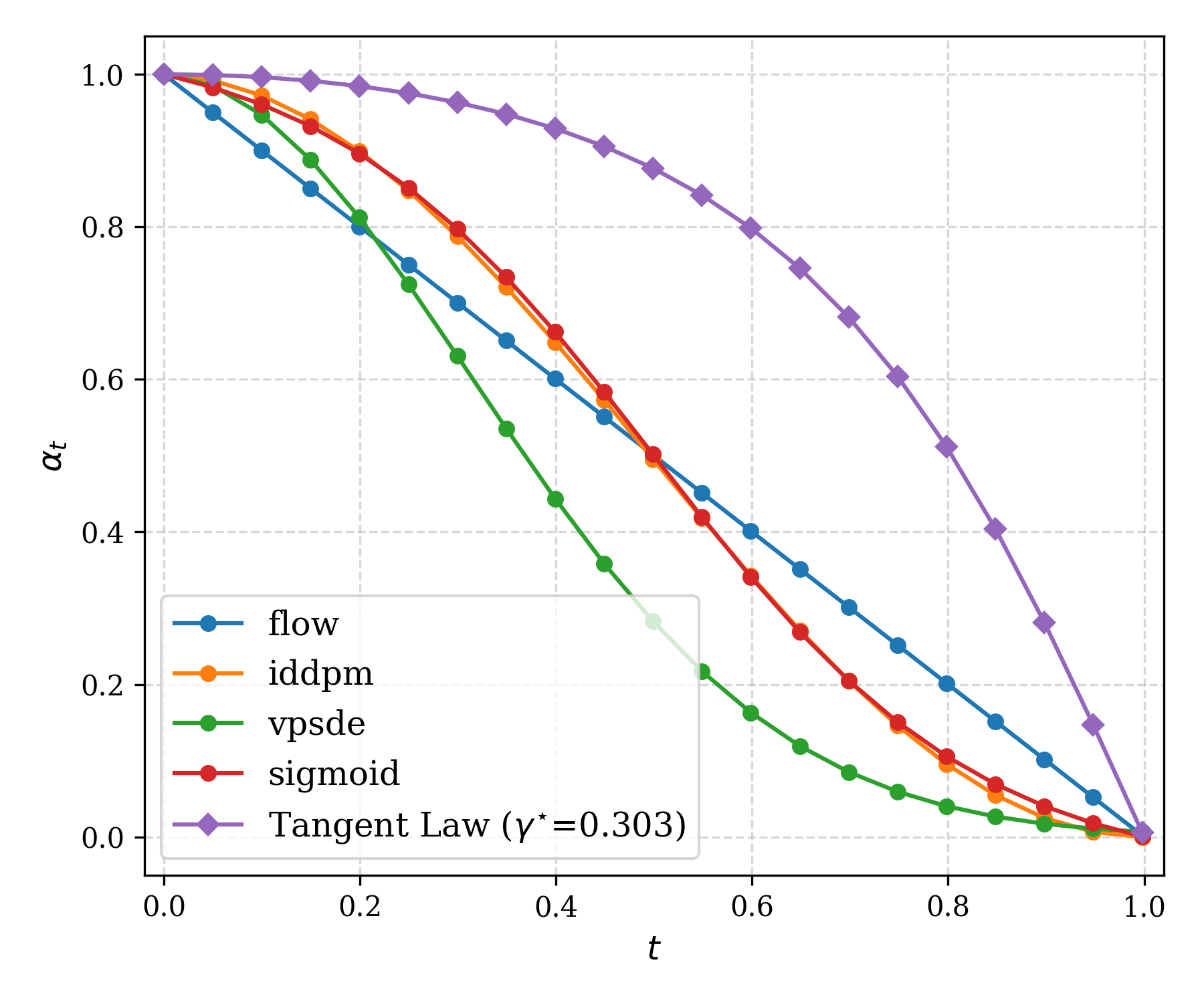} 
        \caption{Schedule curves ($\alpha_t$)}
        \label{subfig:gaussian_alpha}
    \end{subfigure}
    \hfill
    \begin{subfigure}[t]{0.31\columnwidth} 
        \centering
        \includegraphics[width=1\textwidth]{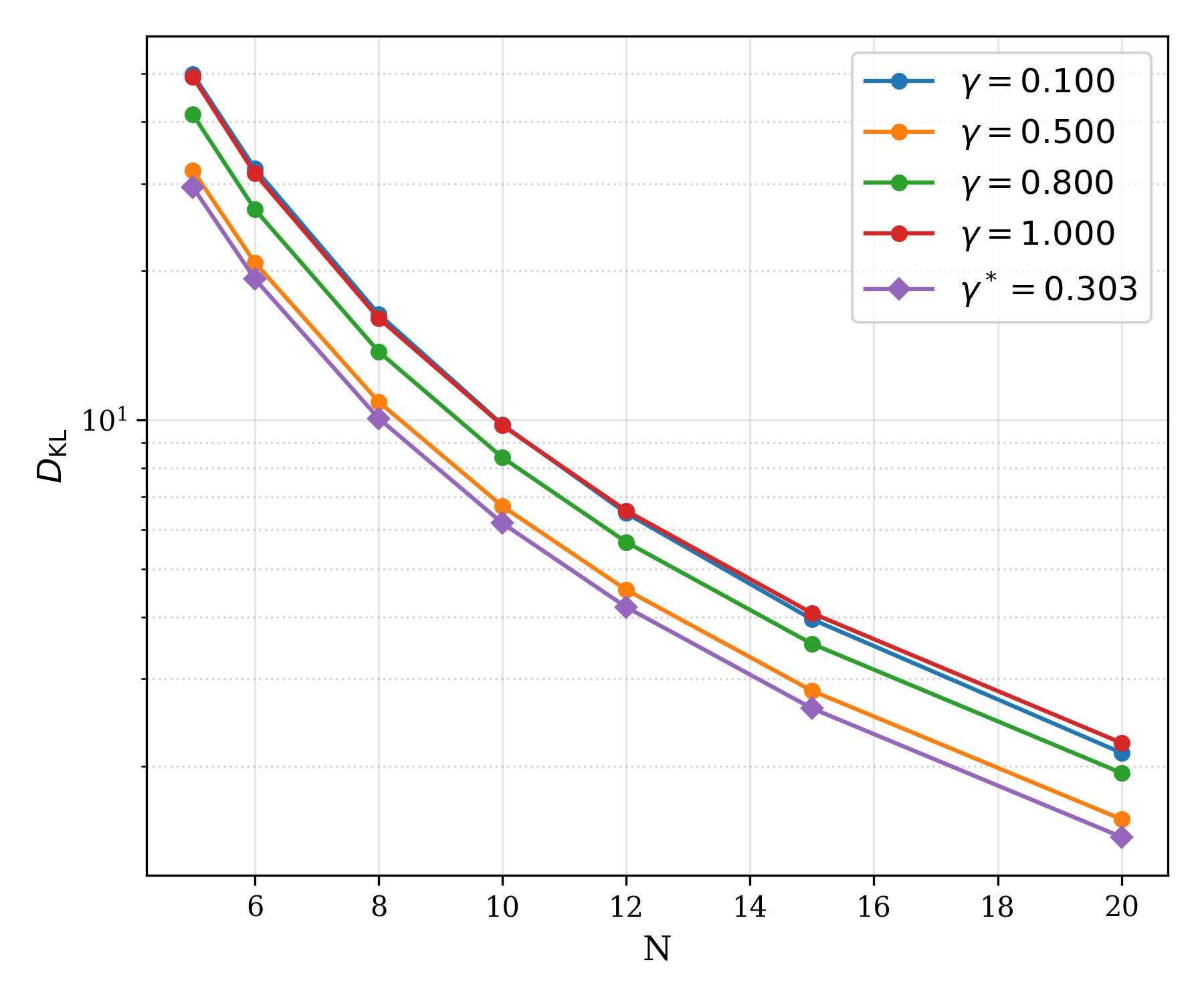} 
        \caption{Ablation on parameter $\gamma$}
        \label{subfig:gaussian_gamma}
    \end{subfigure}
    \caption{Numerical validation on multivariate Gaussian source across different noise schedules. (a) Convergence of the KL divergence. (b) Comparison of $\alpha_t$. (c) Ablation study on the global tangent law parameter $\gamma$.}
    \label{fig:gaussian_validation}
\end{figure}

\begin{figure}[htbp]
    \centering
    \begin{subfigure}[t]{0.48\columnwidth} 
        \centering
        \includegraphics[width=1\textwidth]{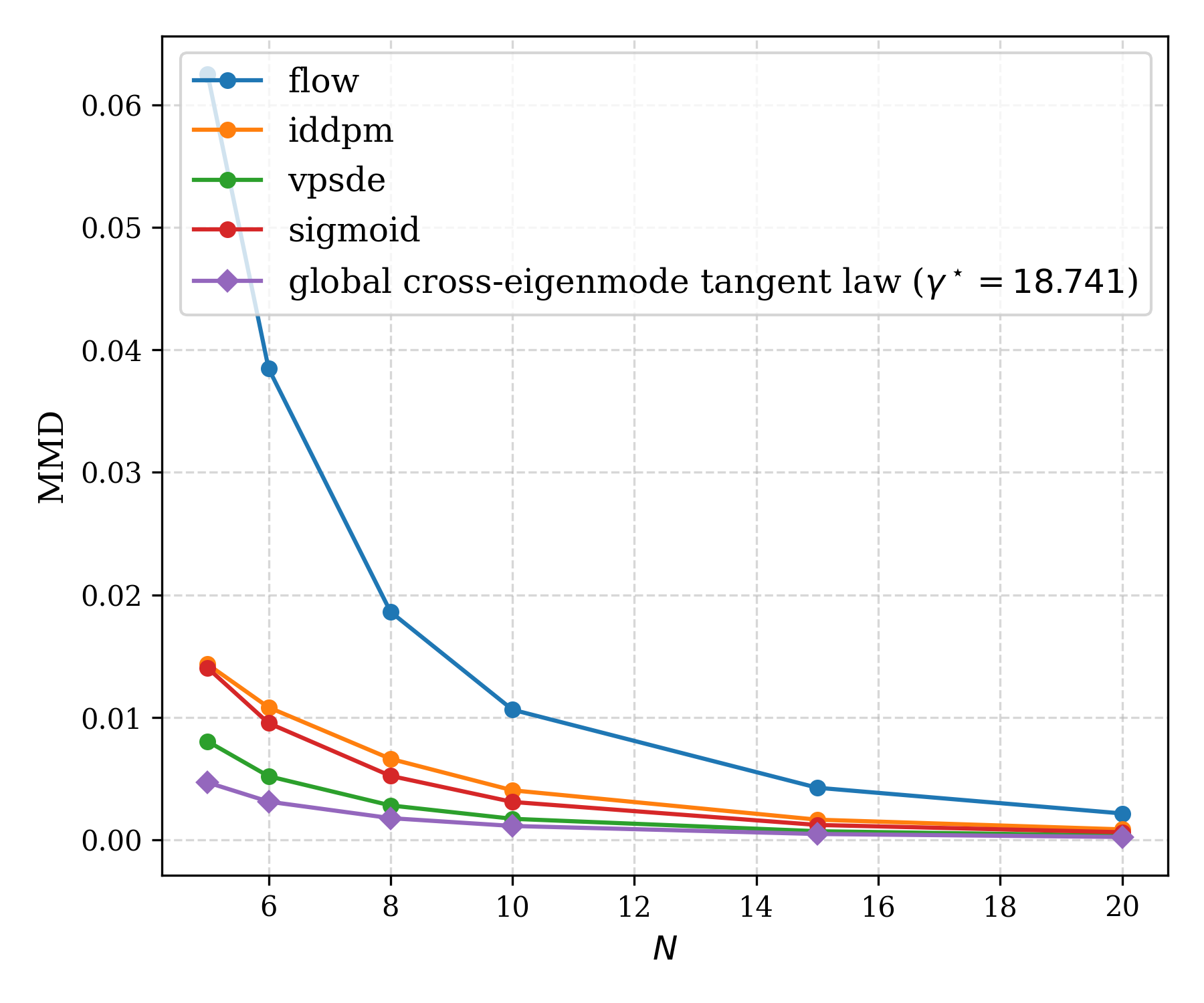} 
        \caption{MMD vs. reverse sampling steps}
    \end{subfigure}
    \hfill 
    \begin{subfigure}[t]{0.48\columnwidth} 
        \centering
        \includegraphics[width=1\textwidth]{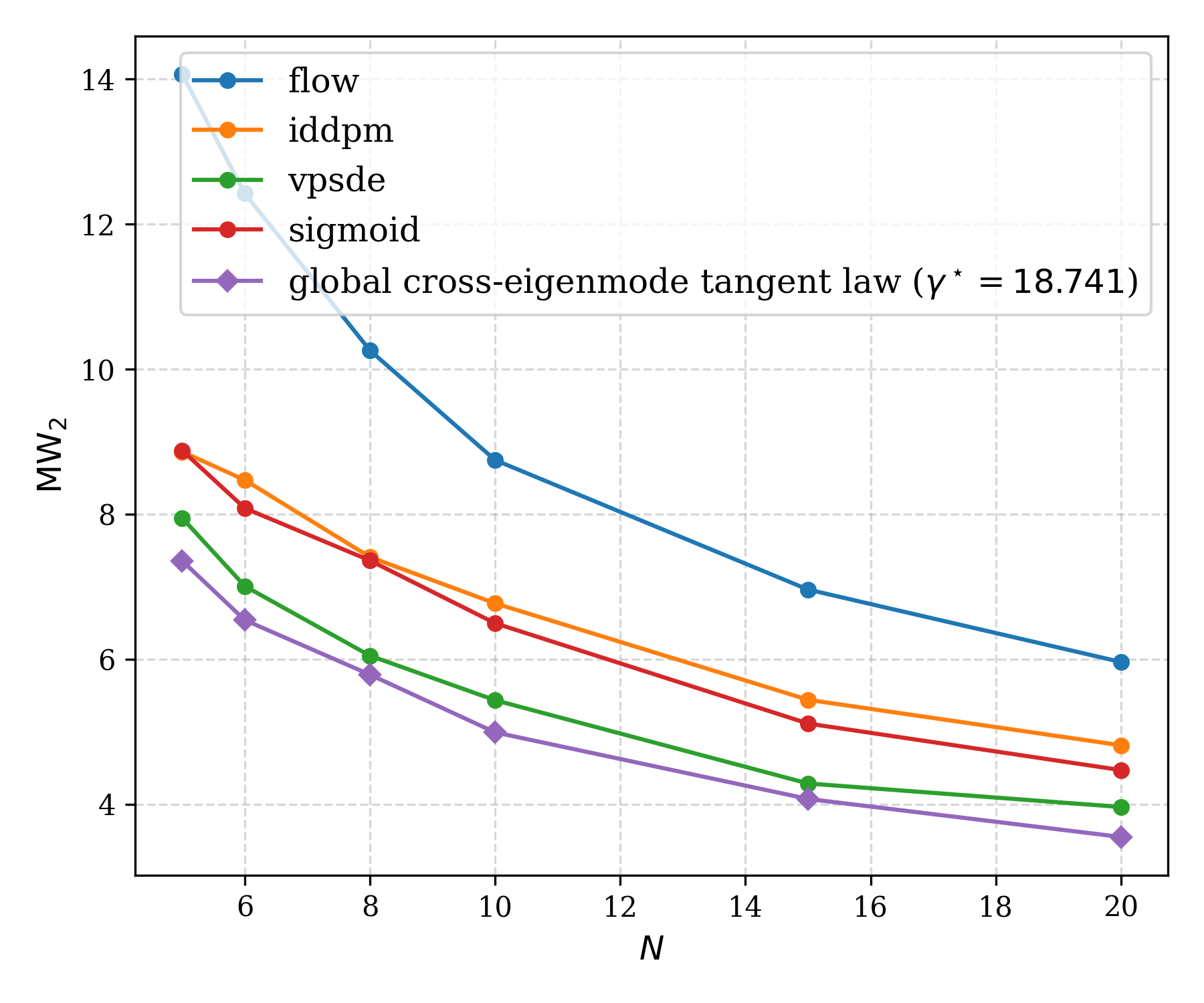} 
        \caption{$MW_2$ vs. reverse sampling steps}
    \end{subfigure}
    \caption{Numerical validation on GMM source across different noise schedules.}
    \label{fig:gmm_compare}
\end{figure}

\begin{figure}[htbp]
    \centering
    \begin{subfigure}[t]{0.31\columnwidth} 
        \centering
        \includegraphics[width=1\textwidth]{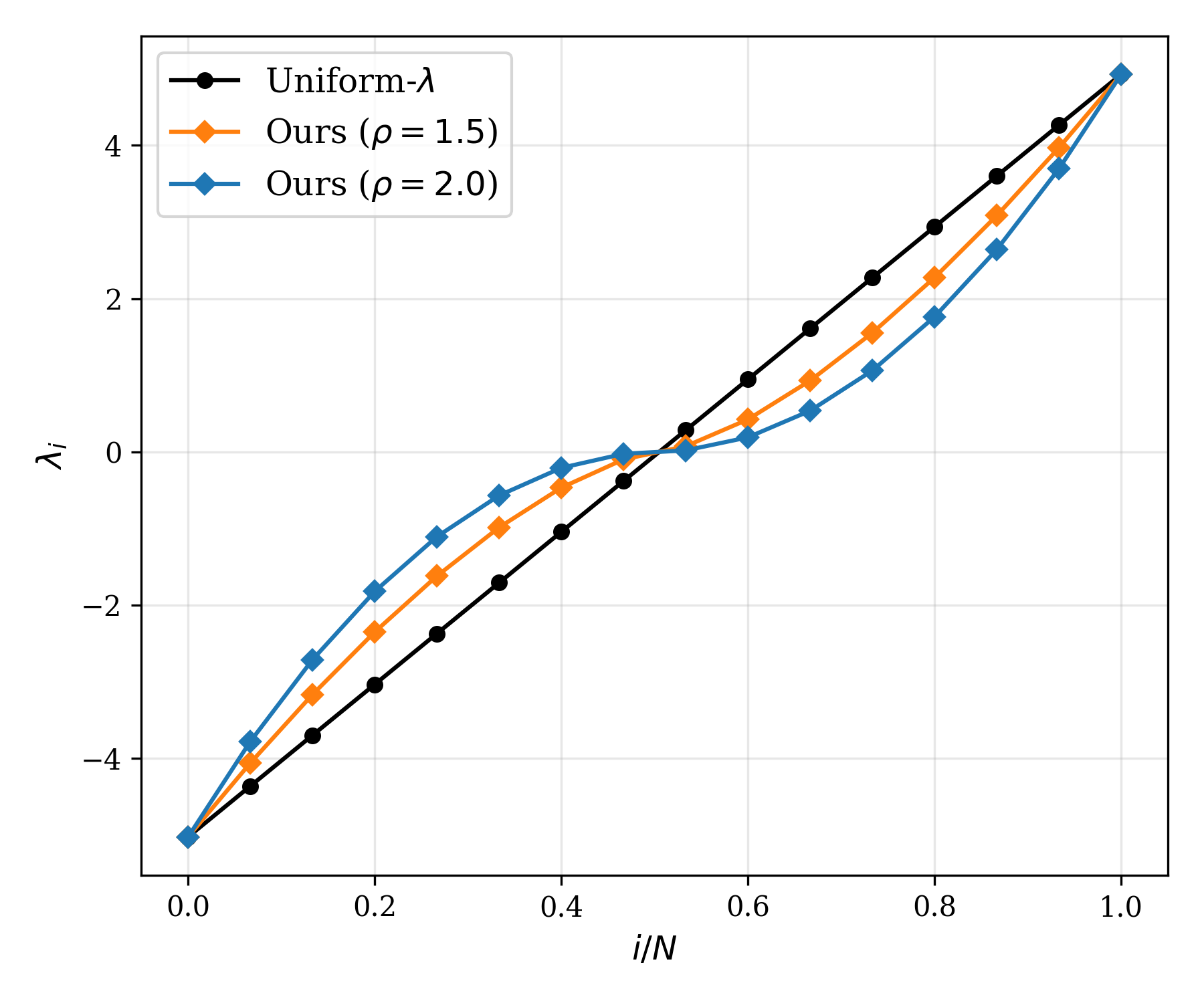} 
        \caption{$\lambda$}
        \label{subfig:lambdasch}
    \end{subfigure}
    \hfill 
    \begin{subfigure}[t]{0.31\columnwidth} 
        \centering
        \includegraphics[width=1\textwidth]{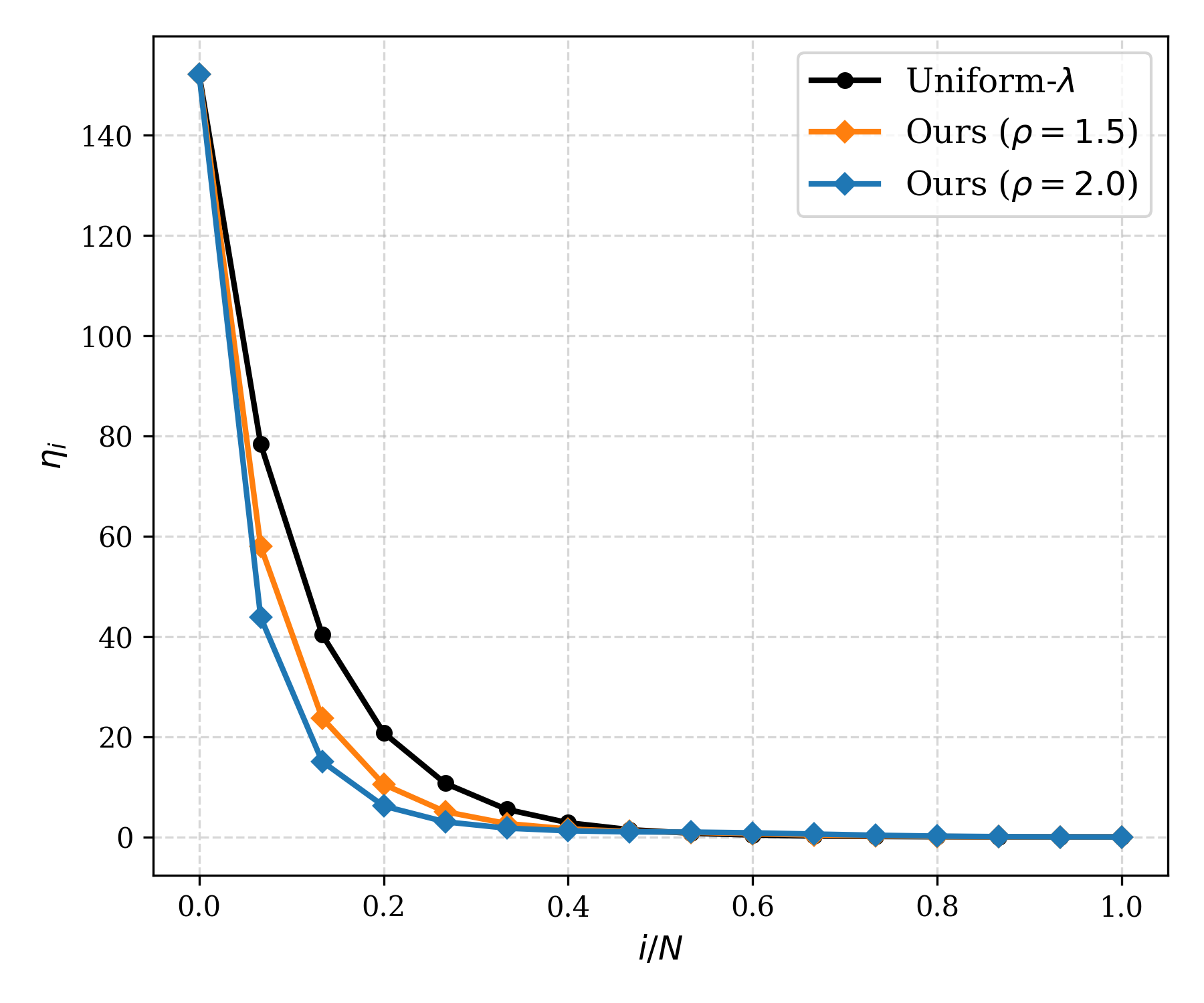} 
        \caption{$\eta$}
        \label{subfig:etasch}
    \end{subfigure}
    \hfill
    \begin{subfigure}[t]{0.31\columnwidth} 
        \centering
        \includegraphics[width=1\textwidth]{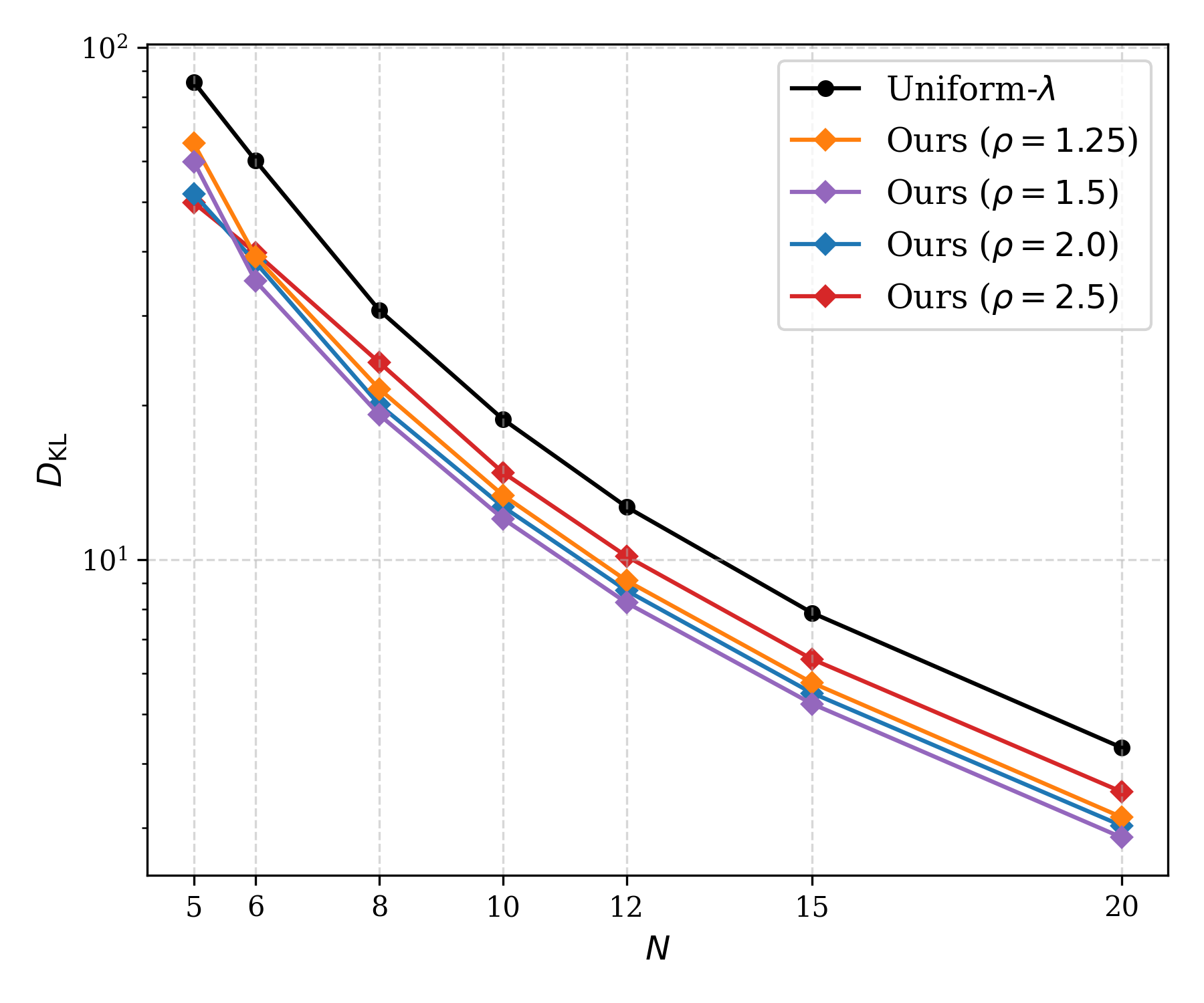} 
        \caption{$D_{\mathrm{KL}}$ vs. Discretization steps}
        \label{subfig:timediscgaussian_kl}
    \end{subfigure}
    \caption{Numerical results across different time discretization strategies}
    \label{fig:gausstimedisc}
\end{figure}

Specifically, for the synthetic multivariate Gaussian source, we construct a covariance matrix $\boldsymbol{\Sigma}_{\mathbf{x}} \in \mathbb{R}^{L \times L}$ with dimension $L=256$. To emulate the heavy-tailed spectral characteristics often observed in natural images~\cite{torralba2003statistics}, the eigenvalues of $\boldsymbol{\Sigma}_{\mathbf{x}}$ are generated according to a power law as
$
    \mu_\ell = c \cdot (\ell + i_0)^{-p},
$
where $p=1.3$ controls the decay rate, $i_0=5.0$ introduces head smoothing, and $c$ is chosen such that the largest eigenvalue equals a predefined value $\mu_{\max}$. The numerical results for this synthetic Gaussian setting are presented in \cref{fig:gaussian_validation}. In \cref{subfig:gaussian_kl}, our global cross-eigenmode tangent law achieves faster convergence than competing methods under the closed-form KL divergence in~\eqref{eq:kl_closed_t0}. We further report the theoretical lower bound induced by the anisotropic noise schedule in~\eqref{eq:matrix_schedule}, where each eigenmode follows its own optimal schedule. \cref{subfig:gaussian_alpha} and \cref{subfig:gaussian_gamma} compare the noise schedule $\alpha_t$ and present ablation studies for the choice of $\gamma$, respectively.

For the GMM source, we construct a synthetic dataset using a $K$-component GMM:
$
\label{eq:gmm_def}
q_{\mathrm{gmm}}(\mathbf{x}_0)
=
\sum_{i=1}^{K}
\pi_i \,
\mathcal{N}(\mathbf{x}_0 \mid \boldsymbol{\tau}_i, \boldsymbol{\Sigma}_i),
$
with $K=32$ and dimension $L=16$. The mixture weights $\{\pi_i\}_{i=1}^{K}$ are sampled uniformly from the probability simplex, the component centers $\{\boldsymbol{\tau}_i\}_{i=1}^{K}$ are drawn uniformly from $[-10,10]^L$, and the covariance matrices are generated as
$
\boldsymbol{\Sigma}_i = \mathbf{P}_i \mathbf{P}_i^\top + \epsilon \mathbf{I},
$
where $\mathbf{P}_i$ are Gaussian random matrices and $\epsilon$ is a small regulatory parameter introduced to avoid possible singularity.
The overall covariance matrix of the GMM is thus given by
\begin{equation}
\label{eq:gmm_cov}
\boldsymbol{\Sigma}_{\mathrm{total}}
=
\sum_{i=1}^{K}
\pi_i
\left(
\boldsymbol{\Sigma}_i
+
(\boldsymbol{\tau}_i - \bar{\boldsymbol{\tau}})
(\boldsymbol{\tau}_i - \bar{\boldsymbol{\tau}})^\top
\right),
\quad
\bar{\boldsymbol{\tau}}
=
\sum_{i=1}^{K}
\pi_i \boldsymbol{\tau}_i.
\end{equation}
which is required for computing the optimal parameter $\gamma^{\star}$ in the global cross-eigenmode tangent law schedule.
Under the forward diffusion process, the exact score function of this GMM admits a closed-form expression~\cite{wang2024unreasonable}:
\begin{equation} \label{eq:gmm_score}
    \nabla_{\mathbf{x}_t} \log q_{\mathrm{gmm},t}(\mathbf{x}_t) = \sum_{i=1}^{K} w_i(\mathbf{x}_t,t) \left[ -\boldsymbol{\Sigma}_{i,t}^{-1} (\mathbf{x}_t - \boldsymbol{\tau}_{i,t}) \right],
\end{equation}
where the time-evolved parameters are defined as $\boldsymbol{\tau}_{i,t} = \alpha_t \boldsymbol{\tau}_i$ and $\boldsymbol{\Sigma}_{i,t} = \alpha_t^2 \boldsymbol{\Sigma}_i + \sigma_t^2 \mathbf{I}$. The posterior weights are given by
$
    w_i(\mathbf{x}_t,t) = \frac{\pi_i \mathcal{N}(\mathbf{x}_t; \boldsymbol{\tau}_{i,t}, \boldsymbol{\Sigma}_{i,t})}{\sum_{j=1}^{K} \pi_j \mathcal{N}(\mathbf{x}_t; \boldsymbol{\tau}_{j,t}, \boldsymbol{\Sigma}_{j,t})}.
$
This provides an oracle score function for evaluating the reverse sampling process~\eqref{eq:ddim}. While our theoretical framework is rigorously established through the minimization of the KL divergence, numerically estimating the exact KL divergence between the target GMM and the generated distribution is highly demanding; sample-based KL estimators often suffer from severe instability and high variance. To enable a robust and sample-efficient evaluation of generative fidelity, we instead adopt maximum mean discrepancy (MMD)~\cite{gretton2012kernel}, which measures the distance between mean embeddings in a reproducing kernel Hilbert space, together with the mixture Wasserstein distance ($MW_2$)~\cite{delon2020wasserstein}, which provides a tractable upper bound for the exact $W_2$ distance between Gaussian mixtures. As shown in \cref{fig:gmm_compare}, the proposed tangent law schedule achieves superior performance under both metrics.

\subsection{Time Discretization Strategy Evaluation} \label{sub:timedisceval}
Given that training DMs for large datasets from scratch is computationally prohibitive, practitioners frequently rely on pretrained models with prescribed noise schedules to conduct inference tasks. Consequently, the generation fidelity is fundamentally affected by the choice of time discretization strategy.

Motivated by this practical consideration, in this and the next subsection, we demonstrate that our Gaussian-based analysis provides an effective tool for time discretization design as well. Instead of relying on costly empirical grid searches, we leverage the closed-form KL divergence in~\eqref{eq:kl_closed_t0} as a tractable objective for the rapid evaluation and principled selection of time discretization strategies.

To this end, we investigate a specific discretization grid, termed the power companding strategy, governed by a tunable parameter $\rho > 0$:
\begin{equation} \label{eq:lam_power_uni}
    \lambda_{i} = \left( \lambda_{t_N}^{\langle 1/\rho \rangle} + \frac{i}{N} \left( \lambda_{t_0}^{\langle 1/\rho \rangle} - \lambda_{t_N}^{\langle 1/\rho \rangle} \right) \right)^{\!\langle \rho \rangle}, \quad i=0,\dots,N,
\end{equation}
where $\lambda(t) \coloneqq \log(\alpha(t)/\sigma(t))$ denotes the half-logSNR, and we introduce the sign-preserving power operator $x^{\langle p \rangle} \coloneqq \operatorname{sgn}(x)|x|^p$ to accommodate the negative domains of $\lambda$ during the initial high-noise generation phase. Notably, this parameterization seamlessly degenerates to the standard uniform-$\lambda$ rule when $\rho=1$~\cite{lu2022dpm}. As illustrated in \cref{subfig:lambdasch} and \cref{subfig:etasch}, the power companding strategy with $\rho>1$ induces a more rapid decay of noise-to-signal ratio $\eta$ in the high-noise regime ($\lambda < 0$). Physically, this allocates coarser steps during the initial phase of the reverse sampling process.

The time sequence for the reverse sampling process~\eqref{eq:ddim} is subsequently recovered via the inverse mapping $t_{i} = \lambda^{-1}(\lambda_{i})$, which depends exclusively on the prescribed noise schedule described by $\alpha(t)$ and $\sigma(t)$.  
To benchmark this behavior, we leverage the synthetic Gaussian source introduced in \cref{sub:gmm} and evaluate its corresponding closed-form KL divergence under the VP formulation~\cite{song2020score}. As demonstrated in \cref{subfig:timediscgaussian_kl}, under an appropriately tuned companding parameter $\rho>1$, our proposed schedule achieves a superior decay rate in the generative discrepancy compared to the widely adopted uniform-$\lambda$ baseline. Applying the optimized power companding strategy, the next subsection extends our evaluation to representative real-world image generation tasks.

\subsection{Real-world Image Generation} \label{sub:realimage}
To evaluate our strategy in practical scenarios, we integrate it into off-the-shelf ODE solvers, including DPM-Solver++~\cite{lu2025dpm}, UniPC~\cite{zhao2023unipc}, and DPM-Solver-v3~\cite{zheng2023dpm}. Our evaluation encompasses various pretrained checkpoints across multiple generative tasks. Specifically, we utilize the standard VP-SDE~\cite{song2020score} and EDM~\cite{karras2022elucidating} models for CIFAR-10 ($32\times32$), alongside EDM checkpoints for larger-scale synthesis on the FFHQ and AFHQv2 datasets (both $64\times64$).

For comparative analysis, we evaluate our discretization strategy against two representative baselines: the widely adopted uniform-$\lambda$ heuristic and the recently developed DMN method~\cite{xue2024accelerating}, a state-of-the-art search approach that attempts to balance optimization overhead and generation performance. Both baselines are evaluated using their official implementation.
To assess the performance-compute trade-off, generation fidelity across varying NFEs is quantified via the FID~\cite{heusel2017fid}. Theoretically, this metric corresponds to the 2-Wasserstein distance between the Gaussian approximations of the extracted feature embeddings, echoing our analytical framework. To ensure statistical significance, all FID scores are computed utilizing 50,000 generated samples against the entire reference datasets.

\begin{figure}[htbp]
    \centering
    \begin{subfigure}[t]{0.48\columnwidth} 
        \centering
        \includegraphics[width=1\textwidth]{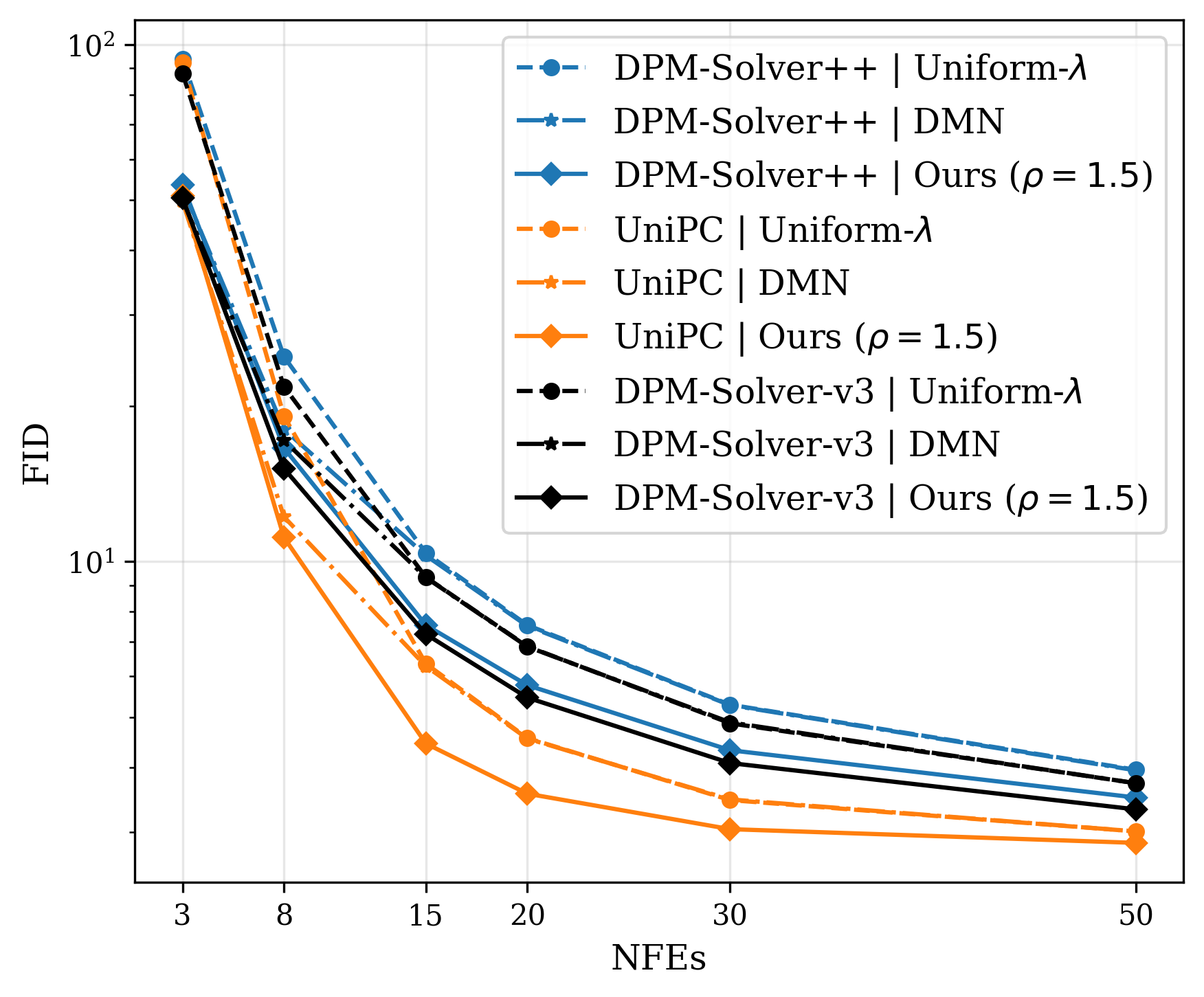} 
        \caption{VP-SDE}
    \end{subfigure}
    \hfill 
    \begin{subfigure}[t]{0.48\columnwidth} 
        \centering
        \includegraphics[width=1\textwidth]{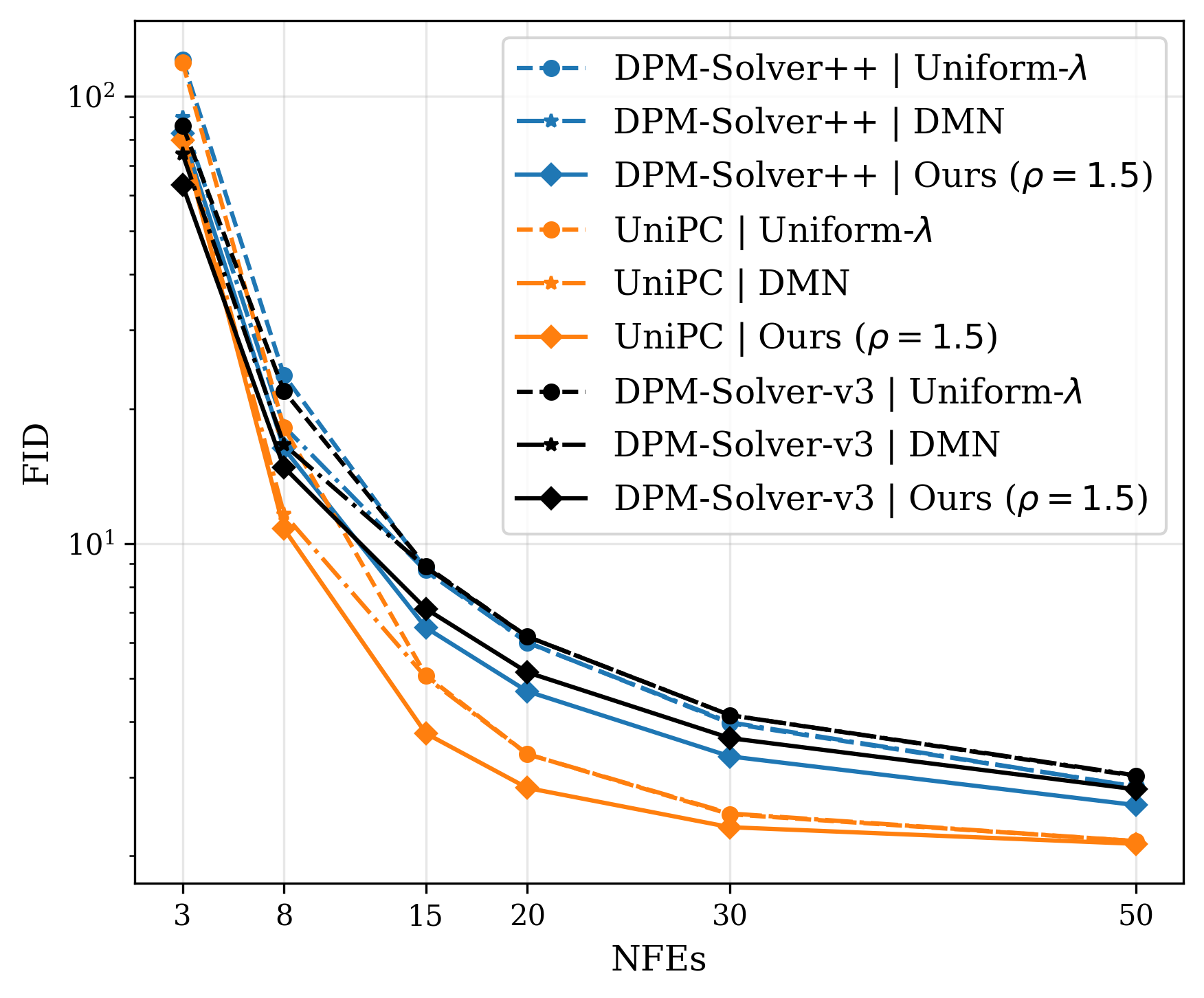} 
        \caption{EDM}
    \end{subfigure}
    \caption{Quantitative comparison of FID scores ($\downarrow$) on the CIFAR-10 dataset. The evaluation is conducted using pretrained EDM and VP-SDE architectures, coupled with three ODE solvers.}
    \label{fig:cifar_comp_fig}
\end{figure}

As illustrated in \cref{fig:cifar_comp_fig}, our discretization strategy consistently outperforms the default baseline on CIFAR-10 across all evaluated solvers and model architectures. Compared with the search-based DMN method, our approach maintains superior performance across most NFE regimes. We further validate the generality of the our strategy on FFHQ and AFHQv2, where the results in \cref{tab:ffhq_comp_tab} and \cref{tab:afhqv2_comp_tab} demonstrate clear improvements under a wide range of NFE budgets. Additional experiments with extended NFE settings are reported in Appendix~\ref{appen:imp_details}.

\begin{table}[htbp]
    \centering
    \caption{FID evaluation ($\downarrow$) on FFHQ using pretrained EDM checkpoint. Best results are in \textbf{bold}.}
    \label{tab:ffhq_comp_tab}
    \begin{tabular}{l | c c c c c }
        \toprule
        \multirow{2}{*}{FFHQ} & \multicolumn{5}{c}{NFEs} \\
        \cmidrule(lr){2-6} & 5 & 8 & 12 & 15 & 30 \\
        \midrule
        \rowcolor{gray!15} \multicolumn{6}{l}{DPM-Solver++} \\
        \quad Uniform-$\lambda$ & 50.69 & 27.63 & 16.24 & 12.28 & 5.90 \\
        \quad DMN & 41.09 & 22.13 & 15.99 & 12.14 & 5.89 \\
        \quad \textbf{Ours} ($\rho=1.5$) & \textbf{37.54} & \textbf{20.95} &\textbf{12.34} & \textbf{9.46} & \textbf{4.89} \\
        \midrule
        \rowcolor{gray!15} \multicolumn{6}{l}{UniPC} \\
        \quad Uniform-$\lambda$ & 47.19 & 22.04 & 10.77 & 7.47 & 3.46 \\
        \quad DMN & 33.37 & 14.99 & 10.51 & 7.36 & 3.47 \\
        \quad \textbf{Ours} ($\rho=1.5$) & \textbf{31.69} & \textbf{14.41} & \textbf{7.32} & \textbf{5.35} & \textbf{2.99} \\
        \bottomrule
    \end{tabular}
\end{table}

\begin{table}[htbp]
    \centering
    \caption{FID evaluation ($\downarrow$) on AFHQv2 using pretrained EDM checkpoint. Best results are in \textbf{bold}.}
    \begin{tabular}{l | c c c c c }
        \toprule
        \multirow{2}{*}{AFHQv2} & \multicolumn{5}{c}{NFEs} \\
        \cmidrule(lr){2-6} & 5 & 8 & 12 & 15 & 30 \\
        \midrule
        \rowcolor{gray!15} \multicolumn{6}{l}{DPM-Solver++} \\
        \quad Uniform-$\lambda$ & 31.55 & 13.95 & 7.97 & 6.05 & 3.27 \\
        \quad DMN & 33.12 & 16.01 & 7.76 & 5.99 & 3.28 \\
        \quad \textbf{Ours} ($\rho=1.5$) & \textbf{23.52} & \textbf{12.74} & \textbf{7.41} & \textbf{5.69} & \textbf{3.14} \\
        \midrule
        \rowcolor{gray!15} \multicolumn{6}{l}{UniPC} \\
        \quad Uniform-$\lambda$ & 29.20 & 11.07 & 5.45 & 3.96 & 2.34 \\
        \quad DMN & 29.81 & 12.68 & 5.20 & 3.91 & 2.34 \\
        \quad \textbf{Ours} ($\rho=1.5$) & \textbf{20.27} & \textbf{9.50} & \textbf{4.82} & \textbf{3.62} & \textbf{2.26} \\
        \bottomrule
    \end{tabular}
    \label{tab:afhqv2_comp_tab}
\end{table}

\section{Conclusion} \label{sec:conclu}
This paper analyzes the behavior of reverse sampling in DMs, revealing a clear relationship between noise scheduling and distributional discrepancy. We establish that for the analytically tractable Gaussian setting, the optimization of noise scheduling can be formulated and solved, leading to the tangent law whose coefficients are determined by the source covariance spectrum. Furthermore, we prove, combining stochastic analysis and information-theoretic arguments, that the Gaussian reverse sampling trajectory asymptotically achieves an extremal property on the KL divergence among all source distributions subject to given covariance constraints. This justifies that reducing the distributional discrepancy in the Gaussian setting is necessary to achieve desirable performance for general generative tasks. Finally, empirical validations across both synthetic and real-world image datasets substantiate the practical efficacy of our theoretical findings. Compelling avenues for future research include generalizing our analytical framework to guidance sampling, solving inverse problems, and other relevant fields in generative models.

\appendices

\section{Proof of \cref{prop:terminal_kl}} \label{appendix:termi_kl_proof}
Starting from the deterministic reverse sampling formulation in~\eqref{eq:ddim}, we have
\begin{align}
    \hat{\mathbf{x}}_{t_{j-1}} 
    &= \frac{\alpha_{t_{j-1}}}{\alpha_{t_{j}}} \hat{\mathbf{x}}_{t_j} + \left( \sigma_{t_{j-1}} - \frac{\alpha_{t_{j-1}}}{\alpha_{t_{j}}} \sigma_{t_j} \right) \boldsymbol{\epsilon}_{\theta}(\hat{\mathbf{x}}_{t_j}, t_j) \notag \\
    &\overset{(a)}{=} \frac{\alpha_{t_{j-1}}}{\alpha_{t_j}}\hat{\mathbf{x}}_{t_j} + \left(\sigma_{t_{j-1}}-\frac{\alpha_{t_{j-1}}}{\alpha_{t_j}}\sigma_{t_j}\right) \frac{\hat{\mathbf{x}}_{t_j}-\alpha_{t_j}\mathbb{E}_{q_G}[\mathbf{x}_0| \hat{\mathbf{x}}_{t_j}]}{\sigma_{t_j}} \notag \\
    &= \frac{\sigma_{t_{j-1}}}{\sigma_{t_j}}\hat{\mathbf{x}}_{t_j} + \left(\alpha_{t_{j-1}}-\frac{\sigma_{t_{j-1}}}{\sigma_{t_j}}\alpha_{t_j}\right) \mathbb{E}_{q_G}[\mathbf{x}_0| \hat{\mathbf{x}}_{t_j}] \notag \\
    &\overset{(b)}{=} \left[ \frac{\sigma_{t_{j-1}}}{\sigma_{t_j}} \mathbf{I} + \left(\alpha_{t_{j-1}}-\frac{\sigma_{t_{j-1}}}{\sigma_{t_j}}\alpha_{t_j}\right) \alpha_{t_j}\boldsymbol{\Sigma}_{\mathbf{x}} \left(\alpha_{t_j}^2\boldsymbol{\Sigma}_{\mathbf{x}}+\sigma_{t_j}^2\mathbf{I}\right)^{-1} \right] \hat{\mathbf{x}}_{t_j} \notag \\
    &= \left[ \frac{\sigma_{t_{j-1}}}{\sigma_{t_j}} \left(\alpha_{t_j}^2\boldsymbol{\Sigma}_{\mathbf{x}}+\sigma_{t_j}^2\mathbf{I}\right) + \left(\alpha_{t_{j-1}}\alpha_{t_j}\boldsymbol{\Sigma}_{\mathbf{x}}-\frac{\sigma_{t_{j-1}}}{\sigma_{t_j}}\alpha_{t_j}^{2}\boldsymbol{\Sigma}_{\mathbf{x}}\right) \right] \left(\alpha_{t_j}^2\boldsymbol{\Sigma}_{\mathbf{x}}+\sigma_{t_j}^2\mathbf{I}\right)^{-1} \hat{\mathbf{x}}_{t_j} \notag \\
    &= \left(\sigma_{t_{j-1}}\sigma_{t_j}\mathbf{I} +\alpha_{t_{j-1}}\alpha_{t_j}\boldsymbol{\Sigma}_{\mathbf{x}}\right) \left(\alpha_{t_j}^2\boldsymbol{\Sigma}_{\mathbf{x}}+\sigma_{t_j}^2\mathbf{I}\right)^{-1} \hat{\mathbf{x}}_{t_j} \notag \\
    &\overset{(c)}{=} \mathbf{U} \left(\sigma_{t_{j-1}}\sigma_{t_j}\mathbf{I} + \alpha_{t_{j-1}}\alpha_{t_j}\boldsymbol{\Lambda}\right) (\alpha_{t_j}^2\boldsymbol{\Lambda}+\sigma_{t_j}^2\mathbf{I})^{-1}\mathbf{U}^{\top} \hat{\mathbf{x}}_{t_j}, \label{eq:app_linear_update}
\end{align}
where (a) follows by substituting the optimal denoising estimator under Gaussian source in~\eqref{eq:optepi}, (b) applies the closed-form linear MMSE estimator in~\eqref{eq:mmse}, and (c) utilizes the orthogonal eigendecomposition of the covariance matrix $\boldsymbol{\Sigma}_{\mathbf{x}}=\mathbf{U}\boldsymbol{\Lambda}\mathbf{U}^{\top}$ with $\boldsymbol{\Lambda}=\mathrm{diag}(\mu_1,\ldots,\mu_L)$, thereby completely reducing the original update in~\eqref{eq:ddim} to a linear process.

By recursively applying the one-step update in~\eqref{eq:app_linear_update} over the entire time discretization sequence $\{t_i\}_{i=0}^{N}$, we obtain the direct mapping from the initial noise to the terminal state
\begin{equation} \label{eq:app_x_t0_recursive}
    \hat{\mathbf{x}}_{t_0} = \mathbf{U} \left( \prod_{j=1}^{N} \left(\sigma_{t_{j-1}}\sigma_{t_j}\mathbf{I} + \alpha_{t_{j-1}}\alpha_{t_j}\boldsymbol{\Lambda}\right) (\alpha_{t_j}^2\boldsymbol{\Lambda}+\sigma_{t_j}^2\mathbf{I})^{-1} \right) \mathbf{U}^{\top} \hat{\mathbf{x}}_{t_N}.
\end{equation}
Recall from \cref{sub:pre_dms} that the reverse sampling is initialized with Gaussian noise $p_{\mathrm{prior}}$
\begin{equation} \label{eq:app_init_noise}
    \hat{\mathbf{x}}_{t_N} \sim \mathcal{N}(\mathbf{0}, c^2 \mathbf{I}), \quad \text{with} \quad
    c =
    \begin{cases}
        1 & \text{for VP setting}, \\
        \sigma_{\max} & \text{for VE setting}.
    \end{cases}
\end{equation}
Since the recursive relation in~\eqref{eq:app_x_t0_recursive} defines a linear transformation applied to the Gaussian random vector $\hat{\mathbf{x}}_{t_N}$, the resulting terminal marginal distribution remains Gaussian. So, we have
\begin{equation} \label{eq:p_xt0_gauss_M}
    \hat{p}_{G,t_0}(\hat{\mathbf{x}}_{t_0}) = \mathcal{N}\!\left(\mathbf{0},\, \mathbf{U}\mathbf{M}\mathbf{U}^{\top}\right),
\end{equation}
where $\mathbf{M}$ is a diagonal matrix whose $\ell$-th entry is explicitly given by
\begin{equation} \label{eq:m_ell_app}
    m_\ell = c^2 \prod_{j=1}^{N} \left( \frac{\alpha_{t_{j-1}}\alpha_{t_j}\mu_\ell+\sigma_{t_{j-1}}\sigma_{t_j}}{\alpha_{t_j}^2\mu_\ell+\sigma_{t_j}^2} \right)^{\!2}, \qquad \ell=1,\ldots,L.
\end{equation}

Recall from~\eqref{eq:forward_prob} that the forward marginal at $t_0$ is given by $q_{G,t_0}(\mathbf{x}_{t_0}) = \mathcal{N}(\mathbf{0}, \boldsymbol{\Sigma}_{t_0})$, where $\boldsymbol{\Sigma}_{t_0} = \alpha_{t_0}^2 \boldsymbol{\Sigma}_{\mathbf{x}} + \sigma_{t_0}^2 \mathbf{I}$. Utilizing the eigen-decomposition $\boldsymbol{\Sigma}_{\mathbf{x}} = \mathbf{U}\boldsymbol{\Lambda}\mathbf{U}^{\top}$, we can diagonalize this covariance matrix as
\begin{equation} \label{eq:app_q_t0_diag}
    q_{G,t_0}(\mathbf{x}_{t_0}) = \mathcal{N}(\mathbf{0}, \mathbf{U}\mathbf{N}\mathbf{U}^{\top}), \quad \text{with} \quad n_\ell = \alpha_{t_0}^2 \mu_\ell + \sigma_{t_0}^2.
\end{equation}

Finally, substituting the exact covariance $\boldsymbol{\Sigma}_1 = \mathbf{U}\mathbf{N}\mathbf{U}^{\top}$ and the generated covariance $\boldsymbol{\Sigma}_2 = \mathbf{U}\mathbf{M}\mathbf{U}^{\top}$ into the closed-form KL divergence formula presented in~\eqref{eq:kl_div}, we obtain
\begin{align} 
    D_{\mathrm{KL}}\!\left( q_{G,t_0}(\mathbf{x}_{t_0})\,\|\,\hat{p}_{G,t_0,}(\hat{\mathbf{x}}_{t_0})\right)  
    &= D_{\mathrm{KL}}\!\left(\mathcal{N}(\mathbf{0}, \mathbf{U}\mathbf{N}\mathbf{U}^{\top}) \,\|\, \mathcal{N}(\mathbf{0}, \mathbf{U}\mathbf{M}\mathbf{U}^{\top})\right) \notag\\
    &= \frac{1}{2} \left[ \mathrm{Tr}(\mathbf{M}^{-1}\mathbf{N}) - L + \ln \frac{\det \mathbf{M}}{\det \mathbf{N}} \right] \notag\\
    &= \frac{1}{2} \sum_{\ell=1}^{L} \left( \frac{n_\ell}{m_\ell} - \log \frac{n_\ell}{m_\ell} - 1 \right). \label{eq:app_kl_final}
\end{align}
This concludes the proof of \cref{prop:terminal_kl}. \qed

\section{Proof of \cref{prop:asymptotic_kl}} \label{appendix:asy_kl_proof}

To begin, taking the logarithm of the eigenvalue $m_\ell$ converts it into a summation over sampling steps. Specifically, under the VP setting, this relationship is given by
\begin{equation} \label{eq:S_ell}
     \frac{1}{2} \log m_\ell \coloneqq S_{\ell}^{N}  = \sum_{j=1}^{N} \log\!\left( \frac{\alpha_{t_{j-1}}\alpha_{t_j}\mu_\ell+\sigma_{t_{j-1}}\sigma_{t_j}}{\alpha_{t_j}^{2}\mu_\ell+\sigma_{t_j}^{2}} \right).
\end{equation}

Then we have the following lemma to characterize the asymptotic behavior of this sum $S_{\ell}^{N}$:
\begin{lemma} \label{lem:S_ell_expansion}
    Let $\{t_i\}_{i=0}^N$ be a uniform time grid on $[0, 1]$ defined by $t_i \coloneqq i/N$, with a constant step size $h \coloneqq 1/N$. Provided that the schedule functions $\alpha_t$ and $\sigma_t$ are sufficiently smooth, the sum $S_{\ell}^{N}$ admits the following asymptotic expansion as $N \to \infty$ (i.e., $h \to 0$):
    \begin{equation} \label{eq:S_ell_em_form}
        S_{\ell}^{N} = I_\ell + E_\ell h + \mathcal{O}(h^{2}),
    \end{equation}
    where
    \begin{equation} \label{eq:E_ell_def}
        I_\ell=\int_{0}^{1} F_\ell(t)\,\mathrm{d}t, \quad E_{\ell} = \frac{1}{2}\bigl(F_{\ell}(1) - F_{\ell}(0)\bigr) + \int_{0}^{1} G_{\ell}(t)\,\mathrm{d}t,
    \end{equation}
    and the functions $F_{\ell}(t)$ and $G_{\ell}(t)$ are defined as:
    \begin{align}
        F_{\ell}(t) &= - \frac{\alpha_t\dot{\alpha}_t\mu_\ell+\sigma_t\dot{\sigma}_t}{\alpha_t^{2}\mu_\ell+\sigma_t^{2}}, \label{eq:F_ell_def} \\
        G_{\ell}(t) &= \frac{\alpha_t\ddot{\alpha}_t\mu_\ell+\sigma_t\ddot{\sigma}_t}{2(\alpha_t^{2}\mu_\ell+\sigma_t^{2})} - \frac{(\alpha_t\dot{\alpha}_t\mu_\ell+\sigma_t\dot{\sigma}_t)^{2}}{2(\alpha_t^{2}\mu_\ell+\sigma_t^{2})^{2}}. \label{eq:G_ell_def}
    \end{align}
\end{lemma}

\begin{proof}
    In the continuous-time limit with smooth noise schedule functions, we perform a Taylor expansion at $t_j$ (with $t_{j-1}=t_j-h$). The schedule functions at the previous time step can be expressed as:
    \begin{align} \label{eq:alpha_tj1_expansion}
        \alpha_{t_{j-1}} &= \alpha(t_j-h) = \alpha_{t_j} - h\,\dot{\alpha}_{t_j} + \frac{h^2}{2}\,\ddot{\alpha}_{t_j} + \mathcal{O}(h^3), \notag\\
        \sigma_{t_{j-1}} &= \sigma(t_j-h) = \sigma_{t_j} - h\,\dot{\sigma}_{t_j} + \frac{h^2}{2}\,\ddot{\sigma}_{t_j} + \mathcal{O}(h^3),
    \end{align}
where $\dot{\alpha}_{t_j}\triangleq (\mathrm{d}\alpha/\mathrm{d}t)(t_j)$ and $\ddot{\alpha}_{t_j}\triangleq (\mathrm{d}^2\alpha/\mathrm{d}t^2)(t_j)$, with analogous definitions for $\sigma_t$. 

Substituting~\eqref{eq:alpha_tj1_expansion} into the numerator of the summand in~\eqref{eq:S_ell}, we obtain:
\begin{align} \label{eq:numerator_expansion}
    \alpha_{t_{j-1}}\alpha_{t_j}\mu_\ell+\sigma_{t_{j-1}}\sigma_{t_j}
    &= \left(\alpha_{t_j}-h\,\dot{\alpha}_{t_j}+\frac{h^2}{2}\,\ddot{\alpha}_{t_j} \right)\alpha_{t_j}\mu_\ell + \left(\sigma_{t_j}-h\,\dot{\sigma}_{t_j}+\frac{h^2}{2}\,\ddot{\sigma}_{t_j} \right)\sigma_{t_j} + \mathcal{O}(h^3) \notag\\
    &= \alpha_{t_j}^2 \mu_\ell + \sigma_{t_j}^2 - h\bigl(\mu_\ell \alpha_{t_j}\dot{\alpha}_{t_j} + \sigma_{t_j}\dot{\sigma}_{t_j}\bigr) + \frac{h^2}{2}\bigl(\mu_\ell \alpha_{t_j}\ddot{\alpha}_{t_j} + \sigma_{t_j}\ddot{\sigma}_{t_j}\bigr) + \mathcal{O}(h^3).
\end{align}
Dividing the expanded numerator by the denominator $(\alpha_{t_j}^{2}\mu_\ell+\sigma_{t_j}^{2})$, the sum $S_{\ell}^{N}$ can be rewritten as:
\begin{align} \label{eq:sn_log_delta}
    S_{\ell}^{N} &= \sum_{j=1}^{N} \log \left( \frac{\alpha_{t_{j-1}}\alpha_{t_j}\mu_\ell+\sigma_{t_{j-1}}\sigma_{t_j}}{\alpha_{t_j}^{2}\mu_\ell+\sigma_{t_j}^{2}} \right) \notag\\
    &\coloneqq \sum_{j=1}^{N} \log \left( 1 + \delta_{j} \right),
\end{align}
where the term $\delta_j$ can be asymptotically expanded as:
\begin{align} \label{eq:delta_j_def}
    \delta_{j} &= \frac{- h\bigl(\mu_\ell \alpha_{t_j}\dot{\alpha}_{t_j} + \sigma_{t_j}\dot{\sigma}_{t_j}\bigr) + \frac{h^2}{2}\bigl(\mu_\ell \alpha_{t_j}\ddot{\alpha}_{t_j} + \sigma_{t_j}\ddot{\sigma}_{t_j}\bigr)}{\alpha_{t_j}^{2}\mu_\ell+\sigma_{t_j}^{2}} + \mathcal{O}(h^{3}) \notag \\
    &= -\frac{\mu_\ell \alpha_{t_j}\dot{\alpha}_{t_j} + \sigma_{t_j}\dot{\sigma}_{t_j}}{\alpha_{t_j}^{2}\mu_\ell+\sigma_{t_j}^{2}}\,h + \frac{\mu_\ell \alpha_{t_j}\ddot{\alpha}_{t_j} + \sigma_{t_j}\ddot{\sigma}_{t_j}}{2(\alpha_{t_j}^{2}\mu_\ell+\sigma_{t_j}^{2})}\,h^2 +\mathcal{O}(h^3) \notag \\
    &\coloneqq a_j h + b_j h^2 + \mathcal{O}(h^3).
\end{align}

Noting from~\eqref{eq:delta_j_def} that the term $\delta_j$ satisfies $\delta_j = \mathcal{O}(h)$, we square it to obtain $\delta_j^2 = a_j^2 h^2 + \mathcal{O}(h^3)$. Applying the standard Taylor expansion $\log(1+x)=x-\frac{x^2}{2}+\mathcal{O}(x^3)$, we deduce that:
\begin{align} \label{eq:log_expansion}
    \log(1+\delta_j) &= \delta_j - \frac{1}{2}\delta_j^2 + \mathcal{O}(\delta_j^3) \notag \\
    &= \left( a_j h + b_j h^2 \right) - \frac{1}{2}  a_j^2 h^2 + \mathcal{O}(h^3) \notag \\
    &\coloneqq F_\ell(t_j) h + G_\ell(t_j) h^{2} + \mathcal{O}(h^{3}),
\end{align}
where the derived coefficients for $h$ and $h^2$ precisely match the definitions of $F_\ell(t_j)$ and $G_\ell(t_j)$ presented in~\eqref{eq:F_ell_def} and~\eqref{eq:G_ell_def}. Summing over all $N$ steps, the accumulation of local $\mathcal{O}(h^3)$ errors yields a global bound of $N \cdot \mathcal{O}(h^3) = \mathcal{O}(h^2)$, which establishes the following formula:
\begin{equation} \label{eq:discrete_sum_conclusion}
    S_{\ell}^{N} = \sum_{j=1}^{N} \Bigl( F_\ell(t_j)\,h + G_\ell(t_j)\,h^{2} \Bigr) + \mathcal{O}(h^{2}).
\end{equation}

To further evaluate the summation in~\eqref{eq:discrete_sum_conclusion}, we invoke the Euler-Maclaurin formula~\cite{zwillinger2002crc}. Specifically, for a sufficiently smooth function $f$ evaluated on the uniform grid $t_j \coloneqq j h$ over $[0, 1]$, the sum admits the following asymptotic expansion:
\begin{equation} \label{eq:emformula}
    \sum_{j=1}^{N} f(t_j)\, h = \int_{0}^{1} f(t)\,\mathrm{d}t + \frac{h}{2}\bigl(f(1)-f(0)\bigr) + \frac{h^2}{12}\bigl(f'(1)-f'(0)\bigr) + \mathcal{O}(h^4).
\end{equation}
    
Applying the Euler-Maclaurin expansion~\eqref{eq:emformula} to the respective discrete sums of $F_\ell(t)$ and $G_\ell(t)$, we obtain:
\begin{align}
    \sum_{j=1}^{N} F_{\ell}(t_{j})\,h &= \int_{0}^{1} F_{\ell}(t)\,\mathrm{d}t + \frac{h}{2}\bigl(F_{\ell}(1)-F_{\ell}(0)\bigr) + \mathcal{O}(h^{2}), \label{eq:em_F_term} \\
    \sum_{j=1}^{N} G_{\ell}(t_{j})\,h &= \int_{0}^{1} G_{\ell}(t)\,\mathrm{d}t + \mathcal{O}(h). \label{eq:em_G_term}
\end{align}
    
Recalling the sum expansion from~\eqref{eq:discrete_sum_conclusion} and substituting the integral approximations from~\eqref{eq:em_F_term} and~\eqref{eq:em_G_term}, we explicitly evaluate the sum $S_{\ell}^{N}$ as:
\begin{align}
      S_{\ell}^{N} &= \sum_{j=1}^{N} \Bigl( F_{\ell}(t_j)\,h + G_{\ell}(t_j)\,h^{2} \Bigr) + \mathcal{O}(h^{2}) \notag \\
      &= \sum_{j=1}^{N}F_{\ell}(t_j)\,h + h \left( \sum_{j=1}^{N}G_{\ell}(t_j)\,h \right) + \mathcal{O}(h^{2}) \notag \\
      &= \left[ \int_{0}^{1} F_{\ell}(t)\,\mathrm{d}t + \frac{h}{2}\bigl(F_{\ell}(1)-F_{\ell}(0)\bigr) \right] + h\left[ \int_{0}^{1} G_{\ell}(t)\,\mathrm{d}t + \mathcal{O}(h) \right] + \mathcal{O}(h^2) \notag \\
      &= I_\ell + \left[ \frac{1}{2}\bigl(F_{\ell}(1)-F_{\ell}(0)\bigr) + \int_{0}^{1} G_{\ell}(t)\,\mathrm{d}t \right] h + \mathcal{O}(h^{2}). \label{eq:SN_expand_app}
\end{align}
    
This rearrangement recovers our asymptotic form in~\eqref{eq:S_ell_em_form}, thereby yielding the expression for the first-order error coefficient:
\begin{equation} \label{eq:E1ell_app}
      E_{\ell} = \frac{1}{2}\bigl(F_{\ell}(1)-F_{\ell}(0)\bigr) + \int_{0}^{1} G_{\ell}(t)\,\mathrm{d}t.
\end{equation}
This concludes the proof of \cref{lem:S_ell_expansion}.
\end{proof}

Furthermore, we reduce the first-order error coefficient on the right-hand side of~\eqref{eq:S_ell_em_form} to the following closed-form expressions:

\begin{lemma} \label{lem:integral_simplification}
    The first-order error coefficient $E_\ell$ reduces to
    \begin{equation} \label{eq:E_ell_app}
        E_\ell = -\frac{\mu_{\ell}}{2} \int_{0}^{1} \left(\frac{\alpha_t\dot{\sigma}_t-\sigma_t\dot{\alpha}_t}{\alpha_t^{2}\mu_{\ell}+\sigma_t^{2}}\right)^2 \,\mathrm{d}t.
    \end{equation}
\end{lemma}
\begin{proof}
To derive the simplified expression for $E_\ell$, we analyze the integral of $G_\ell(t)$. For clarity, we split $G_\ell(t)$ defined in~\eqref{eq:G_ell_def} into two components, $G_\ell(t) = A_\ell(t) - B_\ell(t)$, where:
\begin{align}
    A_\ell(t) = \frac{\alpha_t\ddot{\alpha}_t\mu_\ell+\sigma_t\ddot{\sigma}_t}{2(\alpha_t^{2}\mu_\ell+\sigma_t^{2})}, 
    \,\, B_\ell(t) = \frac{(\alpha_t\dot{\alpha}_t\mu_\ell+\sigma_t\dot{\sigma}_t)^{2}}{2(\alpha_t^{2}\mu_\ell+\sigma_t^{2})^{2}}.
\end{align}

We apply integration by parts exclusively to the integral of $A_\ell(t)$:
\begin{align} \label{eq:A_ibp}
    \int_{0}^{1} A_{\ell}(t)\,\mathrm{d}t 
    &= \frac{1}{2} \int_{0}^{1} \frac{\alpha_t\mu_\ell}{\alpha_t^{2}\mu_\ell+\sigma_t^{2}} \,\mathrm{d}(\dot{\alpha}_t) + \frac{1}{2} \int_{0}^{1} \frac{\sigma_t}{\alpha_t^{2}\mu_\ell+\sigma_t^{2}} \,\mathrm{d}(\dot{\sigma}_t) \notag \\
    &= \frac{1}{2} \left[ \frac{\alpha_t\dot{\alpha}_t\mu_\ell + \sigma_t\dot{\sigma}_t}{\alpha_t^{2}\mu_\ell+\sigma_t^{2}} \right]_{0}^{1} - \frac{1}{2}\int_{0}^{1} \left( \dot{\alpha}_t\frac{\mathrm{d}}{\mathrm{d}t}\frac{\alpha_t\mu_\ell}{\alpha_t^2\mu_\ell+\sigma_t^2} + \dot{\sigma}_t\frac{\mathrm{d}}{\mathrm{d}t}\frac{\sigma_t}{\alpha_t^2\mu_\ell+\sigma_t^2} \right)\mathrm{d}t \notag \\
    &= -\frac{1}{2}\bigl(F_{\ell}(1)-F_{\ell}(0)\bigr) - \frac{1}{2}\int_{0}^{1} \frac{(\dot{\alpha}_t^{2}\mu_\ell+\dot{\sigma}_t^{2})(\alpha_t^2\mu_\ell+\sigma_t^2) - 2 (\alpha_t\dot{\alpha}_t\mu_\ell+\sigma_t\dot{\sigma}_t)^2}{(\alpha_t^2\mu_\ell+\sigma_t^2)^{2}} \,\mathrm{d}t,
\end{align}
where we identify the boundary term explicitly as $-\frac{1}{2}\bigl(F_{\ell}(1)-F_{\ell}(0)\bigr)$ based on the definition of $F_{\ell}(t)$ in~\eqref{eq:F_ell_def}. 

Substituting the result from~\eqref{eq:A_ibp} back into the definition of $E_{\ell}$ in~\eqref{eq:E_ell_def}, the boundary evaluation of the integral precisely cancels the inherent boundary terms of the Euler-Maclaurin formulation:
\begin{align} \label{eq:E_final_derivation}
    E_{\ell} &= \frac{1}{2}\bigl(F_{\ell}(1)-F_{\ell}(0)\bigr) + \int_{0}^{1} A_{\ell}(t)\,\mathrm{d}t - \int_{0}^{1} B_{\ell}(t)\,\mathrm{d}t \notag \\
    &= -\frac{1}{2}\int_{0}^{1} \frac{(\dot{\alpha}_t^{2}\mu_\ell+\dot{\sigma}_t^{2}) (\alpha_t^2\mu_\ell+\sigma_t^2) - 2(\alpha_t\dot{\alpha}_t\mu_\ell+\sigma_t\dot{\sigma}_t)^2}{(\alpha_t^2\mu_\ell+\sigma_t^2)^2}\,\mathrm{d}t - \int_{0}^{1} \frac{(\alpha_t\dot{\alpha}_t\mu_\ell+\sigma_t\dot{\sigma}_t)^2}{2(\alpha_t^2\mu_\ell+\sigma_t^2)^2}\,\mathrm{d}t \notag \\
    &= -\frac{1}{2}\int_{0}^{1} \frac{ (\dot{\alpha}_t^{2}\mu_\ell+\dot{\sigma}_t^{2})(\alpha_t^2\mu_\ell+\sigma_t^2) - (\alpha_t\dot{\alpha}_t\mu_\ell+\sigma_t\dot{\sigma}_t)^{2} }{(\alpha_t^2\mu_\ell+\sigma_t^2)^{2}}\,\mathrm{d}t \notag \\
    &= -\frac{\mu_{\ell}}{2}\int_{0}^{1} \left(\frac{\alpha_t\dot{\sigma}_t-\sigma_t\dot{\alpha}_t}{\alpha_t^{2}\mu_{\ell}+\sigma_t^{2}}\right)^2 \,\mathrm{d}t.
\end{align}
The last step above follows directly from Lagrange's identity, specifically $(a^2+b^2)(c^2+d^2) - (ac+bd)^2 = (ad-bc)^2$. This establishes the exact expression presented in~\eqref{eq:E_ell_app}.
\end{proof}

Building upon \cref{lem:S_ell_expansion} and \cref{lem:integral_simplification}, we now present the main proposition that establishes the relationship between the true and generated eigenvalues, $n_\ell$ and $m_\ell$:

\begin{proposition} \label{prop:log_expansion}
    Under the uniform discretization setting with step size $h$, the scaled logarithmic ratio of the corresponding eigenvalues expands asymptotically as:
    \begin{equation} \label{eq:log_ratio_expansion}
        r_\ell \coloneqq \frac{1}{2}\log \frac{n_\ell}{m_\ell} = -E_{\ell} h + \mathcal{O}(h^{2}),
    \end{equation}
    where the coefficient $E_\ell$ is given by~\eqref{eq:E_ell_app}.
\end{proposition}
    
\begin{proof}
    Under the boundary limit conditions of the continuous VP schedule, the integral $I_{\ell}$ admits a closed-form solution
    \begin{align} 
        I_{\ell} &= \int_{0}^{1} F_{\ell}(t)\,\mathrm{d}t = \int_{0}^{1} - \frac{\alpha_t\dot{\alpha}_t\mu_\ell+\sigma_t\dot{\sigma}_t}{\alpha_t^{2}\mu_\ell+\sigma_t^{2}}\,\mathrm{d}t \notag \\
        &= -\frac{1}{2}\int_{0}^{1}\frac{\mathrm{d}(\alpha_t^{2}\mu_\ell+\sigma_t^{2})}{\alpha_t^{2}\mu_\ell+\sigma_t^{2}} = -\frac{1}{2} \log(\alpha^{2}_{t}\mu_\ell+\sigma^{2}_{t})\bigg|_{t=0}^{t=1} \label{eq:Iellvp} \\
        &= \frac{1}{2} \log n_\ell,\label{eq:Iell_closed_proof}
    \end{align}
    where the last equality follows from the definition of $n_\ell$ in~\eqref{eq:app_q_t0_diag} and the boundary condition of the continuous VP schedule (i.e., $\lim_{t\to1} \mathrm{SNR}_t = 0$, implying $\lim_{t\to1} \alpha_t=0$ and $\lim_{t\to1} \sigma_t=1$).

    Recall the definition of the sum $S_\ell^N \coloneqq \frac{1}{2} \log m_\ell$ from~\eqref{eq:S_ell}. Substituting these identities into the Euler-Maclaurin expansion $S_\ell^N = I_\ell + E_\ell h + \mathcal{O}(h^2)$ established in \cref{lem:S_ell_expansion}, we have
    \begin{equation} \label{eq:I_minus_S_relation}
        \frac{1}{2} \log n_\ell - \frac{1}{2} \log m_\ell = I_\ell - S_\ell^{N} = -E_\ell h + \mathcal{O}(h^2),
    \end{equation}
    which concludes the proof.
\end{proof}

Building upon the key relationship established in \cref{prop:log_expansion}, we now arrive at our goal. The property $r_\ell = \mathcal{O}(1/N)$ justifies the application of a local Taylor expansion. Substituting the definition $2r_\ell = \log(n_\ell/m_\ell)$ into the closed-form KL divergence~\eqref{eq:kl_closed_t0} established in \cref{prop:terminal_kl}, we obtain:
\begin{align} \label{eq:KL_E_ell}
    D_{\mathrm{KL}}\!\left(q_{G,0} \,\|\, \hat{p}_{G,0}\right)
    &= \frac{1}{2} \sum_{\ell=1}^{L} \bigl(e^{2r_\ell} - 2r_\ell - 1\bigr) \notag \\
    &= \sum_{\ell=1}^{L} r_{\ell}^{2} + \mathcal{O}\!\left(\sum_{\ell=1}^{L} |r_\ell|^3\right) \notag \\
    &= \frac{1}{N^2} \sum_{\ell=1}^{L} E_\ell^2 + \mathcal{O}\!\left(\frac{1}{N^{3}}\right). 
\end{align}
Here, the second equality follows from the Taylor series $e^{x}-x-1 = \frac{1}{2}x^2 + \mathcal{O}(|x|^3)$ evaluated at $x = 2r_\ell$. So the dominant error term is the squared sum of $E_\ell$ given in \cref{lem:integral_simplification}. This concludes the proof of Lemma~\ref{prop:asymptotic_kl}.

\section{Proof of \cref{prop:w2_asymptotic}} \label{appendix:asy_w2_proof}
\begin{proof}
    Substituting $q_{G,t_0}(\mathbf{x}_{t_0}) = \mathcal{N}(\mathbf{0}, \boldsymbol{\Sigma}_{t_0})$ with $\boldsymbol{\Sigma}_{t_0} = \mathbf{U}\mathbf{N}\mathbf{U}^{\top}$, and $\hat{p}_{G,t_0}(\hat{\mathbf{x}}_{t_0}) = \mathcal{N}(\mathbf{0}, \mathbf{U}\mathbf{M}\mathbf{U}^{\top})$ into the closed-form $W_2$ distance formula presented in~\eqref{eq:w2_dist} with $t_0=0$, we obtain:
    \begin{equation} \label{eq:w22_gauss}
        W_2^2\!\left(q_{0, G}, \hat{p}_{0, G}\right) = \mathrm{tr} \left(\mathbf{M} + \mathbf{N} - 2(\mathbf{N}^{1/2}\mathbf{M}\mathbf{N}^{1/2})^{1/2}\right) = \sum_{\ell=1}^{L} (\sqrt{m_\ell} - \sqrt{n_\ell})^2.
    \end{equation}
    
    Recall from~\eqref{eq:log_ratio_expansion} the definition of the logarithmic ratio $r_\ell \coloneqq \frac{1}{2}\log(n_\ell/m_\ell)$. Factoring out $\sqrt{n_\ell}$ and substituting this exponential relation into~\eqref{eq:w22_gauss}, we have the following asymptotic expression
\begin{align}
    W_2^2\!\left(q_{0, G}, \hat{p}_{0, G}\right)
    &= \sum_{\ell=1}^{L} n_\ell (e^{-r_\ell} - 1)^2 \notag \\
    &= \sum_{\ell=1}^{L} n_\ell \left(r_\ell^2 + \mathcal{O}(|r_\ell|^3)\right) \notag \\
    &= \frac{1}{N^2} \sum_{\ell=1}^{L} n_{\ell}E_\ell^2 + \mathcal{O}\!\left(\frac{1}{N^{3}}\right)  \label{eq:w2_proof_step} \\
    &\le \mu_{\max} \left( \frac{1}{N^2} \sum_{\ell=1}^{L} E_\ell^2 \right) + \mathcal{O}\!\left(\frac{1}{N^{3}}\right) \notag \\
    &= \mu_{\max} D_{\mathrm{KL}}\!\left(q_{0, G} \,\|\, \hat{p}_{0, G}\right) + \mathcal{O}\!\left(\frac{1}{N^{3}}\right).
\end{align}
Here, the second equality follows from squaring the Taylor expansion $e^{-x} - 1 = -x + \mathcal{O}(x^2)$ evaluated at $x=r_\ell$, and the third equality results from substituting the result $r_\ell = -E_\ell/N + \mathcal{O}(1/N^2)$ in \cref{prop:log_expansion}. To establish the upper bound, the inequality applies the maximum eigenvalue condition $\mu_\ell \le \mu_{\max}$. Finally, the last equality recovers the asymptotic expansion of the KL divergence derived in \cref{prop:asymptotic_kl}, thus concluding the proof.
\end{proof}

\section{Results in the VE setting} \label{appendix:veresults}
In the VE setting, the forward diffusion process is characterized by a constant scale factor $\alpha_t \equiv 1$ alongside the noise boundaries $\sigma_0 = 0$ and $\sigma_{1} = \sigma_{\max} \gg 1$. We now extend the asymptotic error analysis of the scaled logarithmic eigenvalue ratio $r_\ell$ as introduced in \cref{prop:log_expansion} to this setting.

As detailed in Appendix~\ref{appendix:termi_kl_proof}, initializing the reverse sampling process with $\hat{\mathbf{x}}_{t_N} \sim \mathcal{N}(\mathbf{0}, \sigma_{\max}^2 \mathbf{I})$ yields the corresponding terminal eigenvalue:
\begin{equation}
    m_\ell^{(ve)} = \sigma_{\max}^2 \prod_{j=1}^{N} \left( \frac{\alpha_{t_{j-1}}\alpha_{t_j}\mu_\ell+\sigma_{t_{j-1}}\sigma_{t_j}}{\alpha_{t_j}^2\mu_\ell+\sigma_{t_j}^2} \right)^{\!2}.
\end{equation}
Analogous to \cref{lem:S_ell_expansion}, taking the logarithm leads to the asymptotic form:
\begin{equation} \label{eq:mell_ve}
    \frac{1}{2} \log m_\ell^{(ve)} \coloneqq S_\ell^N + \log \sigma_{\max} = I_\ell + E_\ell h + \mathcal{O}(h^{2}) + \log \sigma_{\max}.
\end{equation}
Using~\eqref{eq:Iellvp}, the integral $I_\ell$ under the VE boundary conditions evaluates to
\begin{equation} \label{eq:Iell_VE_exact}
    I_\ell = \frac{1}{2} \log\!\left( \frac{\alpha_{0}^2\mu_\ell + \sigma_{0}^2}{\alpha_{1}^2\mu_\ell + \sigma_{1}^2} \right) = \frac{1}{2} \log\!\left( \frac{n_\ell}{\mu_\ell + \sigma_{\max}^2} \right).
\end{equation}
In standard DMs, the terminal noise strictly dominates the source covariance spectrum (i.e., $\sigma_{\max}^2 \gg \mu_\ell$). By factoring out $\sigma_{\max}^2$ and applying the Taylor expansion $\log(1+x) = x + \mathcal{O}(x^2)$, we isolate the target logarithmic term $\frac{1}{2}\log n_\ell$ alongside its leading-order residual:
\begin{align} \label{eq:Iell_VE_taylor}
    I_\ell + \log \sigma_{\max} 
    &= \frac{1}{2} \log\!\left( \frac{n_\ell \sigma_{\max}^2}{\mu_\ell + \sigma_{\max}^2} \right) \notag \\
    &= \frac{1}{2} \log n_\ell - \frac{1}{2}\log\!\left( 1 + \frac{\mu_\ell}{\sigma_{\max}^2} \right) \notag \\
    &= \frac{1}{2} \log n_\ell - \frac{\mu_\ell}{2\sigma_{\max}^2} + \mathcal{O}\!\left( \left(\frac{\mu_\ell}{\sigma_{\max}^2}\right)^{\!\!2} \, \right).
\end{align}

Subtracting~\eqref{eq:Iell_VE_taylor} from the asymptotic relation in~\eqref{eq:mell_ve} yields a residual error formulation strictly analogous to \cref{prop:log_expansion}, as:
\begin{align} \label{eq:r_ell_VE_final}
    r_\ell^{(ve)} \coloneqq \frac{1}{2}\log \frac{n_\ell}{m_\ell^{(ve)}} 
    &= I_\ell - S_\ell^N - \frac{\mu_\ell}{2\sigma_{\max}^2} + \mathcal{O}\!\left( \left(\frac{\mu_\ell}{\sigma_{\max}^2}\right)^{\!\!2} \, \right) \notag \\
    &= -E_{\ell} h + \mathcal{O}(h^{2}) - \frac{\mu_\ell}{2\sigma_{\max}^2} + \mathcal{O}\!\left( \left(\frac{\mu_\ell}{\sigma_{\max}^2}\right)^{\!\!2} \, \right).
\end{align}
Following a derivation analogous to~\eqref{eq:KL_E_ell}, we obtain
\begin{align} \label{eq:KL_E_ell_VE}
    D_{\mathrm{KL}}\!\left(q_{G,0} \,\|\, \hat{p}_{G,0}^{(ve)}\right)
    &= \frac{1}{2} \sum_{\ell=1}^{L} \bigl(e^{2r_\ell^{(ve)}} - 2r_\ell^{(ve)} - 1\bigr) \notag \\
    &= \sum_{\ell=1}^{L} \bigl(r_{\ell}^{(ve)}\bigr)^{2} + \mathcal{O}\!\left(\sum_{\ell=1}^{L} \bigl|r_\ell^{(ve)}\bigr|^3\right) \notag \\
    &= \frac{1}{N^2} \sum_{\ell=1}^{L} E_\ell^2
    +\frac{1}{N\sigma_{\max}^2} \sum_{\ell=1}^{L} E_\ell \mu_\ell 
    + \frac{1}{4\sigma_{\max}^4} \sum_{\ell=1}^{L} \mu_\ell^2
    + \mathcal{O}\!\left( \frac{1}{N^3} + \frac{1}{\sigma_{\max}^6} \right).
\end{align}
Notably, the third term represents an error that is independent of the noise schedule. Furthermore, in the practical accelerated sampling regime where $N \ll \sigma_{\max}^2$, the first term dominates the distributional discrepancy. Consequently, the optimization objective remains the minimization of $\sum_{\ell=1}^{L} E_\ell^2$. This result confirms that our asymptotic theoretical framework remains applicable to the VE setting.

\section{Proof of \cref{thm:global_opt_gamma}} \label{appen:proofofthm2}

\begin{lemma} \label{lem:gamma_E_ell_closed_form}
    Under the global cross-eigenmode tangent law schedule $\eta_\gamma(t) = \sqrt{\gamma}\tan\bigl(\frac{\pi}{2} t\bigr)$ for a scalar $\gamma > 0$, the dominant error coefficient $E_{\ell}(\gamma)$ associated with the $\ell$-th eigenvalue $\mu_\ell > 0$ admits a closed-form expression as:
    \begin{equation} \label{eq:E_gamma_appen}
        E_{\ell}(\gamma) = -\frac{\pi^2}{16}\left(\sqrt{\frac{\mu_\ell}{\gamma}}+\sqrt{\frac{\gamma}{\mu_\ell}}\right).
    \end{equation}
\end{lemma}

\begin{proof}
    Substituting the parameterized tangent law schedule into the defining integral of $E_{\ell}$ given in~\eqref{eq:E_ell_app}, we evaluate the discretization error as follows:
    \begin{align}
        E_{\ell}(\gamma)
        &= -\frac{\mu_{\ell}}{2}\int_{0}^{1} \frac{(\alpha\dot{\sigma}-\sigma\dot{\alpha})^{2}} {(\alpha^{2}\mu_{\ell}+\sigma^{2})^{2}} \,\mathrm{d}t \notag \\
        &= -\frac{\mu_{\ell}}{2}\int_{0}^{1} \frac{\dot{\eta}(t)^{2}}{(\mu_\ell+\eta(t)^{2})^{2}} \,\mathrm{d}t \notag \\
        &= - \frac{\mu_{\ell}}{2} \int_0^1 \frac{\left[ \frac{\mathrm{d}}{\mathrm{d}t}\left(\sqrt{\gamma}\tan\bigl(\frac{\pi}{2} t \bigr)\right) \right]^2}{\left(\mu_{\ell}+\gamma\tan^{2}\bigl(\frac{\pi}{2} t \bigr) \right)^{2}} \,\mathrm{d} t \notag \\
        &= -\frac{\mu_{\ell}}{2}\int_0^1 \frac{\gamma \bigl(\frac{\pi}{2}\bigr)^{2}\sec^{4}\bigl(\frac{\pi}{2} t\bigr)}{\left[ \frac{\mu_{\ell}\cos^{2}\bigl(\frac{\pi}{2} t\bigr)+\gamma\sin^{2}\bigl(\frac{\pi}{2} t\bigr)}{\cos^{2}\bigl( \frac{\pi}{2} t \bigr)} \right]^{2}} \,\mathrm{d} t \notag \\
        &= -\frac{\mu_\ell\gamma \pi^2}{8} \int_{0}^{1} \frac{1}{\bigl(\mu_\ell\cos^{2}\bigl(\frac{\pi}{2} t\bigr)+\gamma\sin^{2}\bigl(\frac{\pi}{2} t\bigr)\bigr)^{2}} \,\mathrm{d}t. \label{eq:E1_pre_sub}
    \end{align}
    By applying the change of variables $x = \frac{\pi}{2}t$, the integral in~\eqref{eq:E1_pre_sub} simplifies to:
    \begin{equation} \label{eq:E1_integral_form}
        E_{\ell}(\gamma) = -\frac{\mu_\ell\gamma \pi}{4} \int_{0}^{\frac{\pi}{2}} \frac{1}{\bigl(\mu_\ell + (\gamma-\mu_\ell)\sin^{2}x\bigr)^{2}} \,\mathrm{d}x.
    \end{equation}
    To evaluate~\eqref{eq:E1_integral_form}, we invoke the standard integral reduction formula~\cite[Eq.~2.563.1 and 2.562.1, p.~177]{gradshteyn2014table}:
    \begin{equation}
        \int \frac{\mathrm{d}x}{(a+b\sin^2 x)^2} = \frac{1}{2a(a+b)} \left[ \frac{2a+b}{\sqrt{a(a+b)}} \arctan\left( \sqrt{\frac{a+b}{a}}\tan x \right) + \frac{b\sin x \cos x}{a+b\sin^2 x} \right],
        \label{eq:reduction_step}
    \end{equation}
    which holds for $a > 0$ and $a + b > 0$. Substituting $a=\mu_\ell$ and $b=\gamma - \mu_{\ell}$ into~\eqref{eq:reduction_step}, we evaluate the integral by taking the limit as $x \to (\pi/2)^{-}$:
    \begin{align*}
        E_{\ell}(\gamma)
        &= -\frac{\mu_\ell\gamma\pi}{4} \cdot \frac{1}{2\mu_\ell\gamma} \lim_{x \to (\pi/2)^{-}} \left[ \frac{\mu_\ell+\gamma}{\sqrt{\mu_\ell\gamma}} \arctan\left( \sqrt{\frac{\gamma}{\mu_\ell}}\tan x \right) + \frac{(\gamma-\mu_\ell)\sin x \cos x}{\mu_\ell\cos^2 x + \gamma\sin^2 x} \right] \notag \\
        &= -\frac{\pi}{8} \left[ \frac{\mu_\ell+\gamma}{\sqrt{\mu_\ell\gamma}} \left( \frac{\pi}{2} \right) + 0 \right] \notag \\
        &= -\frac{\pi^2}{16}\left(\sqrt{\frac{\mu_\ell}{\gamma}}+\sqrt{\frac{\gamma}{\mu_\ell}}\right).
    \end{align*}
\end{proof}
    
    Substituting~\eqref{eq:E_gamma_appen} in \cref{lem:gamma_E_ell_closed_form} into the objective function $\mathcal{L}(\gamma) = \sum_{\ell=1}^{L} \left(E_\ell(\gamma)\right)^2$, we expand the squared terms to obtain:
    \begin{equation} \label{eq:vns_gamma_epa}
        \mathcal{L}(\gamma) = \left(\frac{\pi^2}{16}\right)^2 \sum_{\ell=1}^{L} \left( \frac{\mu_\ell}{\gamma} + \frac{\gamma}{\mu_\ell} + 2 \right).
    \end{equation}
    Recognizing the trace identities $\sum_{\ell=1}^L \mu_\ell = \mathrm{tr}(\boldsymbol{\Sigma}_{\mathbf{x}})$ and $\sum_{\ell=1}^L \mu_\ell^{-1} = \mathrm{tr}(\boldsymbol{\Sigma}_{\mathbf{x}}^{-1})$, and omitting the positive scaling factor alongside the additive constant $2L$, minimizing the objective $\mathcal{L}(\gamma)$ is strictly equivalent to minimizing the simplified function $J(\gamma)$ defined as:
    \begin{equation}
        J(\gamma) \coloneqq \gamma^{-1} \mathrm{tr}(\boldsymbol{\Sigma}_{\mathbf{x}}) + \gamma \, \mathrm{tr}(\boldsymbol{\Sigma}_{\mathbf{x}}^{-1}).
    \end{equation}
    The first- and second-order derivatives of $J(\gamma)$ with respect to $\gamma$ are:
    \begin{align}
        J'(\gamma) &= -\gamma^{-2} \mathrm{tr}(\boldsymbol{\Sigma}_{\mathbf{x}}) + \mathrm{tr}(\boldsymbol{\Sigma}_{\mathbf{x}}^{-1}), \\
        J''(\gamma) &= 2 \gamma^{-3} \mathrm{tr}(\boldsymbol{\Sigma}_{\mathbf{x}}).
    \end{align}
    Since the source covariance matrix $\boldsymbol{\Sigma}_{\mathbf{x}}$ is positive definite, its trace is strictly positive. Consequently, $J''(\gamma) > 0$ for all $\gamma > 0$, which inherently guarantees that $J(\gamma)$ (and thus $\mathcal{L}(\gamma)$) is strictly convex. Finally, setting the first derivative $J'(\gamma) = 0$ yields:
    \begin{equation}
        \gamma^2 = \frac{\mathrm{tr}(\boldsymbol{\Sigma}_{\mathbf{x}})}{\mathrm{tr}(\boldsymbol{\Sigma}_{\mathbf{x}}^{-1})},
    \end{equation}
    which naturally leads to the unique closed-form minimizer $\gamma^\star$ presented in~\eqref{eq:gamma_star}, thereby concluding the proof.

\section{Proof of \cref{prop:w2_optimal_gamma}} \label{appen:w2_gamma_proof}
As established in~\eqref{eq:E_gamma_appen} in \cref{lem:gamma_E_ell_closed_form}, the parameterized error coefficient is $E_\ell(\gamma) = -\frac{\pi^2}{16} \left( \sqrt{\frac{\mu_\ell}{\gamma}} + \sqrt{\frac{\gamma}{\mu_\ell}} \right)$. Substituting this into the $W_2$-weighted objective, we expand the squared terms to obtain:
    \begin{align} \label{eq:weighted_error_sum_expanded}
        \mathcal{L}_{W_2}(\gamma) = \sum_{\ell=1}^{L} \mu_{\ell} \left(E_{\ell}(\gamma)\right)^2 
        &= \frac{\pi^4}{256} \sum_{\ell=1}^{L} \left( \frac{\mu_\ell^2}{\gamma} + \gamma + 2\mu_\ell \right) \notag\\
        &= \frac{\pi^4}{256} \left[ \gamma^{-1} \mathrm{tr}(\boldsymbol{\Sigma}_{\mathbf{x}}^2) + L\gamma + 2\mathrm{tr}(\boldsymbol{\Sigma}_{\mathbf{x}}) \right].
    \end{align}
    
    We take the first derivative of $\mathcal{L}_{W_2}(\gamma)$ with respect to $\gamma$ and set it to zero:
    \begin{equation}
        -\gamma^{-2} \mathrm{tr}(\boldsymbol{\Sigma}_{\mathbf{x}}^2) + L = 0.
    \end{equation}
    Solving for $\gamma$ yields the unique global minimizer presented in~\eqref{eq:w2_gamma_star}.
    
    Next, substituting this optimal parameter $\gamma_{W_2}^\star = \sqrt{\mathrm{tr}(\boldsymbol{\Sigma}_{\mathbf{x}}^2) / L}$ back into the expanded objective~\eqref{eq:weighted_error_sum_expanded}, the minimum achievable $W_2$ error evaluates to:
    \begin{align} \label{eq:w2_min_value}
        \mathcal{L}_{W_2}(\gamma_{W_2}^\star) 
        &= \frac{\pi^4}{256} \left[ \sqrt{\frac{L}{\mathrm{tr}(\boldsymbol{\Sigma}_{\mathbf{x}}^2)}} \mathrm{tr}(\boldsymbol{\Sigma}_{\mathbf{x}}^2) + L \sqrt{\frac{\mathrm{tr}(\boldsymbol{\Sigma}_{\mathbf{x}}^2)}{L}} + 2\mathrm{tr}(\boldsymbol{\Sigma}_{\mathbf{x}}) \right] \notag\\
        &= \frac{\pi^4}{128} \left[ \sqrt{L \cdot \mathrm{tr}(\boldsymbol{\Sigma}_{\mathbf{x}}^2)} + \mathrm{tr}(\boldsymbol{\Sigma}_{\mathbf{x}}) \right].
    \end{align}
    
    On the other hand, the absolute theoretical lower bound under the $W_2$ metric is attained when each eigenmode independently follows its optimal profile. Recall from \eqref{eq:E_mu_ell} that the minimum mode-specific squared error is $(E_{\ell}^{\mathrm{(a)}}(\eta_\ell))^2 = \frac{\pi^4}{64}$. Thus, the lower bound evaluates to:
    \begin{equation} \label{eq:w2_lower_bound}
        \mathcal{L}_{W_2,\mathrm{lb}} = \sum_{\ell=1}^{L} \mu_\ell \left(E_{\ell}^{\mathrm{(a)}}(\eta_\ell)\right)^2 = \frac{\pi^4}{64} \mathrm{tr}(\boldsymbol{\Sigma}_{\mathbf{x}}).
    \end{equation}
    
    Finally, subtracting $\mathcal{L}_{W_2,\mathrm{lb}}$ in~\eqref{eq:w2_lower_bound} from the minimum $\mathcal{L}_{W_2}(\gamma_{W_2}^\star)$ in~\eqref{eq:w2_min_value} directly yields the gap $\Delta \mathcal{L}_{W_2}$ defined in~\eqref{eq:w2_performgap}, which concludes the proof.

\section{Proof of \cref{prop:piecewise_sde_kl}} \label{appen:sdediserror}

The proof proceeds in two steps. First, we apply Girsanov's theorem to derive the exact integral form of the KL divergence. Second, we analyze the local discretization error using It\^o calculus to extract the dominant leading-order term. Before proceeding with the formal derivation, we recall several fundamental results from stochastic analysis~\cite{oksendal2003stochastic}.

\begin{lemma}[It\^o Isometry] \label{lem:ito_isometry}
    Let $\mathbf{H}_t \in \mathbb{R}^{n \times m}$ be a stochastic process adapted to the filtration generated by $\mathbf{w}_t$, which is an $m$-dimensional standard Brownian motion. Assuming the square-integrability condition $\mathbb{E} \!\left[ \int_S^T \|\mathbf{H}_t\|_F^2 \,\mathrm{d}t \right] < \infty$ holds, we have
    \begin{equation} \label{eq:ito_isometry_matrix}
        \mathbb{E} \!\left[ \left\| \int_{S}^{T} \mathbf{H}_t \,\mathrm{d}\mathbf{w}_t \right\|^2 \right] = \mathbb{E} \!\left[ \int_{S}^{T} \| \mathbf{H}_t \|_F^2 \,\mathrm{d}t \right].
    \end{equation}
\end{lemma}

\begin{lemma}[It\^o's Formula] \label{thm:general_ito}
    Let $\mathbf{Y}_t \in \mathbb{R}^m$ be an It\^o process governed by $\mathrm{d}\mathbf{Y}_t = \mathbf{r}(t, \mathbf{Y}_t) \mathrm{d}t + g_t \mathrm{d}\mathbf{w}_t$, where $\mathbf{w}_t$ is an $m$-dimensional standard Brownian motion, $\mathbf{r}(t, \mathbf{Y}_t) \in \mathbb{R}^m$ is the drift, and $g_t \in \mathbb{R}$ is a scalar diffusion coefficient. 
    For any vector-valued function $\mathbf{z}(t, \mathbf{x}) \in C^{1,2}([0, \infty) \times \mathbb{R}^m, \mathbb{R}^n)$, the composite process $\mathbf{Z}_t \coloneqq \mathbf{z}(t, \mathbf{Y}_t)$ satisfies~\cite[Theorems 4.1.2 and 4.2.1]{oksendal2003stochastic}:
    \begin{equation} \label{eq:ito_formula_vector}
        \mathrm{d}\mathbf{Z}_t = \left( \frac{\partial \mathbf{z}}{\partial t} + \mathbf{J}_{\mathbf{z}} \, \mathbf{r} + \frac{1}{2} g_t^2 \Delta \mathbf{z} \right) \mathrm{d}t + g_t \mathbf{J}_{\mathbf{z}} \mathrm{d}\mathbf{w}_t, 
    \end{equation}
    where $\mathbf{J}_{\mathbf{z}}(t, \mathbf{Y}_t) \in \mathbb{R}^{n \times m}$ denotes the Jacobian matrix of $\mathbf{z}$ with respect to $\mathbf{x}$, and $\Delta \mathbf{z}(t, \mathbf{Y}_t) \in \mathbb{R}^n$ represents the vector Laplacian.
\end{lemma}

\begin{lemma}[Chain Rule for Radon-Nikodym Derivatives] \label{lem:rn_product}
    Let $\mathbb{P}$ and $\mathbb{Q}$ be two probability measures defined on the same path space $\Omega = \mathcal{C}([0,1], \mathbb{R}^d)$. Assuming that $\mathbb{Q}$ is absolutely continuous with respect to $\mathbb{P}$ (i.e., $\mathbb{Q} \ll \mathbb{P}$), their Radon-Nikodym derivative admits the disintegration~\cite{leonard2013properties}:
    \begin{equation}
        \frac{\mathrm{d}\mathbb{Q}}{\mathrm{d}\mathbb{P}}(\mathbf{x}_{[0,1]}) = \frac{\mathrm{d}\mathbb{Q}_{1}}{\mathrm{d}\mathbb{P}_{1}}(\mathbf{x}_1) \cdot \frac{\mathrm{d}\mathbb{Q}(\cdot \mid \mathbf{x}_1)}{\mathrm{d}\mathbb{P}(\cdot \mid \mathbf{x}_1)}(\mathbf{x}_{[0,1]}),
    \end{equation}
    where $\mathbb{P}_1, \mathbb{Q}_1$ denote the marginal distributions at time $t=1$, and $\mathbb{P}(\cdot \mid \mathbf{x}_1), \mathbb{Q}(\cdot \mid \mathbf{x}_1)$ denote the conditional path measures given the state $\mathbf{x}_1$.
\end{lemma}
\begin{lemma}[Girsanov's Theorem] \label{thm:girsanov}
    Let $\mathbf{Y}_t \in \mathbb{R}^m$ be an It\^o process satisfying $\mathrm{d}\mathbf{Y}_t = \mathbf{r}(t, \mathbf{Y}_t) \mathrm{d}t + g_t \mathrm{d}\mathbf{w}_t, \, t\in[0,T]$ under a probability measure $\mathbb{P}$, where $\mathbf{w}_t$ is an $m$-dimensional standard Brownian motion and $g_t \in \mathbb{R}$ is a scalar diffusion coefficient. Suppose there exist adapted processes $\mathbf{u}(t, \mathbf{Y}_t)$ and $\boldsymbol{\alpha}(t, \mathbf{Y}_t)$ such that $g_t \mathbf{u}(t, \mathbf{Y}_t) = \mathbf{r}(t, \mathbf{Y}_t) - \boldsymbol{\alpha}(t, \mathbf{Y}_t)$. Assuming that the kernel $\mathbf{u}$ satisfies Novikov's condition:
    \begin{equation} \label{eq:novikov}
        \mathbb{E}_{\mathbb{P}} \!\left[ \exp \!\left( \frac{1}{2} \int_{0}^{T} \| \mathbf{u}(s, \mathbf{Y}_s) \|^2 \mathrm{d}s \right) \right] < \infty,
    \end{equation}
    the exponential process $M_t = \exp\!\left( -\int_{0}^{t} \mathbf{u}^\top \mathrm{d}\mathbf{w}_s - \frac{1}{2} \int_{0}^{t} \| \mathbf{u} \|^2 \mathrm{d}s \right)$ is a true martingale. This defines an equivalent probability measure $\mathbb{Q}$ via $\mathrm{d}\mathbb{Q} = M_T \mathrm{d}\mathbb{P}$. Under $\mathbb{Q}$, the shifted process $\tilde{\mathbf{w}}_t = \mathbf{w}_t + \int_{0}^{t} \mathbf{u}(s, \mathbf{Y}_s) \mathrm{d}s$ is a standard Brownian motion, yielding the modified dynamics~\cite[Theorem 8.6.6]{oksendal2003stochastic}:
    \begin{equation} \label{eq:girsanov_dynamics}
        \mathrm{d}\mathbf{Y}_t = \boldsymbol{\alpha}(t, \mathbf{Y}_t) \mathrm{d}t + g_t \mathrm{d}\tilde{\mathbf{w}}_t.
    \end{equation}
\end{lemma}

Now, we begin the proof and first recall the continuous-time reverse SDE and its piecewise approximation counterpart as formulated in \cref{prop:piecewise_sde_kl}:
\begin{subequations} \label{eq:two_sdes_app}
    \begin{align} 
        \mathbb{P}_{\theta}: \quad & \mathrm{d}\hat{\mathbf{x}}_t = \left( f_t \hat{\mathbf{x}}_t - h_t \boldsymbol{\epsilon}_{\theta}(\hat{\mathbf{x}}_{t},t) \right)\mathrm{d}t + g_t \mathrm{d} \bar{\mathbf{w}}_t, \quad \hat{\mathbf{x}}_1 \sim q_1, \label{eq:twosdes1_app} \\
        \hat{\mathbb{P}}_{\theta}: \quad & \mathrm{d}\hat{\mathbf{x}}_t = \left( f_t \hat{\mathbf{x}}_t - h_t \boldsymbol{\epsilon}_{\theta}(\hat{\mathbf{x}}_{t_j},t_j) \right)\mathrm{d}t + g_t \mathrm{d} \bar{\mathbf{w}}_t, \quad \hat{\mathbf{x}}_{1} \sim \mathcal{N}(\mathbf{0}, c^2 \mathbf{I}). \label{eq:twosdes2_app}
    \end{align}
\end{subequations}

\begin{lemma} \label{lem:kl_discretization_gir}
    Suppose that Novikov's condition~\eqref{eq:novikov} is satisfied for the Girsanov kernel defined below, the KL divergence between $\mathbb{P}_{\theta}$ and $\hat{\mathbb{P}}_{\theta}$ admits the decomposition:
    \begin{equation} \label{eq:pathkl_gir}
        D_{\mathrm{KL}}(\mathbb{P}_{\theta} \parallel \hat{\mathbb{P}}_{\theta}) = D_{\mathrm{KL}}(q_1 \parallel \mathcal{N}(\mathbf{0}, c^2 \mathbf{I})) + \frac{1}{2} \sum_{j=1}^{N} \int_{t_{j-1}}^{t_j} \frac{h_t^2}{g_t^2} \, \mathbb{E}_{\mathbb{P}_{\theta}} \!\left[ \| \boldsymbol{\epsilon}_{\theta}(\hat{\mathbf{x}}_t, t) - \boldsymbol{\epsilon}_{\theta}(\hat{\mathbf{x}}_{t_j}, t_j) \|^2 \right] \mathrm{d}t.
    \end{equation}
\end{lemma}
\begin{proof}
    By invoking the chain rule for Radon-Nikodym derivatives (\cref{lem:rn_product}), the total measure discrepancy decouples into contributions from the initial marginal distribution and the conditional path dynamics. Applying Girsanov's theorem (\cref{thm:girsanov}) to the conditional path measure yields the  density ratio:
    \begin{equation}
        \frac{\mathrm{d}\hat{\mathbb{P}}_\theta}{\mathrm{d}\mathbb{P}_\theta}(\hat{\mathbf{x}}_{[0,1]})
        =\frac{\mathcal{N}(\hat{\mathbf{x}}_1;\mathbf{0},c^2\mathbf I)}{q_1(\hat{\mathbf{x}}_1)}\cdot M_1 ,
    \end{equation}
    where the exponential martingale evaluated at the terminal time $t=1$ is given by:
    \begin{equation}
        M_1 = \exp\!\left( -\int_{0}^{1} \mathbf{u}_t^\top \mathrm{d}\bar{\mathbf{w}}_t - \frac{1}{2} \int_{0}^{1} \| \mathbf{u}_t \|^2 \mathrm{d}t \right),
    \end{equation}
    with the piecewise Girsanov kernel explicitly defined as $\mathbf{u}_t = \frac{h_t}{g_t} \bigl( \boldsymbol{\epsilon}_\theta(\hat{\mathbf{x}}_{t_j}, t_j) - \boldsymbol{\epsilon}_\theta(\hat{\mathbf{x}}_t, t) \bigr)$ for $t \in (t_{j-1}, t_j]$.
    
    By definition, the KL divergence between the path measures evaluates to:
    \begin{align} \label{eq:path_kl_formula}
        D_{\mathrm{KL}}(\mathbb{P}_{\theta} \parallel \hat{\mathbb{P}}_{\theta}) 
        &= \mathbb{E}_{\mathbb{P}_{\theta}} \!\left[ -\log \frac{\mathrm{d}\hat{\mathbb{P}}_\theta}{\mathrm{d}\mathbb{P}_\theta}(\hat{\mathbf{x}}_{[0,1]}) \right] \notag \\
        &= \mathbb{E}_{q_1} \!\left[ \log \frac{q_1}{\mathcal{N}(\mathbf{0}, c^2 \mathbf{I})} \right] + \mathbb{E}_{\mathbb{P}_{\theta}} \!\left[ -\log M_1 \right] \notag \\
        &= D_{\mathrm{KL}}(q_1 \parallel \mathcal{N}(\mathbf{0}, c^2 \mathbf{I})) + \mathbb{E}_{\mathbb{P}_{\theta}} \!\left[ \int_{0}^{1} \mathbf{u}_t^\top \mathrm{d}\bar{\mathbf{w}}_t + \frac{1}{2} \int_{0}^{1} \| \mathbf{u}_t \|^2 \mathrm{d}t \right].
    \end{align}
    By Novikov's condition, the Girsanov exponential is a true martingale, and the stochastic integral has zero expectation under $\mathbb P_\theta$. Furthermore, since the integrand is non-negative, we can interchange the time integral and the expectation. Finally, decomposing the global time integral into a summation over the discrete grid intervals yields the exact formulation:
    \begin{equation*}
        D_{\mathrm{KL}}(\mathbb{P}_{\theta} \parallel \hat{\mathbb{P}}_{\theta}) = D_{\mathrm{KL}}(q_1 \parallel \mathcal{N}(\mathbf{0}, c^2 \mathbf{I})) + \frac{1}{2} \sum_{j=1}^{N} \int_{t_{j-1}}^{t_j} \frac{h_t^2}{g_t^2} \, \mathbb{E}_{\mathbb{P}_{\theta}} \!\left[ \| \boldsymbol{\epsilon}_{\theta}(\hat{\mathbf{x}}_t, t) - \boldsymbol{\epsilon}_{\theta}(\hat{\mathbf{x}}_{t_j}, t_j) \|^2 \right] \mathrm{d}t.
    \end{equation*}
    This completes the proof.
\end{proof}

We conduct our analysis in the asymptotic regime where the maximum step size $\delta \coloneqq \max_j(t_j-t_{j-1}) \to 0$. To  quantify the integral in \cref{lem:kl_discretization_gir}, we perform a local stochastic expansion of the integrand. Applying It\^o's formula (\cref{thm:general_ito}) to the neural estimator $\boldsymbol{\epsilon}_{\theta}(\hat{\mathbf{x}}_t,t)$ under $\mathbb{P}_{\theta}$ yields:
\begin{align} \label{eq:ito_phi}
    \mathrm{d}\boldsymbol{\epsilon}_{\theta}(\hat{\mathbf{x}}_t,t) 
    &= \left( \frac{\partial \boldsymbol{\epsilon}_{\theta}}{\partial t} + \mathbf{J}_{\boldsymbol{\epsilon}}(\hat{\mathbf{x}}_t,t) \left( f_t\hat{\mathbf{x}}_t - h_t\boldsymbol{\epsilon}_{\theta}(\hat{\mathbf{x}}_t,t) \right) + \frac{1}{2}g_t^2\Delta_{\mathbf{x}} \boldsymbol{\epsilon}_{\theta}(\hat{\mathbf{x}}_t,t) \right) \mathrm{d}t + g_t\mathbf{J}_{\boldsymbol{\epsilon}}(\hat{\mathbf{x}}_t,t) \mathrm{d}\bar{\mathbf{w}}_t \notag \\
    &\eqqcolon \boldsymbol{\phi}(\hat{\mathbf{x}}_t,t)\mathrm{d}t + g_t\mathbf{J}_{\boldsymbol{\epsilon}}(\hat{\mathbf{x}}_t,t) \mathrm{d}\bar{\mathbf{w}}_t.
\end{align}
Integrating this stochastic differential equation over the interval $[t, t_j]$ yields
\begin{equation}
    \boldsymbol{\epsilon}_{\theta}(\hat{\mathbf{x}}_{t_j}, t_j) - \boldsymbol{\epsilon}_{\theta}(\hat{\mathbf{x}}_t, t) = \int_{t}^{t_j} \boldsymbol{\phi}(\hat{\mathbf{x}}_\tau, \tau) \mathrm{d}\tau + \int_{t}^{t_j} g_\tau \mathbf{J}_{\boldsymbol{\epsilon}}(\hat{\mathbf{x}}_\tau, \tau) \mathrm{d}\bar{\mathbf{w}}_\tau.
\end{equation}
Taking the squared Euclidean norm of both sides and computing the expectation under the path measure $\mathbb{P}_{\theta}$, we decompose the resulting expression into three distinct components. By applying the It\^o isometry (\cref{lem:ito_isometry}) to the stochastic term, we obtain
\begin{align} \label{eq:lemma_proof_diffepi}
    \mathbb{E}_{\mathbb{P}_{\theta}} \!\left[ \| \boldsymbol{\epsilon}_{\theta}(\hat{\mathbf{x}}_{t_j}, t_j) - \boldsymbol{\epsilon}_{\theta}(\hat{\mathbf{x}}_t, t) \|^2 \right] 
    &= \int_{t}^{t_j} g_\tau^2 \, \mathbb{E}_{p_{\theta, \tau}} \!\left[ \| \mathbf{J}_{\boldsymbol{\epsilon}}(\hat{\mathbf{x}}_\tau, \tau) \|_F^2 \right] \mathrm{d}\tau + \mathbb{E}_{\mathbb{P}_{\theta}} \!\left[ \left\| \int_{t}^{t_j} \boldsymbol{\phi}(\hat{\mathbf{x}}_\tau, \tau) \mathrm{d}\tau \right\|^2 \right] \notag \\
    &\quad + 2 \, \mathbb{E}_{\mathbb{P}_{\theta}} \!\left[ \left\langle \int_{t}^{t_j} \boldsymbol{\phi}(\hat{\mathbf{x}}_\tau, \tau) \mathrm{d}\tau, \int_{t}^{t_j} g_\tau \mathbf{J}_{\boldsymbol{\epsilon}}(\hat{\mathbf{x}}_\tau, \tau) \mathrm{d}\bar{\mathbf{w}}_\tau \right\rangle \right].
\end{align}

To control the higher-order residual terms, we assume the scalar function $\varphi(\tau) \coloneqq g_\tau^2 \mathbb{E}_{p_{\theta,\tau}} \!\big[ \|\mathbf{J}_{\boldsymbol{\epsilon}}(\hat{\mathbf{x}}_\tau,\tau)\|_F^2 \big]$ is Lipschitz continuous. Furthermore, we assume the It\^o drift vector $\boldsymbol{\phi}$ has a locally uniformly bounded second moment; i.e., there exists a constant $C_1 > 0$ such that $\sup_{\tau\in[0,1]} \mathbb{E}_{p_{\theta,\tau}} \!\big[ \|\boldsymbol{\phi}(\hat{\mathbf{x}}_\tau,\tau)\|^2 \big] \le C_1 < \infty$.

\begin{lemma} \label{lem:local_truncation}
    Under the above regularity assumptions, for any $t\in(t_{j-1},t_j]$, we have
    \begin{equation} \label{eq:bound_local_expectation}
        \int_{t_{j-1}}^{t_j} \frac{h_t^2}{g_t^2} \, \mathbb{E}_{\mathbb{P}_{\theta}} \!\left[ \| \boldsymbol{\epsilon}_{\theta}(\hat{\mathbf{x}}_t, t) - \boldsymbol{\epsilon}_{\theta}(\hat{\mathbf{x}}_{t_j}, t_j) \|^2 \right] \mathrm{d}t 
        = \frac{1}{2} h_{t_j}^2 \delta_j^2 \, \mathbb{E}_{p_{\theta, t_j}} \!\left[ \| \mathbf{J}_{\boldsymbol{\epsilon}}(\hat{\mathbf{x}}_{t_j}, t_j) \|_F^2 \right] + \mathcal{O}(\delta_j^{5/2}).
    \end{equation}
    where $\delta_j \coloneqq t_j - t_{j-1}$.
\end{lemma}

\begin{proof}
   By the Lipschitz continuity of $\varphi$, for any $\tau\in[t,t_j]$, $\varphi(\tau)=\varphi(t_j)+\mathcal{O}(t_j-\tau)$.
    Integrating this over $[t, t_j]$ gives:
    \begin{equation} \label{eq:expansion_diffusion}
        \int_t^{t_j}\varphi(\tau)\mathrm{d}\tau = \varphi(t_j)(t_j-t) + \mathcal{O}((t_j-t)^2).
    \end{equation}

    For the drift term, applying the Cauchy-Schwarz inequality and leveraging the uniform bound $C_1$ yields:
    \begin{equation} \label{eq:expansion_drift}
        \mathbb{E}_{\mathbb{P}_{\theta}} \!\left[ \left\| \int_t^{t_j} \boldsymbol{\phi}(\hat{\mathbf{x}}_\tau,\tau)\mathrm{d}\tau \right\|^2 \right] \le (t_j-t) \int_t^{t_j} \mathbb{E}_{p_{\theta,\tau}} \!\left[ \|\boldsymbol{\phi}(\hat{\mathbf{x}}_\tau,\tau)\|^2 \right] \mathrm{d}\tau \le C_1 (t_j-t)^2 = \mathcal{O}((t_j-t)^2).
    \end{equation}

    For the cross-correlation term, another application of Cauchy-Schwarz, combined with the asymptotic bounds established in \eqref{eq:expansion_diffusion} and \eqref{eq:expansion_drift}, directly provides:
    \begin{align} \label{eq:cross_term_bound}
        \left| \mathbb{E}_{\mathbb{P}_{\theta}} \!\left[ \left\langle \int_{t}^{t_j} \boldsymbol{\phi}(\hat{\mathbf{x}}_\tau, \tau) \mathrm{d}\tau, \int_{t}^{t_j} g_\tau \mathbf{J}_{\boldsymbol{\epsilon}}(\hat{\mathbf{x}}_\tau, \tau) \mathrm{d}\bar{\mathbf{w}}_\tau \right\rangle \right] \right| 
        &\le \left( \mathbb{E}_{\mathbb{P}_{\theta}} \!\left[ \left\| \int_{t}^{t_j} \boldsymbol{\phi}(\hat{\mathbf{x}}_\tau, \tau) \mathrm{d}\tau \right\|^2 \right] \right)^{\!1/2} \left( \int_{t}^{t_j} \varphi(\tau) \mathrm{d}\tau \right)^{\!1/2} \notag \\
        & = \mathcal{O}((t_j-t)^{3/2}).
    \end{align}
    Finally, substituting \eqref{eq:expansion_diffusion}, \eqref{eq:expansion_drift}, and \eqref{eq:cross_term_bound} back into \eqref{eq:lemma_proof_diffepi}, we have
    \begin{align}
    \int_{t_{j-1}}^{t_j} &\frac{h_t^2}{g_t^2} \, \mathbb{E}_{\mathbb{P}_{\theta}} \!\left[ \| \boldsymbol{\epsilon}_{\theta}(\hat{\mathbf{x}}_t, t) - \boldsymbol{\epsilon}_{\theta}(\hat{\mathbf{x}}_{t_j}, t_j) \|^2 \right] \mathrm{d}t \notag \\
    &= \int_{t_{j-1}}^{t_j} \left( \frac{h_{t_j}^2}{g_{t_j}^2} + \mathcal{O}(t_j - t) \right) \left( g_{t_j}^2 \, \mathbb{E}_{p_{\theta, t_j}} \!\left[ \| \mathbf{J}_{\boldsymbol{\epsilon}}(\hat{\mathbf{x}}_{t_j}, t_j) \|_F^2 \right] (t_j - t) + \mathcal{O}((t_j - t)^{3/2}) \right) \mathrm{d}t \notag \\
    &= \int_{t_{j-1}}^{t_j} \left( h_{t_j}^2 \, \mathbb{E}_{p_{\theta, t_j}} \!\left[ \| \mathbf{J}_{\boldsymbol{\epsilon}}(\hat{\mathbf{x}}_{t_j}, t_j) \|_F^2 \right] (t_j - t) + \mathcal{O}((t_j - t)^{3/2}) \right) \mathrm{d}t \notag \\
    &= \frac{1}{2} h_{t_j}^2 \delta_j^2 \, \mathbb{E}_{p_{\theta, t_j}} \!\left[ \| \mathbf{J}_{\boldsymbol{\epsilon}}(\hat{\mathbf{x}}_{t_j}, t_j) \|_F^2 \right] + \mathcal{O}(\delta_j^{5/2}).
\end{align}
This concludes the proof
\end{proof}

Finally, substituting this result back into the path KL divergence decomposition~\eqref{eq:pathkl_gir} yields
\begin{align}
    D_{\mathrm{KL}}(\mathbb{P}_{\theta} \parallel \hat{\mathbb{P}}_{\theta}) 
    &= D_{\mathrm{KL}}(q_1 \parallel \mathcal{N}(\mathbf{0}, c^2 \mathbf{I})) + \frac{1}{2} \sum_{j=1}^{N} \int_{t_{j-1}}^{t_j} \frac{h_t^2}{g_t^2} \, \mathbb{E}_{\mathbb{P}_{\theta}} \!\left[ \| \boldsymbol{\epsilon}_{\theta}(\hat{\mathbf{x}}_t, t) - \boldsymbol{\epsilon}_{\theta}(\hat{\mathbf{x}}_{t_j}, t_j) \|^2 \right] \mathrm{d}t \notag \\
    &= D_{\mathrm{KL}}(q_1 \parallel \mathcal{N}(\mathbf{0}, c^2 \mathbf{I})) + \frac{1}{2} \sum_{j=1}^{N} \left( \frac{1}{2} h_{t_j}^2 \delta_j^2 \, \mathbb{E}_{p_{\theta, t_j}} \!\left[ \| \mathbf{J}_{\boldsymbol{\epsilon}}(\hat{\mathbf{x}}_{t_j}, t_j) \|_F^2 \right] + \mathcal{O}(\delta_j^{5/2}) \right) \notag \\
    &= D_{\mathrm{KL}}(q_1 \parallel \mathcal{N}(\mathbf{0}, c^2 \mathbf{I})) + \frac{1}{4} \sum_{j=1}^{N} h_{t_j}^2 \delta_j^2 \, \mathbb{E}_{p_{\theta, t_j}} \!\left[ \| \mathbf{J}_{\boldsymbol{\epsilon}}(\hat{\mathbf{x}}_{t_j}, t_j) \|_F^2 \right] + \mathcal{R}(\delta),
\end{align}
Given the total time span $\sum_{j=1}^N \delta_j = 1$, the global remainder simplifies to:
\begin{equation*}
    \mathcal{R}(\delta) \coloneqq \mathcal{O}\bigg( \sum_{j=1}^N \delta_j^{5/2} \bigg) = \mathcal{O}\bigg( \delta^{3/2} \sum_{j=1}^N \delta_j \bigg) = \mathcal{O}(\delta^{3/2}),
\end{equation*}
which directly concludes the derivation of \cref{prop:piecewise_sde_kl}.

\section{Additional Experimental Results} \label{appen:imp_details}

\begin{table*}[htbp]
    \centering
    \scriptsize 
    \setlength{\tabcolsep}{3.5pt} 
    \caption{
    Quantitative comparison of time discretization strategies in terms of FID scores ($\downarrow$).  
    \textbf{Bold} highlights the best result, while \underline{underlining} marks the second best.
    }
    \label{tab:exp_results}
    
    \begin{tabular}{@{}ll ccccccccccccccc@{}}
    \toprule
    \multirow{2}{*}{\textbf{Sampler}} &
    \multirow{2}{*}{\textbf{Step strategy}} &
    \multicolumn{15}{c}{\textbf{NFEs}} \\
    \cmidrule(lr){3-17}
    & & 3 & 4 & 5 & 6 & 7 & 8 & 9 & 10 & 12 & 15 & 20 & 25 & 30 & 50 & 100\\
    \midrule

    \multicolumn{17}{c}{\cellcolor{gray!10}\textit{\textbf{Model: EDM --- Dataset: CIFAR-10}}} \\
    \midrule
    
    \multirow{4}{*}{DPM-Solver++}
    & Uniform-$\lambda$
      & 120.54 & 72.33 & 53.15 & 37.78 & 29.43 & 23.78 & 19.29 & 16.29 & 12.16 & 8.74 & 6.01 & 4.69 & 3.96 & 2.86 & 2.32 \\
    & Xue'24
      & 89.59 & 56.29 & 37.94 & \textbf{26.07} & 21.65 & 18.26 & 18.99 & 15.87 & 11.89 & 8.67 & 6.00 & 4.70 & 3.98 & 2.86 & 2.33 \\
    & \textbf{Ours ($\rho=1.5$)}
      & \textbf{82.52} & \underline{54.95} & \textbf{35.93} & \underline{26.76} & \textbf{20.01} & \textbf{16.31} & \textbf{13.26} & \textbf{11.47} & \textbf{8.70} & \textbf{6.48} & \textbf{4.67} & \textbf{3.83} & \textbf{3.34} & \textbf{2.60} & \textbf{2.24} \\
    & \textbf{Ours ($\rho=2.0$)}
      & \underline{86.27} & \textbf{53.82} & \underline{37.26} & 27.04 & \underline{20.74}& \underline{16.75}& \underline{13.75} &\underline{11.78}  & \underline{9.04} & \underline{6.66} & \underline{4.81} & \underline{3.92} & \underline{3.42} &\underline{2.63} & \underline{2.25} \\
    \addlinespace 
    
    \multirow{4}{*}{UniPC}
    & Uniform-$\lambda$
      & 119.07 & 69.04 & 48.41 & 32.47 & 23.77 & 18.14 & 13.83 & 11.13 & 7.65 & 5.07 & 3.38 & 2.76 & 2.48 & 2.16 & 2.08 \\
    & Xue'24
      & 83.91	& 50.00	& 29.89&	\textbf{18.70}&	14.49	&11.63	&13.23&	10.48&	7.44&	5.02&	3.38&	2.76&	2.49&	2.16&	2.08 \\
    & \textbf{Ours ($\rho=1.5$)}
      & \underline{79.62}&	\underline{49.44}&	\underline{29.50}&	20.16&	\underline{14.11}	&\underline{10.78} &	\textbf{8.37}&	\textbf{6.94}&	\textbf{5.09}&	\textbf{3.76}&	\textbf{2.84}&	\textbf{2.49}&	\textbf{2.32}&	\textbf{2.13}&	2.08\\
    & \textbf{Ours ($\rho=2.0$)}
      & \textbf{78.88}&	\textbf{45.79}&	\textbf{28.27}&	\underline{18.94}&	\textbf{13.84}&	\textbf{10.72}&	\underline{8.67}&	\underline{7.22}&	\underline{5.45}&	\underline{4.08}&	\underline{3.06}&	\underline{2.63}&	\underline{2.42}&	2.16&	2.08 \\

    \midrule
    \multicolumn{17}{c}{\cellcolor{gray!10}\textit{\textbf{Model: EDM --- Dataset: FFHQ-64}}} \\
    \midrule
    
    \multirow{4}{*}{DPM-Solver++}
    & Uniform-$\lambda$
      & 88.98&	63.98&	50.69&	40.48&	32.92&	27.63&	23.69&	20.61&	16.24&	12.28&	8.79&	6.97&	5.90&	4.10&	3.11  \\
    & Xue'24
      & 89.62&	55.70&	41.09&	\textbf{29.80}&	25.77&	22.13&	22.55&	20.33&	15.99&	12.14&	8.75&	6.98&	5.89&	4.10&	3.12 \\
    & \textbf{Ours ($\rho=1.5$)}
      &\textbf{78.71}&	\textbf{50.79}&	\textbf{37.54}&	30.46&	\textbf{24.68}&	\textbf{20.95}&	\textbf{17.64}&	\textbf{15.58}&	\textbf{12.34}&	\textbf{9.46}&	\textbf{6.95}&	\textbf{5.64}&	\textbf{4.89}&	\textbf{3.61}&	\underline{2.91} \\
    & \textbf{Ours ($\rho=2.0$)}
      & \underline{82.31}&	\underline{54.54}&	\underline{38.60}&	\underline{29.96}&	\underline{24.75}&	\underline{21.18}&	\underline{18.07}&	\underline{15.89}&	\underline{12.63}&	\underline{9.67}&	\underline{7.09}&	\underline{5.75}&	\underline{4.95}&	\underline{3.63}&	\textbf{2.90}\\
    \addlinespace
    
    \multirow{4}{*}{UniPC}
    & Uniform-$\lambda$
      & 88.15&	61.85&	47.19&	35.91&	27.71&	22.04&	17.93&	14.85&	10.77&	7.47&	5.02&	3.98&	3.46&	2.78&	2.53 \\
    & Xue'24
      & 83.65&	49.87&	33.37&	\underline{22.35}&	18.38&	14.99&	16.26&	14.25&	10.51&	7.36&	4.99&	3.99&	3.47&	2.79& 2.53 \\
    & \textbf{Ours ($\rho=1.5$)}
      & \underline{75.76}&	\textbf{46.25}&	\underline{31.69}&	23.76&	\underline{18.14}&	\underline{14.41}&	\textbf{11.62}&	\textbf{9.80}&	\textbf{7.32}&	\textbf{5.35}&	\textbf{3.91}&	\textbf{3.29}&	\textbf{2.99}&	\textbf{2.61}&	\underline{2.49} \\
    & \textbf{Ours ($\rho=2.0$)}
      & \textbf{75.49}&	\underline{46.84}&	\textbf{30.18}&	\textbf{21.66}&	\textbf{17.13}&	\textbf{14.05}&	\underline{11.73}&	\underline{10.01}&	\underline{7.68}&	\underline{5.69}&	\underline{4.12}&	\underline{3.42}&	\underline{3.05}&	\textbf{2.61} & \textbf{2.48} \\

    \midrule
    \multicolumn{17}{c}{\cellcolor{gray!10}\textit{\textbf{Model: VP-SDE --- Dataset: CIFAR-10}}} \\
    \midrule

    \multirow{4}{*}{DPM-Solver++}
    & Uniform-$\lambda$
      & 93.66	&71.75	&49.70&	38.16&	30.38&	24.91&	20.93&	17.91&	13.95&	10.39&	7.54&	6.07&	5.29&	3.96&	3.24  \\
    & Xue'24
      & \underline{52.80}&	\underline{38.45}&	\textbf{24.85}&	\textbf{19.13}&	\textbf{16.61}&	18.02&	20.47&	15.11&	12.70&	10.28&	7.51&	6.05&	5.28&	3.95&	3.26 \\
    & \textbf{Ours ($\rho=1.5$)}
      & 53.56	&41.23&	33.83&	25.29&	20.69&	\underline{16.59}&	\underline{14.19}&	\underline{12.25}&	\underline{9.75}&	\underline{7.54}&	\underline{5.77}&	\underline{4.86}&	\underline{4.32}&	\underline{3.50}&	\underline{3.06} \\
    & \textbf{Ours ($\rho=2.0$)}
      & \textbf{51.87}&	\textbf{32.27}&	\underline{28.98}&	\underline{23.09}&	\underline{19.82}&	\textbf{16.28}&	\textbf{14.12}&	\textbf{12.08}&	\textbf{9.57}&	\textbf{7.51}&	\textbf{5.70}&	\textbf{4.79}&	\textbf{4.29}&	\textbf{3.46}&	\textbf{3.03} \\
    \addlinespace
    
    \multirow{4}{*}{UniPC}
    & Uniform-$\lambda$
      & 92.24&	68.91&	45.57&	33.01&	24.71&	19.12&	15.23&	12.42&	9.03&	6.35&	4.56&	3.80&	3.46&	3.01&	2.85 \\
    & Xue'24
      & \underline{49.77}&	\underline{36.07}&	\textbf{21.33}&	\textbf{14.94}&	\textbf{12.39}&	12.21&	14.55&	9.51&	7.98&	6.28&	4.55&	3.80&	3.47&	3.01&	2.85 \\
    & \textbf{Ours ($\rho=1.5$)}
      & 50.91&	38.36&	28.49&	19.75&	14.77&	\underline{11.14}&	\underline{9.00}&	\underline{7.55}&	\textbf{5.76}&	\textbf{4.45}&	\textbf{3.56}&	\underline{3.19}&	\underline{3.04}&	\underline{2.86}&	\underline{2.81} \\
    & \textbf{Ours ($\rho=2.0$)}
      &\textbf{43.55}&	\textbf{28.43}&	\underline{22.24}&	\underline{17.02}&	\underline{13.48}&	\textbf{10.70}&	\textbf{8.89}&	\textbf{7.46}&	\textbf{5.76}&	\underline{4.51}&	\textbf{3.56}&	\textbf{3.17}&	\textbf{3.01}&	\textbf{2.82}&	\textbf{2.80}\\

    \bottomrule
    \end{tabular}
\end{table*}

\bibliographystyle{IEEEtran}
\bibliography{reflong.bib}

\end{document}